\documentclass[useAMS,usenatbib,colorlinks]{mnras}
\usepackage{amsmath}
\usepackage{amssymb}
\usepackage{graphicx}
\usepackage{txfonts}
\usepackage{ae,aecompl}
\title[Grooves and spiral modes]{Spiral eigenmodes triggered by grooves in
  the phase space of disc galaxies}

\author[S. De Rijcke et al.]{S. De Rijcke$^{1}$\thanks{E-mail:
    sven.derijcke@Ugent.be}, I. Voulis$^{1}$\\ $^{1}$Ghent University, Dept. Physics \& Astronomy,
  Krijgslaan 281, S9, B-9000, Ghent, Belgium }

\date{Accepted XXX. Received YYY; in original form ZZZ}

\pubyear{2015}

\begin{document}

\def \aj {Astron. J.}
\def \mnras {Mon. Not. R. Astron. Soc.}
\def \apj {Astrophys. J.}
\def \apjs {Astrophys. J. Suppl.}
\def \apjl {Astrophys. J. Let.}
\def \aap {Astron. \& Astrophys.}
\def \aaps {Astron. \& Astrophys. Suppl. Ser.}
\def \nat {Nature}
\def \pasp {Pub. Astron. Soc. Pac.}
\def \araa {ARA\&A}
\label{firstpage}

\pagerange{\pageref{firstpage}--\pageref{lastpage}} \pubyear{2009}

\maketitle

\begin{abstract}
We use linear perturbation theory to investigate how a groove in the
phase space of a disc galaxy changes the stellar disc's stability
properties. Such a groove is a narrow trough around a fixed angular
momentum from which most stars have been removed, rendering part of
the disc unresponsive to spiral waves. We find that a groove can
dramatically alter a disc's eigenmode spectrum by giving rise to a set
of vigorously growing eigenmodes. These eigenmodes are particular to
the grooved disc and are absent from the original ungrooved disc's
mode spectrum. We discuss the properties and possible origin of the
different families of new modes.

By the very nature of our technique, we prove that a narrow
phase-space groove can be a source of rapidly growing spiral patterns
that are true eigenmodes of the grooved disc and that no non-linear
processes need to be invoked to explain their presence in $N$-body
simulations of disc galaxies. Our results lend support to the idea
that spiral structure can be a recurrent phenomenon, in which one
generation of spiral modes alters a disc galaxy's phase space in such
a way that a following generation of modes is destabilized.
\end{abstract}

\newcommand{\beqn}{\begin{equation}}
\newcommand{\neqn}{ \end{equation}}

\section{Introduction}\label{intro}

Although it is a topic with a venerable history, the study of how disc
galaxies develop their beautiful spiral patterns is far from
finished. Explanations for these patterns range from the large-scale,
quasi-stationary density waves envisaged by \citet{linshu64} to the
amplification of small-scale irregularities in a differentially
rotating stellar disc \citep{goly65,juto66}, which could be caused
e.g. by small density concentrations \citep{ovh13}, via feed-back
cycles \citep{mark77,toomre81}.

An attractive idea for the origin of recurrent spiral patterns as
genuine modes of the stellar disc has been put forward by
\citet{sellwood89} and was further developed by \citet{sellwood91},
using both $N$-body simulations and analytical arguments. These
authors found that a spiral pattern can carve a groove at its outer
Lindblad resonance, or OLR, in a simulated disc's phase space and that
this groove is most likely the cause of the growth of a next
generation of spiral patterns which in turn carve their own
phase-space grooves, etc. This cycle can in principle continue as long
as the disc remains cool enough to support coherent waves  and as
  long as grooves are carved in responsive regions of phase
  space. Each subsequent generation of spirals is radially more
extended than the previous one, transporting angular momentum ever
further away from the galaxy center, in accordance with the second law
of thermodynamics \citep{lyn72}. According to this scenario, each
spiral pattern has a finite lifetime but there are always spirals
present in the stellar disc.

In this paper we further investigate this hypothesis. High-resolution
$N$-body simulations are computationally expensive tools to test how
grooves in different locations of phase space affect a disc's
stability properties. It is, moreover, difficult to proove that a
spiral pattern arrising in a numerical simulation is a true eigenmode
of the disc. We therefore employ {\sc pyStab}, a fast computer code
developed by us that efficiently traces the eigenmodes of a given disc
galaxy model using linear pertubation theory and computes their
properties (density distribution, velocity field, etc.). We present
{\sc pyStab} in section \ref{pystab} and the cored exponential galaxy
disc model whose eigenmodes we will determine in section
\ref{themodel}. In sections \ref{m2modes} and \ref{m4modes}, we
determine the $m=2$ and $m=4$ eigenmode spectra, respectively, of
grooved versions of the disc model and discuss the properties of the
eigenmodes related to the presence of a groove. The effect of the
shape of the phase-space groove is investigated in section
\ref{grooveshape}. We summarize our conclusions in section \ref{conc}.

\section{pyStab:~a stability analysis code} \label{pystab}

We use {\sc pyStab}, a Python computer code, to analyse the stability
of a razor-thin stellar disc with an axisymmetric or spherically
symmetric central bulge and dark-matter halo. To maximize
computational efficiency, {\sc pyStab} relies on {\sc numpy} and {\sc
  scipy}\footnote{http://www.scipy.org/} routines to speed up the pure
Python parts of the code. Moreover, we extended Python with fast C++
modules that interface with Python via the {\sc Boost Python
  Library}\footnote{http://www.boost.org}. These modules in turn employ
routines for minimization, root finding, spline interpolation, and
numerical quadrature from the {\sc GNU Scientific
  Library}\footnote{http://www.gnu.org/software/gsl/}. The routines
for solving linear systems and for matrix eigendecompositions are
taken from the C++ linear algebra library {\sc Armadillo}
\citep{sa10}. The code is controled from a Graphical User Interface
(GUI), implemented in {\sc
  pyQt4}\footnote{http://www.riverbankcomputing.com/software/pyqt/},
and contains a wide variety of numerical checks on the results as well
as plotting options. The mathematical formalism behind this code can
be found in \citet{dury08} and \citet{b9}. For completeness, we
summarize it below.

The halo and bulge are assumed to be dynamically too hot to develop
any instabilities and enter the calculations only via their
contributions to the global gravitational potential. We only consider
the stellar component of the disc and neglect the dynamical influence
of gas and dust. We describe an instability as the superposition of a
time-independent axisymmetric equilibrium configuration and a
perturbation that is sufficiently small to warrant the linearisation
of the Boltzmann equation. The equilibrium configuration is
characterised completely by the global potential $V_0(r)$ and the
distribution function, or DF, $f_0(E,J)$. Here and in the following,
\begin{align}
E & = V_0(r) - \frac{1}{2}\left( v_r^2+v_\theta^2 \right)  \\
J &= r v_\theta
\end{align}
are a star's binding energy and angular momentum, respectively. We use
polar coordinates $(r,\theta)$ in the stellar disc, with associated
radial and tangential velocity components $(v_r,v_\theta)$.

A general perturbing potential can be expanded in a series of normal
modes of the form
\begin{equation}
V_{\rm pert}(r,\theta , t)=V_{\rm pert}(r)e^{i(m\theta -\omega t)}, \label{Vp}
\end{equation}
with multiplicity $m$,  pattern speed $\Re(\omega)/m$,
and  growth rate $\Im(\omega)$, that, owing to the linearity of the
relevant equations, can be studied independently from each other. We
write the response of the DF to a perturbation as:
\begin{equation}
f(r,\theta,v_r,v_{\theta},t)=f_0(E,J)+f_{\rm resp}(r,\theta,v_r,v_{\theta},t).
\end{equation}
The evolution of the perturbed part of the DF is
calculated using the linearised collisionless Boltzmann equation:
\begin{equation}\label{v1}
\left. \frac{Df_{\rm resp}}{Dt} \right|_0 = -\frac{\partial
  f_0}{\partial \vec{v}}.\vec{\nabla} V_{\rm pert}.
\end{equation}
The left-hand side of eq. (\ref{v1}) is the total time derivative of
$f_{\rm resp}$ along an orbit in the unperturbed potential $V_0$. If
we integrate eq. (\ref{v1}) along an unperturbed orbit, we immediately
obtain the response of the DF to the perturbing
potential given by eq. (\ref{Vp}):
\begin{align} \label{v2}
f_{\rm resp}(r,\theta,v_r,v_{\theta},t) &= \frac{\partial f_0}{\partial
  E}V_{\rm pert}(r,\theta,t) + i \left[ \omega\frac{\partial
  f_0}{\partial E}-m\frac{\partial f_0}{\partial J} \right] \nonumber \\
& \hspace*{-4em} 
\times 
\int_{-\infty}^{t}
V_{\rm pert}(r(t'))e^{i(m\theta(t')-\omega t')}{\rm d}t'.
\end{align}
The integral in eq. (\ref{v2}) converges if the perturbation
disappears for $t \rightarrow -\infty$ and is growing sufficiently
fast in time. Changing variables such that
\begin{align}
t' &= t + t'' \nonumber \\
\theta(t') &= \theta + \theta''(t''),
\end{align}
this expression can be brought in the same harmonic form as the
potential perturbation
\begin{equation}
f_{\rm resp}(r,\theta,v_r,v_{\theta},t) = f_{\rm
  resp}(r,v_r,v_{\theta}) e^{i(m\theta -\omega t)} 
\end{equation}
with
\begin{align}
f_{\rm resp}(r,v_r,v_{\theta}) &=\frac{\partial f_0}{\partial E}
V_{\rm pert}(r) + i \left[ \omega \frac{\partial F_0}{\partial E} + m
\frac{\partial F_0}{\partial J} \right]\nonumber \\
& \hspace*{-3em} \times \int_{-\infty}^0 V_{\rm
  pert}(r(t'')){ e}^{i(m\theta''(t'')-\omega t'')}dt''.
\end{align}

Along an unperturbed orbit, the radial coordinate $r$ is a periodic
function of time with angular frequency $\omega_r$, just like $v_r$ en
$v_{\theta}$. $\theta(t)$ is the superposition of a periodic function
$\theta_p(t)$ and a uniform drift, $\omega_{\theta}t$:
\begin{equation}
\theta(t'')=\omega_{\theta}t''+\theta_p(t'').
\end{equation}
Since $\theta''(0)=0$, it follows that $\theta_p(0)=0$. We separate
the part of the integrand in eq. (\ref{v2}) that is periodic with
frequency $\omega_r$ from the aperiodic part and expand it in a
Fourier series,
\begin{equation} 
V_{\rm pert}(r(t''))e^{im\theta_p(t'')} =
\sum_{l=-\infty}^{\infty}I_le^{il\omega_rt''},\label{four}
\end{equation}
with purely real Fourier coefficients $I_l$. 

Instead of using $E$ and $J$, orbits in the unperturbed potential are
catalogued by their apocentre and pericentre distances, denoted by
$r_{\rm apo}$ and $r_{\rm peri}$, respectively. The sense of rotation
is indicated by the sign of $r_{\rm peri}$. For each orbit in a $300
\times 300$ grid in $(r_{\rm peri},r_{\rm apo})$-space, each passing
through its apocenter at $t=0$, we store $\omega_r$ and
$\omega_{\theta}$ and tabulate $t$ and $\theta_p$ as a function of
radius. These two offsets in time and azimuth are necessary if one
wants to compute the response DF for an orbit that doesn't pass
through its apocenter at $t=0$. Thus, the method samples phase space
on a total of 44,850 grid points. We have tested the numerical
convergence of the method in terms of orbital phase-space coverage by
also using $200 \times 200$, and $400 \times 400$ grids in $(r_{\rm
  peri},r_{\rm apo})$-space. While there was still a noticeable
difference between the mode frequencies (see below) when going from a
$200 \times 200$ to a $300 \times 300$ grid, this difference was
negligible when comparing the $400 \times 400$ and $300 \times 300$
grids. In the end, we settled for a $300 \times 300$ grid in $(r_{\rm
  peri},r_{\rm apo})$-space. We checked that this grid offers
sufficient resolution for all models presented in this paper.

With
\begin{align}
\int_{-\infty}^0 V_{\rm pert}(r(t'')){ e}^{i(m\theta''(t'')-\omega
  t'')}dt'' &= \nonumber \\ & \hspace*{-7em} -i \sum_l
\frac{I_l}{m\omega_\theta + l\omega_r -\omega}, \nonumber \\ V_{\rm
  pert}(r) = V_{\rm pert}(r(t''=0)) &= \sum_l I_l,
\end{align}
the response of the DF to the perturbation now
assumes the following concise form:
\begin{eqnarray}
f_{\rm resp}(r,v_r,v_\theta) &=& 
 \nonumber \\ && \hspace{-9em} \sum_l
I_l \left[ (l\omega_r+m\omega_{\theta})\frac{\partial f_0}{\partial
E}-m\frac{\partial f_0}{\partial
J}\right]\frac{e^{i(l\omega_rt(r)-m\theta_p(r))}}{l\omega_r+m\omega_{\theta}-\omega}.
\end{eqnarray}
From the response DF it is in principle possible to compute the
response density, $\rho_{\rm resp}(r,\theta,t)$, and, via the Poisson
equation, the response potential, $V_{\rm resp}(r,\theta,t)$.

\begin{figure}
\includegraphics[trim=0 10 0 0,clip,width=0.48\textwidth]{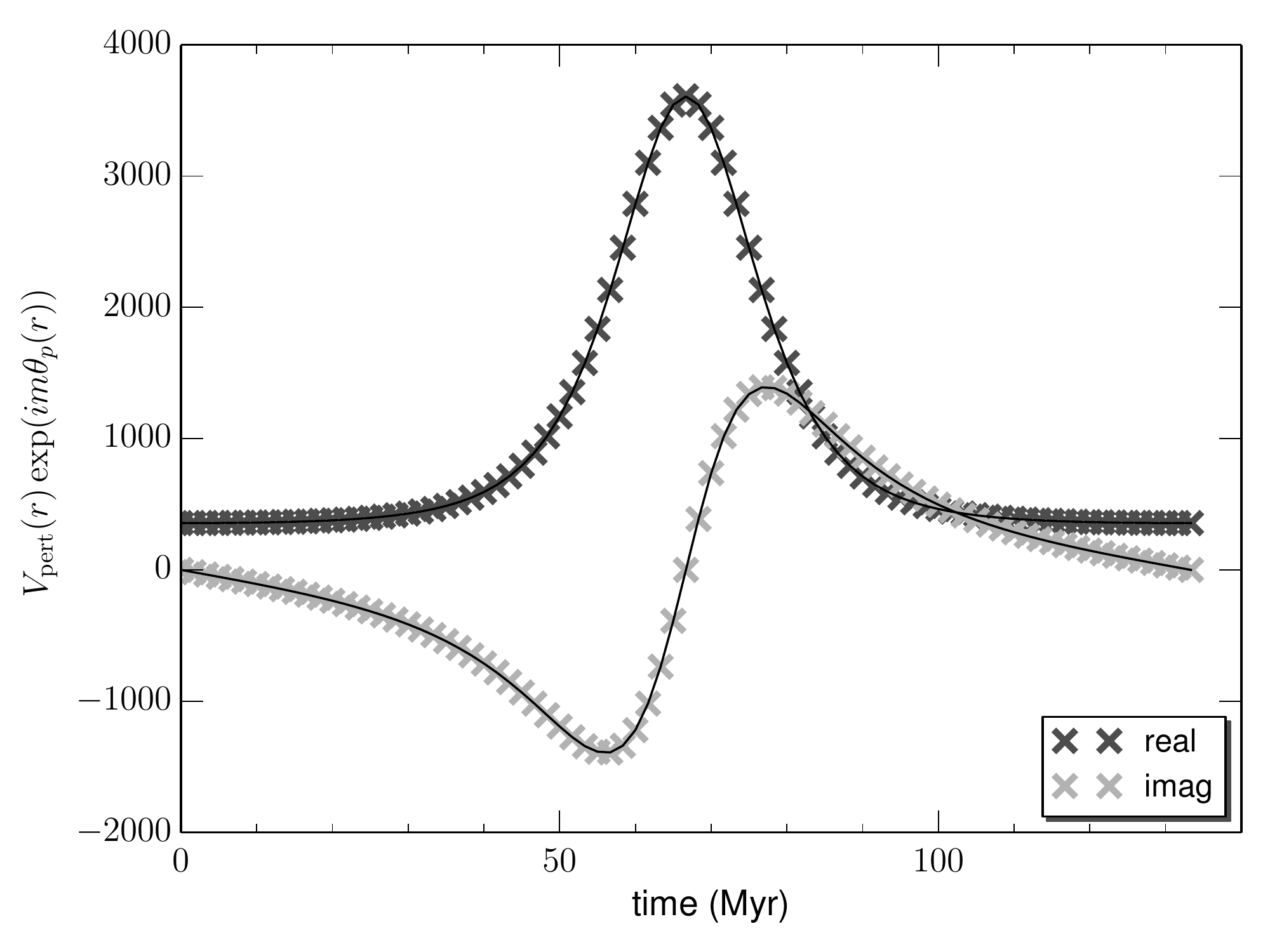}
\caption{The actual (black lines) and Fourier-reconstructed (grey
  crosses) behaviour of the real and imaginary parts of the
  $r$-periodic part of a typical potential basis function $V_l$ over
  one radial oscillation of an orbit.
 \label{fig:Vptp.pdf}}
\end{figure}
In order to cast the search for eigenmodes in the form of a matrix
eigenvalue problem \citep{kalnajs77}, we employ a family of basis
density-potential pairs, $\rho_l(r)$ and $V_l(r)$. We adopt a basis
set of 24 density basis functions of the form \beqn \rho_l(r) =
r^{m-1}\rho_0(r)\exp\left( -\frac{1}{2} \left( \frac{r-r_l}{\sigma_l}
\right)^2 \right) \label{basefunc} \neqn where the average radii $r_l$ cover the
complete stellar disc and are evenly spaced on a logarithmic scale so
the resolution is highest in the inner regions of the disc. The widths
$\sigma_l$ are automatically chosen such that the consecutive basis
functions are unresolved according to the Rayleigh criterion. The
radial part of any perturbation can be expanded in these basis
functions:
\begin{equation}
V_{\rm pert}(r) = \sum_l a_l V_l(r). \label{genVp}
\end{equation}
We denote the response to the perturbation $V_l$ with pattern
frequency $\omega$ by $\rho_{l,\rm resp}(r,\omega)$. The Fourier
coefficients of the expansion (\ref{four}), for each potential basis
function as perturbing potential, and for all orbits in the $(r_{\rm
  peri},r_{\rm apo})$-grid, are computed and stored. The Fourier
expansion is performed from order $l=-40$ up to $l=+40$. In
Fig. \ref{fig:Vptp.pdf}, we show the good agreement between the actual
(black lines) and the Fourier-reconstructed (grey crosses)
behaviour of the $r$-periodic part of such a perturbing potential over
one radial oscillation of a typical orbit. Only the most extreme
radial orbits suffer from the Gibbs phenomenon, inherent to using a
finite Fourier series to reconstruct a sharply varying
function. However, such orbits will be virtually free from stars in
most realistic disc galaxy models.

Expanding the response $\rho_{l,\rm resp}(r,\omega)$ in the basis
functions with coefficients $C_{kl}(\omega)$ yields
\begin{equation}
\rho_{l,\rm resp}(r,\omega) = \sum_k C_{kl}(\omega) \rho_k(r).
\end{equation}
The coefficients $C_{kl}(\omega)$ can easily be obtained via a
least-squares fit on a grid of $r$-values. The response to a general
perturbation (\ref{genVp}) can then be written as
\begin{align}
\rho_{\rm resp}(r,\omega) &= \sum_l a_l \rho_{l,\rm resp}(r,\omega) \nonumber \\
&= \sum_l a_l \sum_k C_{kl}(\omega) \rho_k(r).
\end{align}
Likewise,
\begin{equation}
V_{\rm resp}(r,\omega) = \sum_l a_l \sum_k C_{kl}(\omega) V_k(r). \label{respV}
\end{equation}
For an eigenmode, $V_{\rm resp} = V_{\rm pert}$, and \beqn a_k =
\sum_l C_{kl}(\omega) a_l \,\, \rightarrow \,\, A = C(\omega)A.  \neqn
In other words, the matrix $C$ has a unity eigenvalue for an eigenmode
and the corresponding eigenvector $A$ yields the expansion of the
response density in terms of the basis functions.

Obviously, we need to be able to efficiently compute $\rho_{l,\rm
  resp}(r,\omega)$ for a variety of $\omega$ values. Computing a
response density implies a computationally very expensive double
integral of the response DF over velocity space. However, as shown by
\citet{b9}, the response density $\rho_{l,\rm resp}(r)$ can be written
as a Hilbert transform,
\begin{equation}
\rho_{l, \rm resp}(r) = \int \frac{W_l(r,p)}{p-\omega}dp,
\end{equation}
where the $\omega$-independent functions $W_l(r,p)$ can be
pre-computed from the response DF and stored for a grid of $r$ and $p$
values. Thus, the response densities $\rho_{l,\rm resp}(r,\omega)$,
and hence the matrix $C(\omega)$, can be calculated efficiently for
different values for $\omega$. Thus, {\tt pyStab} is capable of
computing the eigenmode spectrum of any disc galaxy model.

\section{The cored exponential disc model} \label{themodel}

\begin{figure}
\includegraphics[trim=10 10 10 10,clip,width=0.49\textwidth]{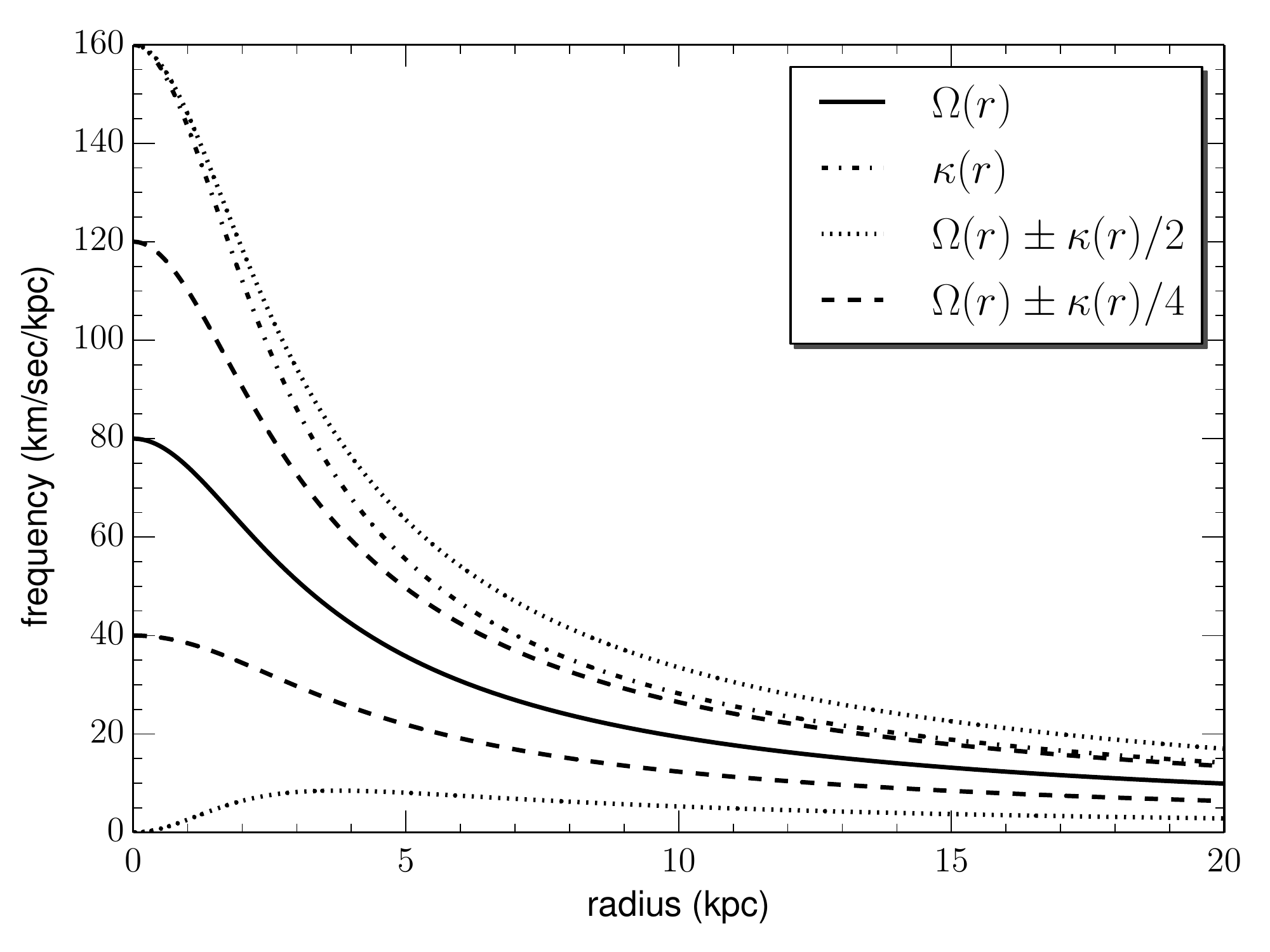}
\caption{The angular velocity, $\Omega(r)$, epicyclic frequency,
  $\kappa(r)$, and $m=2$ and $m=4$ Lindblad frequencies, $\Omega(r)
  \pm \kappa(r)/m$, of the cored exponential disc model with a
  logarithmic potential.
 \label{fig:freqsO.pdf}}
\end{figure}

\begin{figure}
\includegraphics[trim=50 10 0 10,clip,width=0.49\textwidth]{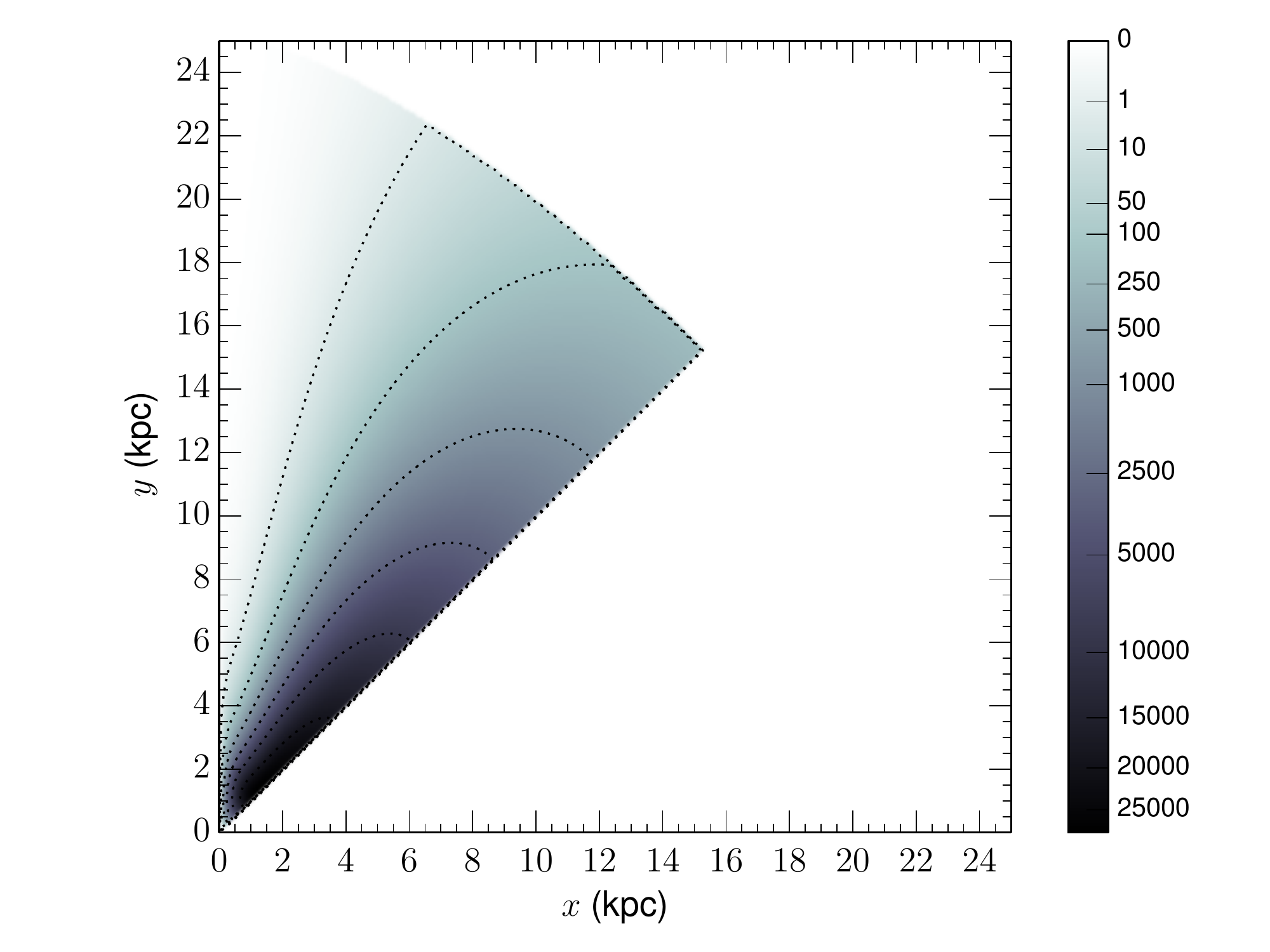}
\includegraphics[trim=50 0 60 1,clip,width=0.49\textwidth]{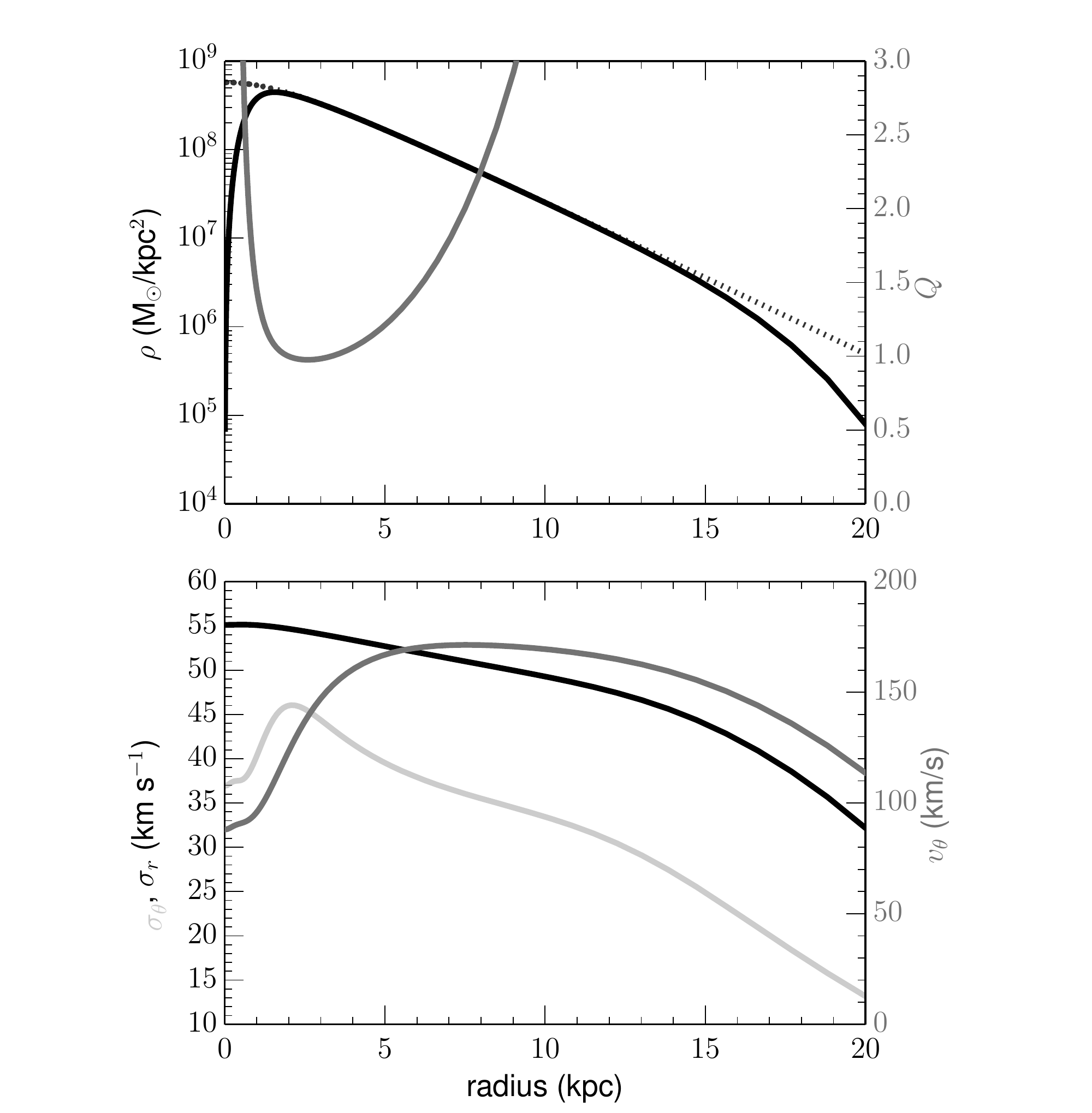}
\caption{Top panel:~the distribution function of the cored exponential
  disc model with a central cutout, shown in turning-point space. The
  colorbar indicates the phase-space density expressed in
  $M_\odot\,\text{kpc}^{-2}\,(\text{km}\,\text{s}^{-1})^{-2}$. Middle
  panel:~the stellar density, $\rho$ (black curve, left axis), and the
  Toomre $Q$-parameter (dark grey curve, right axis). The dotted black
  curve traces the stellar density (\ref{dens0}), without inner cutout
  and outer tapering. Bottom panel:~the radial velocity dispersion,
  $\sigma_r$ (black curve, left axis), tangential velocity dispersion,
  $\sigma_\theta$ (light grey curve, left axis), and the mean rotation
  velocity, $v_\theta$ (dark grey curve, right axis) of the cored
  exponential disc model with an inner cutout.
 \label{fig:DF0.pdf}}
\end{figure}

\begin{figure}
\includegraphics[trim=20 10 0 0,clip,width=0.49\textwidth]{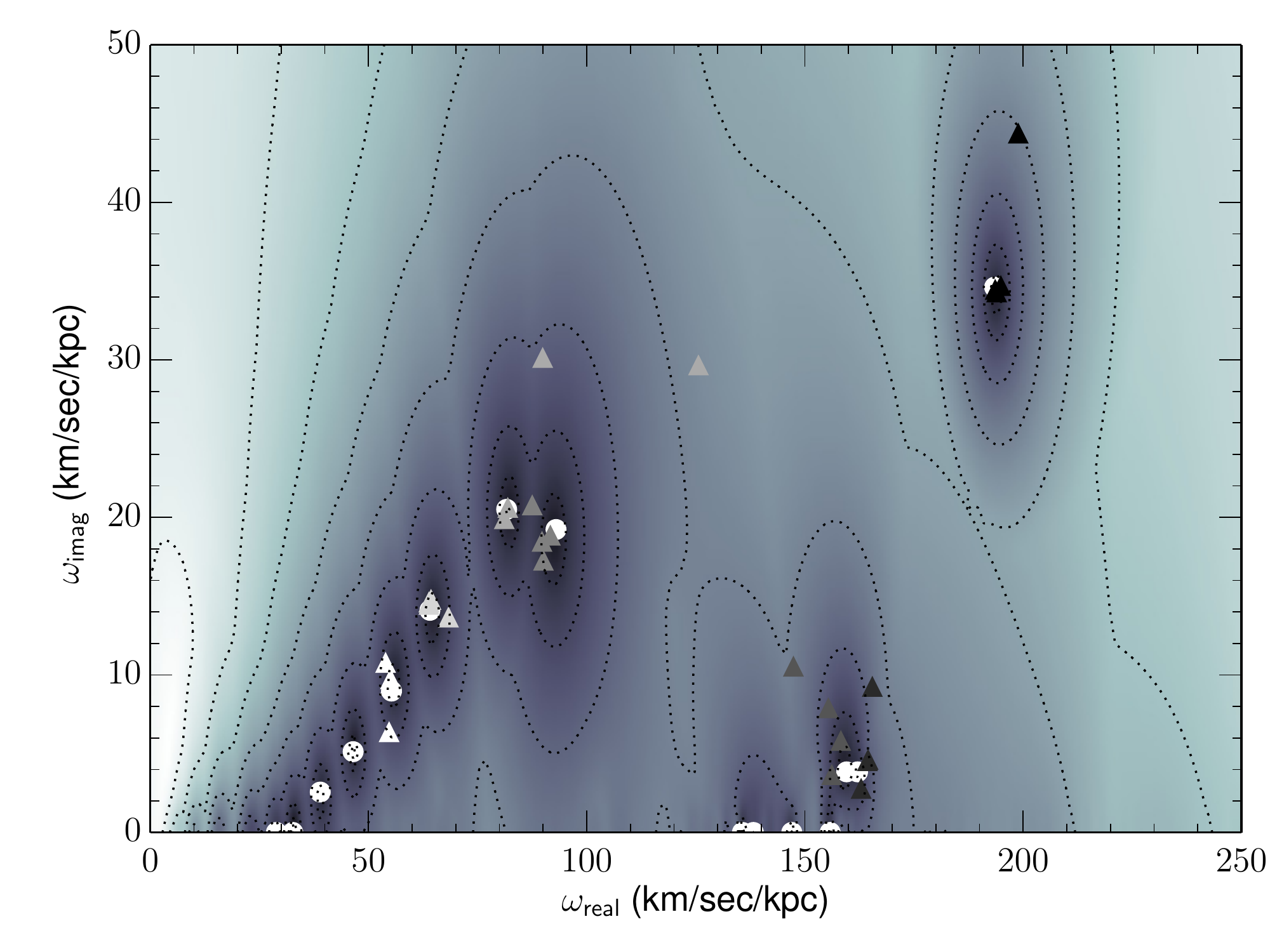}
\caption{The mode spectrum of the cored exponential disc model
  \citep{jalali05} in the complex frequency plane. The white dots
  indicate the location of the eigenmodes, in the centers of the
  dark-hued regions, as found with {\sc pyStab}. The triangular data
  points mark the position of the fastest rotating eigenmodes as found
  by \citet{omurkanov14} using different methods. Dots with the same
  tint are different frequency estimates for the same eigenmode.
 \label{fig:jalali_base.pdf}}
\end{figure}
The disc galaxy model introduced by \citet{jalali05} lives in a
spherically symmetric soft-centered logarithmic binding potential of
the form \beqn V_0(r) = -\frac{v_0^2}{2} \ln \left( 1+
\frac{r^2}{r_c^2} \right). \label{pot0}\neqn Here, $v_0$ is the
asymptotic velocity reached in the flat part of the rotation curve and
$r_c$ is a scale-length. The angular velocity, $\Omega(r)$, epicyclic
frequency, $\kappa(r)$, and $m=2$ and $m=4$ Lindblad frequencies,
$\Omega(r) \pm \kappa(r)/m$, are shown as a function of radius in
Fig. \ref{fig:freqsO.pdf}. Since $\Omega(r)$ does not diverge for zero
radius, it is perfectly possible for modes to have no corotation
resonance, or CR.

The quasi-exponential stellar surface density of the razor-thin
responsive disc is \beqn \rho_0(r) = \rho_s \exp \left( -\lambda\sqrt{
  1+\frac{r^2}{r_c^2} } \right) \label{dens0} \neqn with $\rho_s =
e^\lambda \rho_0(0)$ and $\lambda=r_c/r_D$ with $r_D$ the scale-length
of the exponential disc. The DF that self-consistently generates the
surface density (\ref{dens0}) within the binding potential
(\ref{pot0}) is given by
\begin{align} 
f_0(E,J) &= \rho_s \sum_{n=0}^N
\left( \begin{array}{c} N \\ n \end{array} \right) \left(
\frac{J}{r_c} \right)^{2n} g_n(E), \hspace*{2em}  J>0 \nonumber \\
&=0 \hspace*{13.25em} J \le 0 \label{DF0}
\end{align}
with
\begin{align}
g_n(E) &= \frac{1}{2^n \sqrt{\pi} \Gamma\left( n+\frac{1}{2} \right)} 
\times \nonumber \\
& \hspace*{4em} \frac{d^{n+1}}{dE^{n+1}} \left[ \exp\left( 2N\frac{E}{v_0^2}  -\lambda\exp\left( -\frac{E}{v_0^2} \right) \right) \right].
\end{align}

\begin{figure*}
\includegraphics[trim=40 5 90 1,clip,width=0.33\textwidth]{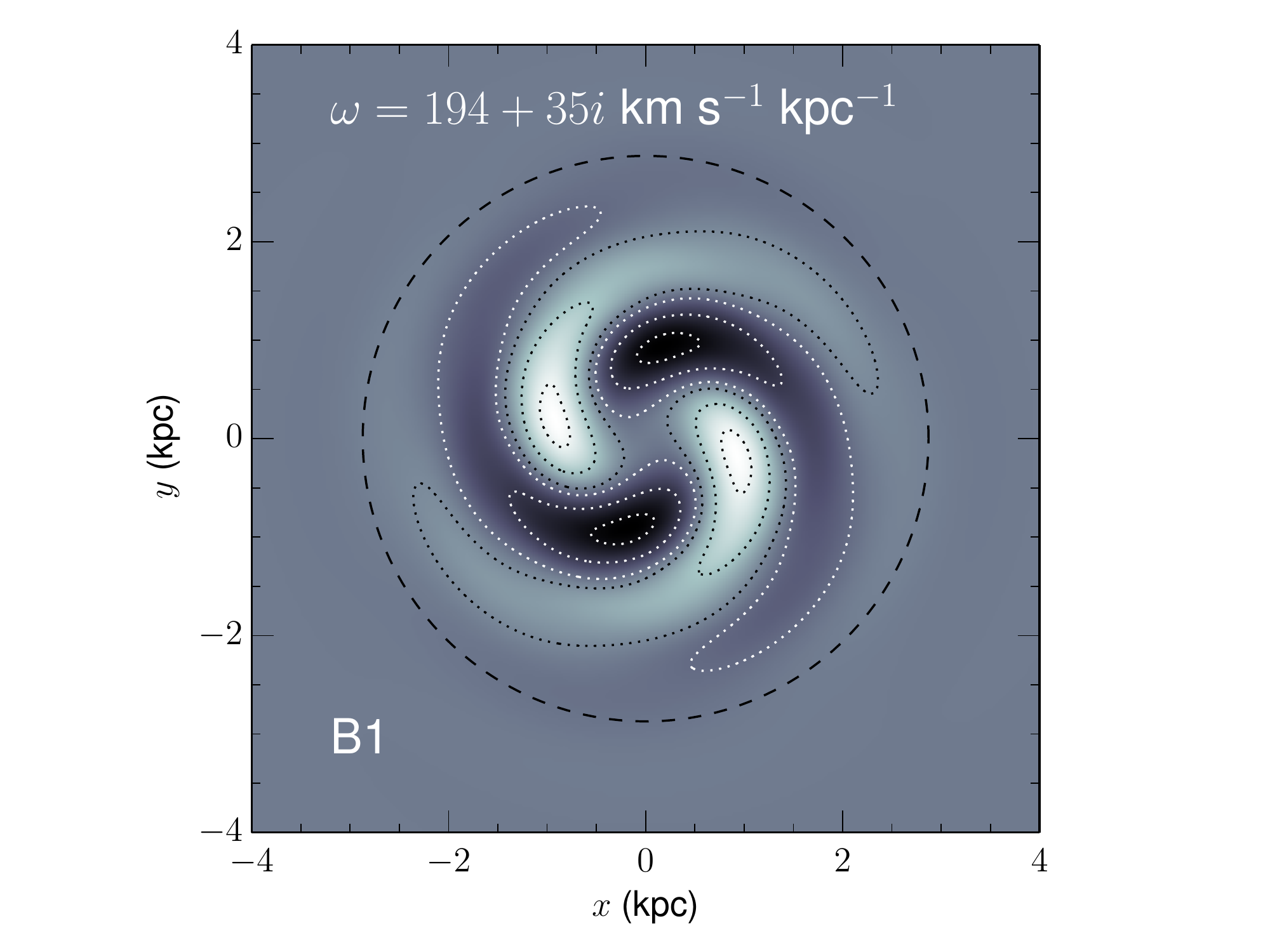}
\includegraphics[trim=40 5 90 1,clip,width=0.33\textwidth]{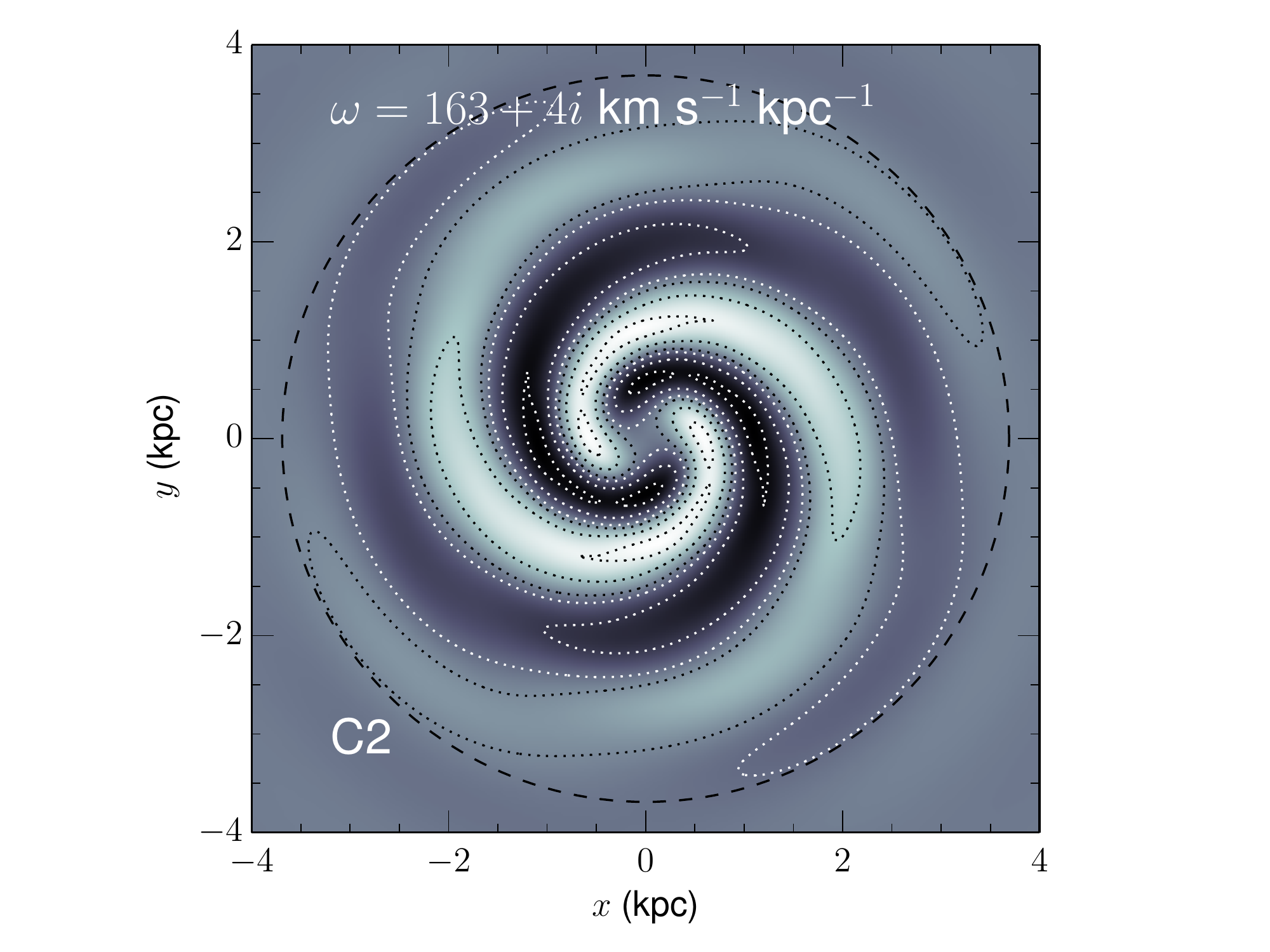}
\includegraphics[trim=40 5 90 1,clip,width=0.33\textwidth]{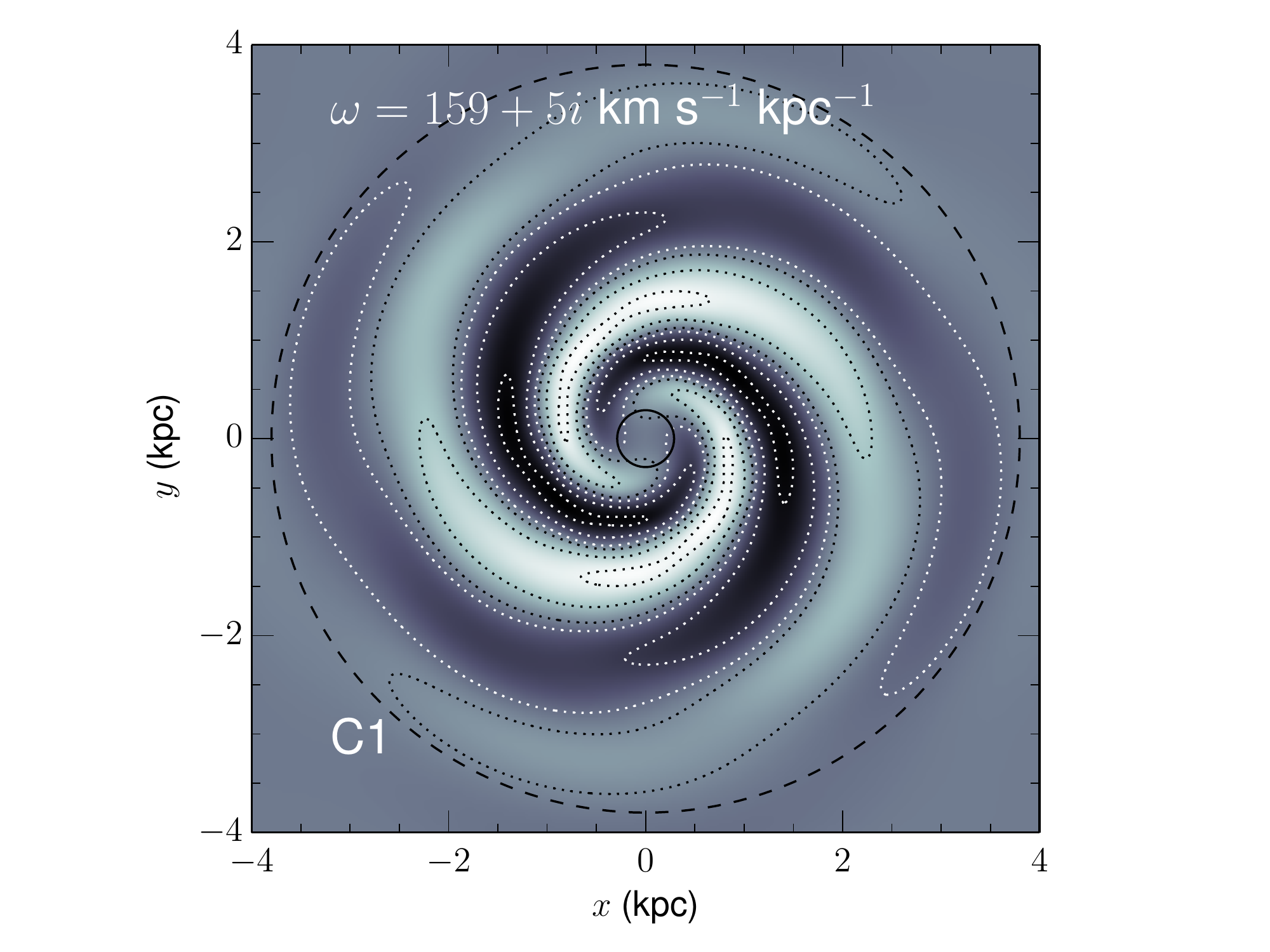}
\includegraphics[trim=40 5 90 1,clip,width=0.325\textwidth]{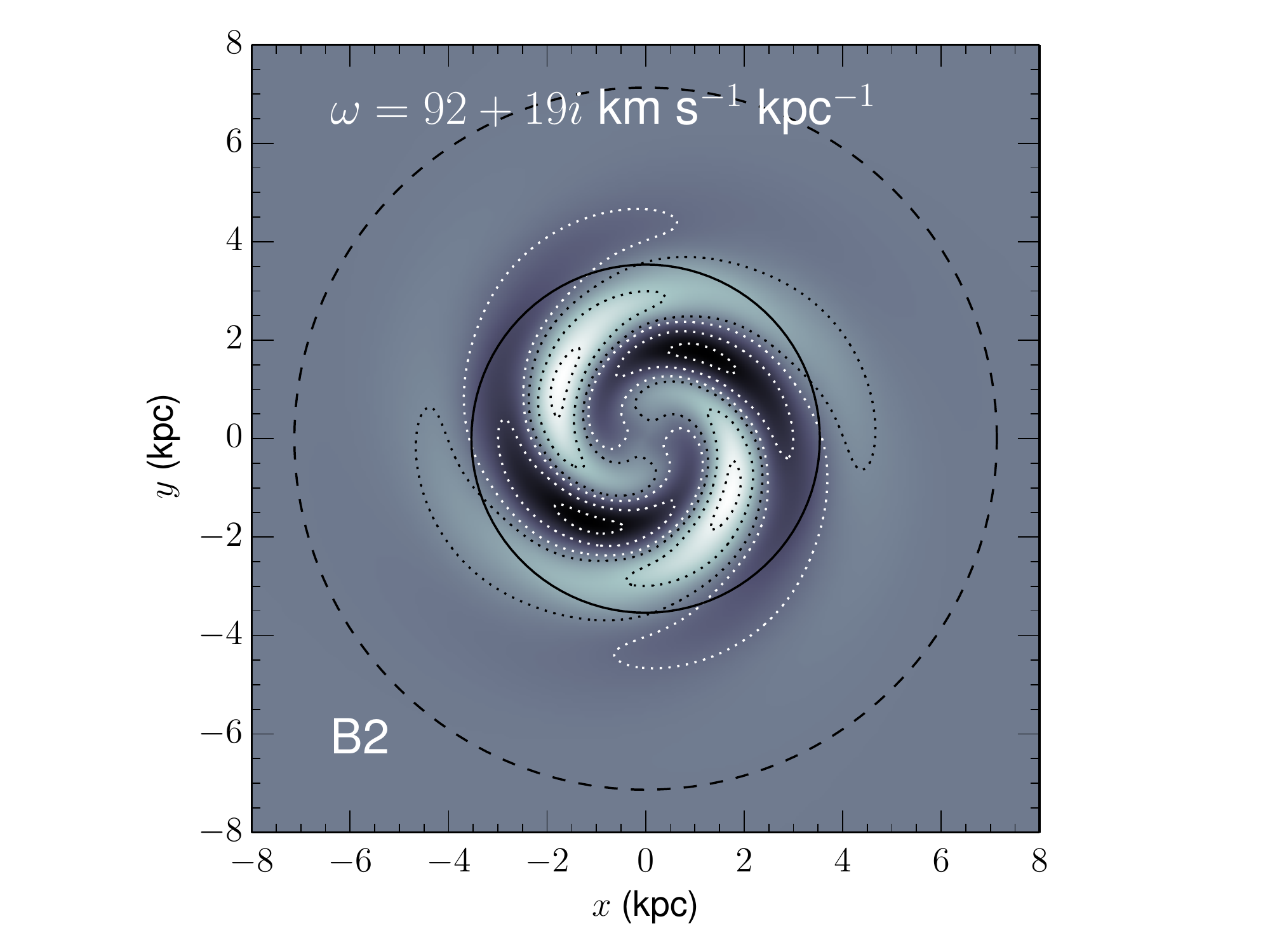}
\includegraphics[trim=40 5 90 1,clip,width=0.325\textwidth]{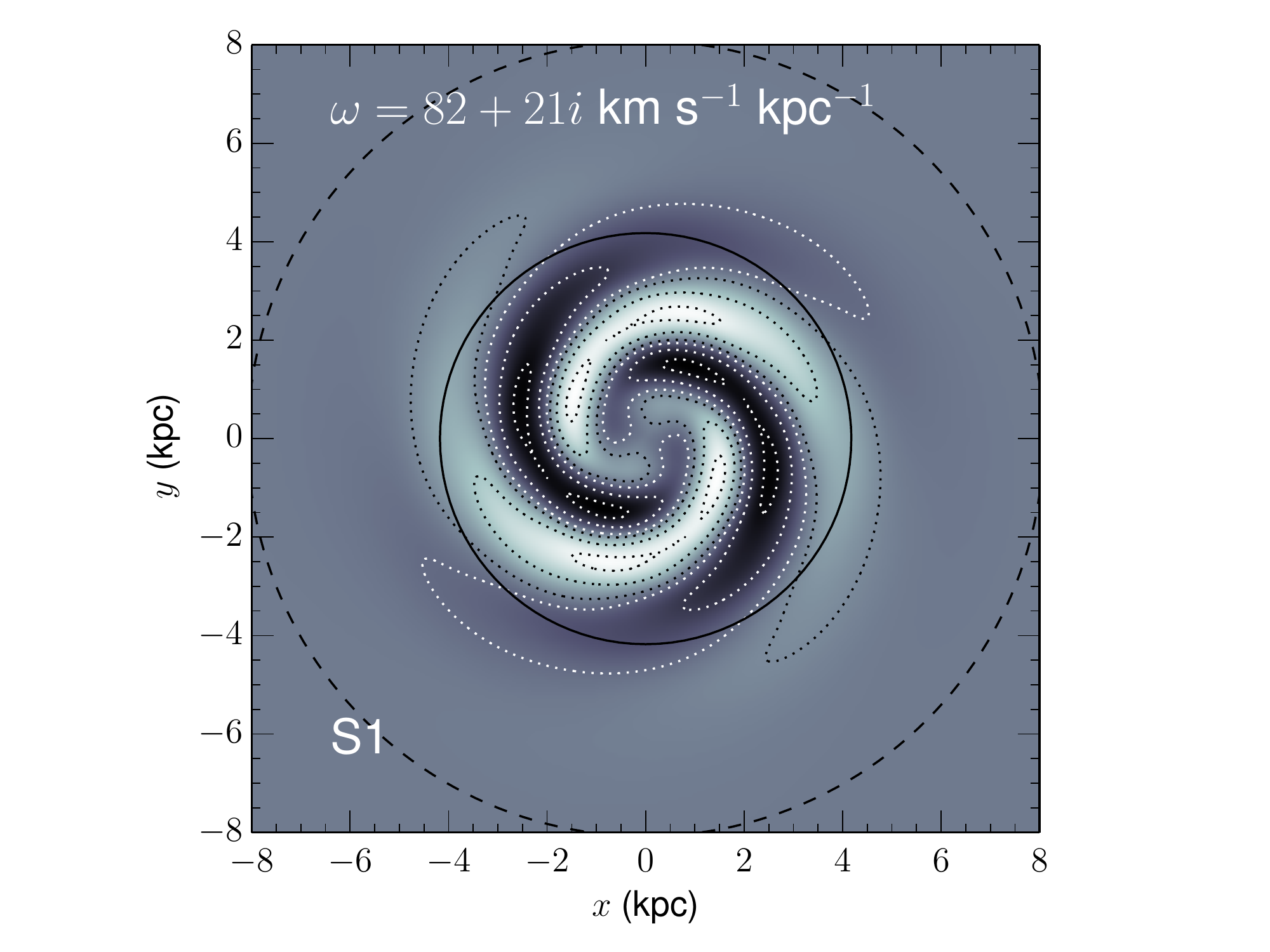}
\includegraphics[trim=40 5 90 1,clip,width=0.325\textwidth]{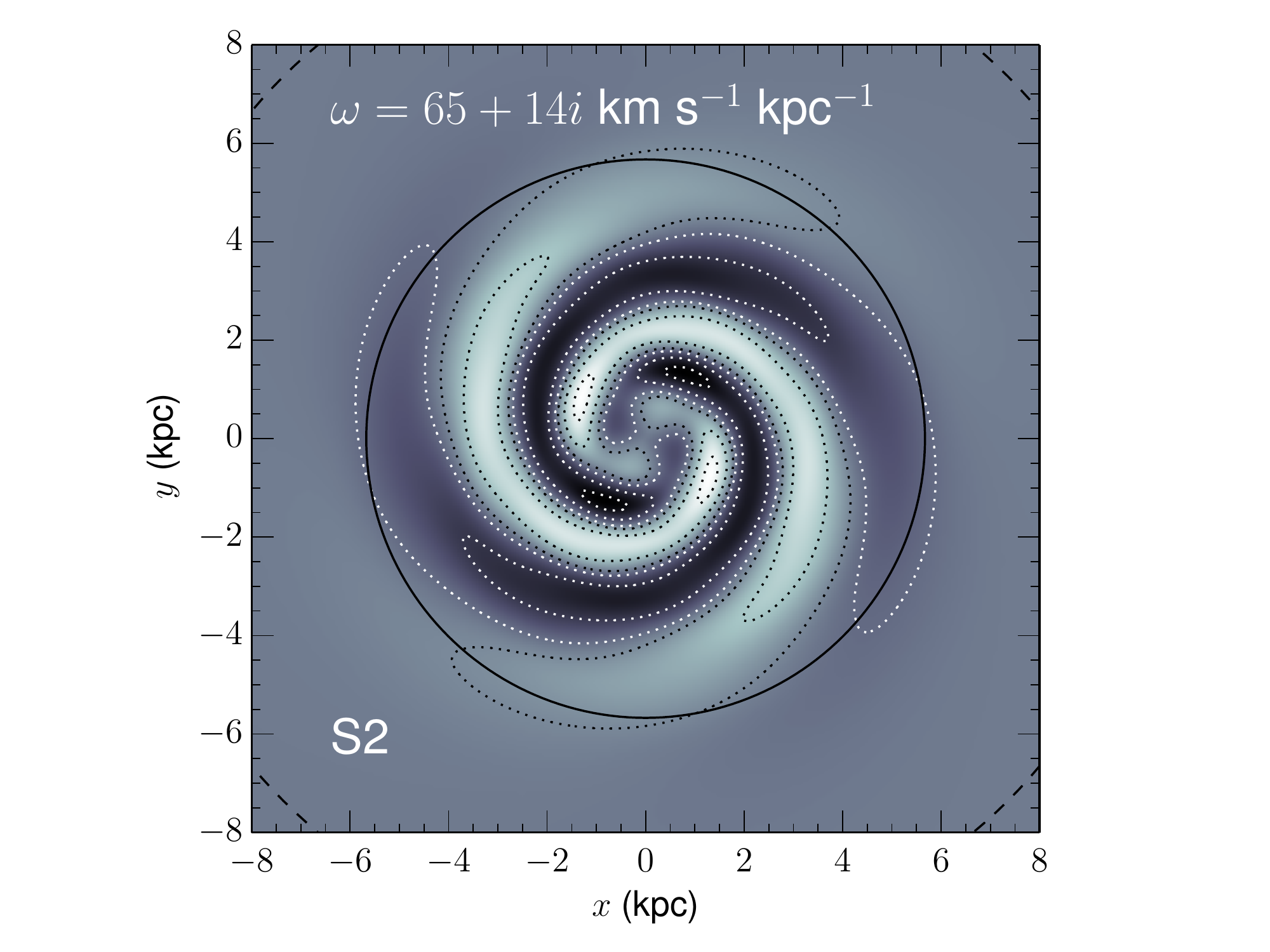}
\includegraphics[trim=40 5 90 1,clip,width=0.325\textwidth]{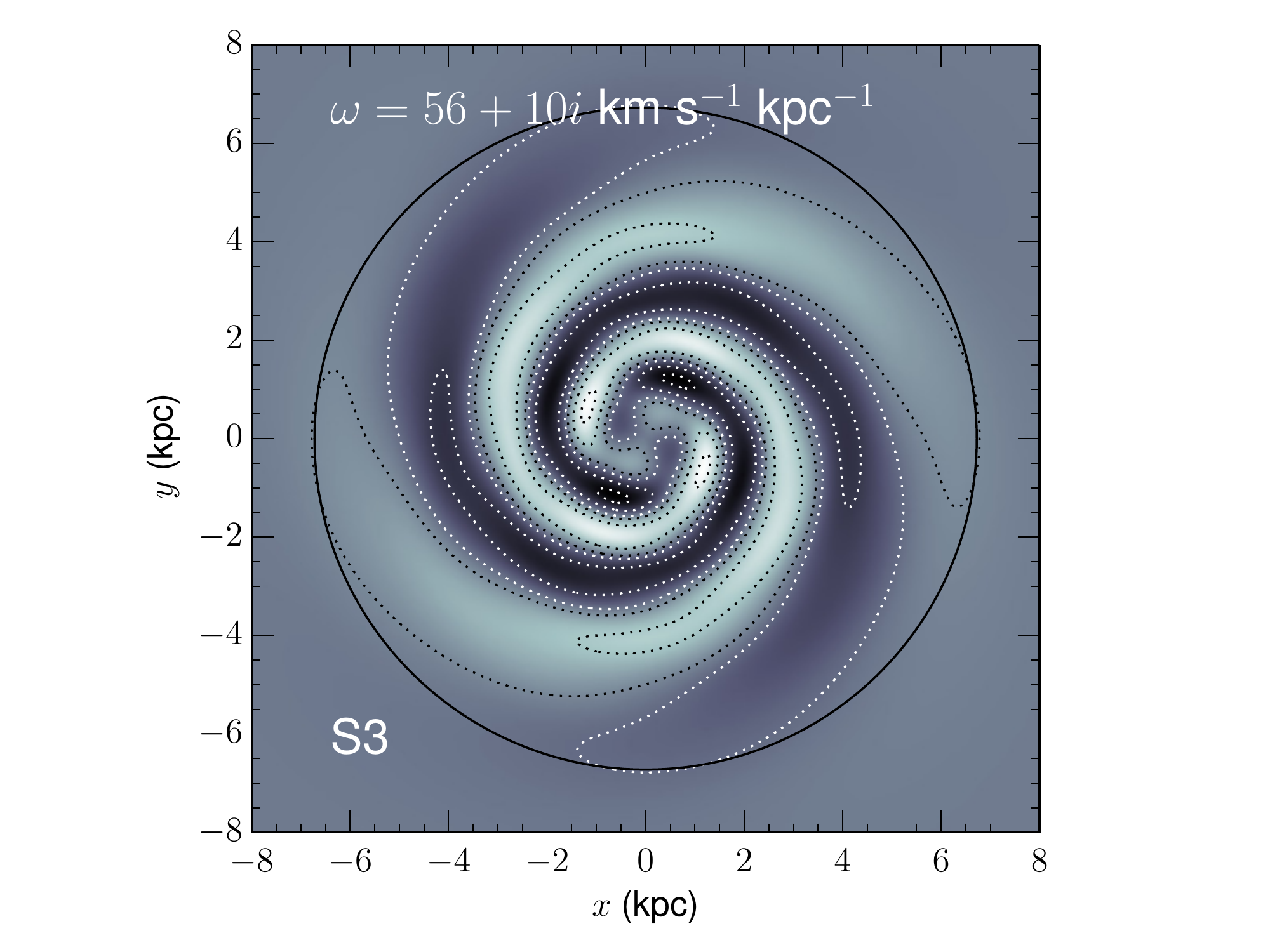}
\includegraphics[trim=40 5 90 1,clip,width=0.325\textwidth]{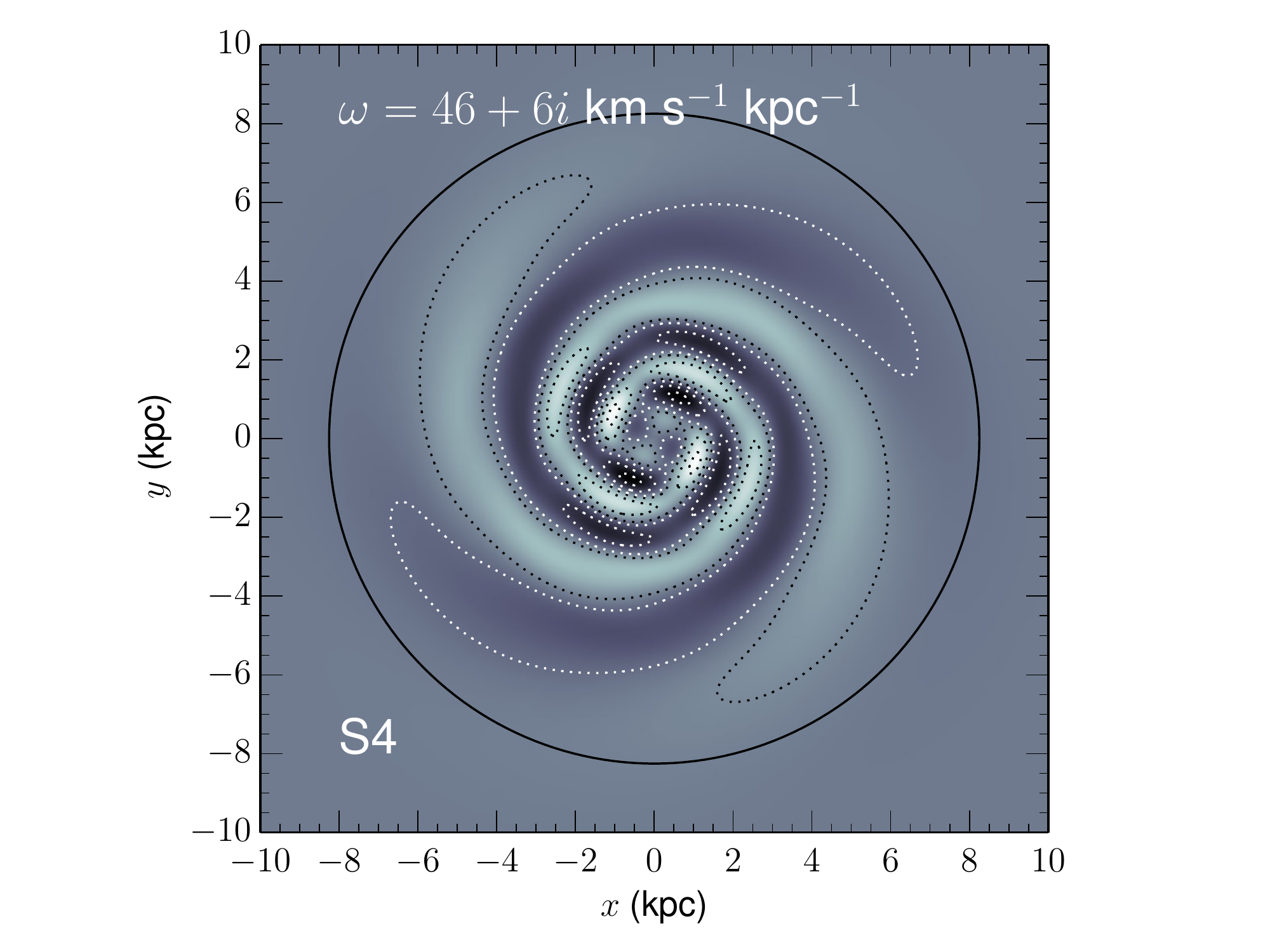}
\includegraphics[trim=40 5 90 1,clip,width=0.325\textwidth]{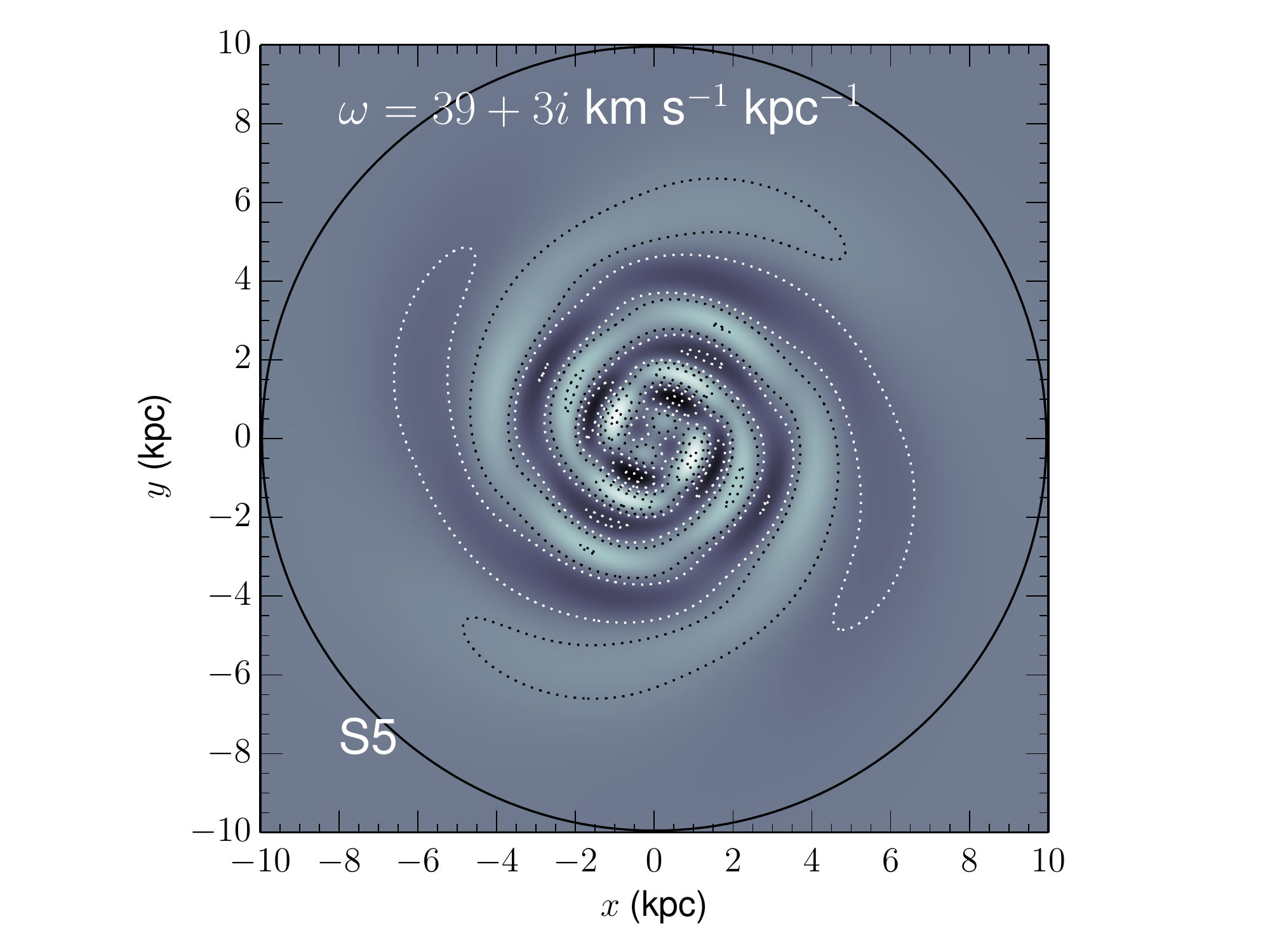}
\caption{Surface density of the most prominent $m=2$ eigenmodes of the
  cored exponential disc model, labeled by their complex frequency
  $\omega$ and their name taken from \citet{jalali07}. Positive
  densities are drawn in light-greys/white; negative ones in
  dark-greys/black. The corotation radius is indicated with a thick full line;
  the outer Lindblad resonance by a dashed line. The thin dotted-line
  contours trace surface density levels at $\pm 10$~\%, $\pm 50$~\%,
  and $\pm 90$~\% of the maximum value.
 \label{fig:modes.pdf}}
\end{figure*}
This DF is smoothly tapered to zero at $J=0$ by multiplying it with a
cutout function of the form \beqn H_{\rm cut}(J) = 1-\exp\left(
-\frac{J^2}{J_0^2} \right). \neqn This obviously removes stars close
to the galaxy center, causing a central hole in the stellar surface
density. The DF of this disc galaxy model in turning-point space is
plotted in the top panel of Fig. \ref{fig:DF0.pdf}. This DF is clearly
weighted towards quasi-circular orbits (the diagonal line in this
diagram) and contains very few stars on radial orbits (the vertical
axis of the diagram). We also smoothly taper the stellar density to
zero at an outer radius of 25~kpc. Here, we choose the following
values for all parameters involved:
\begin{align}
N &= 6 \nonumber\\
v_0 &= 200~{\rm km~s}^{-1} \nonumber\\
r_c &= r_D = 2.5~{\rm kpc} \rightarrow \lambda=1\nonumber\\
J_0 &= 50~{\rm kpc~km~s}^{-1} \nonumber\\
\rho_s &= 0.42 \frac{v_0^2}{G r_c} = 1.56 \times 10^9~M_\odot/{\rm kpc}^2.
\end{align}
We choose $r_c=r_D$ to be able to compare the mode analysis of the
cored exponential disc presented here with the analyses done by
\citet{jalali05}, \citet{jalali07}, and \citet{omurkanov14}.  We show
the stellar kinematics of this model, computed numerically as velocity
moments of the DF, in Fig. \ref{fig:DF0.pdf}.


In Fig. \ref{fig:jalali_base.pdf}, we show the $m=2$ mode spectrum of
this exponential disc model in the complex frequency plane. The grey
scale traces the value of
$\min\left(\left|{\lambda-1}\right|\right)$:~the smallest distance
between 1 and any of the eigenvalues of $C(\omega)$. For eigenmodes,
this distance is zero (dark regions); they are indicated with white
dots in Fig. \ref{fig:jalali_base.pdf}. The triangular data points
mark the position of the eigenmodes as found by \citet{omurkanov14}
using different methods. Dots with the same tint are different
frequency estimates for the same eigenmode; their spread gives an
indication on the intrinsic uncertainty on the eigenmode
frequencies. No growing modes exist with pattern speeds low enough to
have an inner Lindblad resonance, or ILR. The sequence of growing
modes ends with a neutral mode at $\omega_{\rm real}\approx
32$~km~s$^{-1}$~kpc$^{-1}$, well above the limit $\omega_{\rm real} = 17.0$~km~s$^{-1}$~kpc$^{-1}$
for having an ILR (cf. Fig. \ref{fig:freqsO.pdf}).

\begin{table}
\begin{center}
\caption{Real and imaginary parts of the frequencies of the fastest
  growing modes of the cored exponential disc model with a central
  cutout. The naming of the modes is taken from
  \citet{jalali07}. \label{tbl-1}}
\begin{tabular}{ccc} \hline
Name & $\omega_{\rm real}$ (km~s$^{-1}$~kpc$^{-1}$) & $\omega_{\rm imag}$ (km~s$^{-1}$~kpc$^{-1}$) \\ \cline{1-3}
B1   & 193.91 & 34.50 \\
C2   & 162.51 &  4.41 \\
C1   & 158.94 &  4.81 \\
B2   &  92.35 & 19.25 \\
S1   &  82.18 & 20.88 \\
S2   &  64.53 & 14.34 \\
S3   &  55.76 & 10.11 \\
S4   &  46.39 &   5.90 \\
S5   &  38.97 &   2.88 \\\cline{1-3}
\end{tabular}
\end{center}
\end{table}

\citet{omurkanov14} investigated the stability properties of this disc
galaxy model using the linear matrix method (PME) of \citet{poly05},
the linear method using basis function (ECB) of \citet{jalali07}, and
the finite element method (FEM) of \citet{jalali10}, both in its full
form (FEMf) and in its restricted form suitable for models dominated
by quasi-circular orbits (FEMc). Contrary to {\sc pyStab}, which uses
polar coordinates throughout, these methods all employ action-angle
variables in phase space. They all rely on a Fourier expansion of the
perturbing potential, truncated at some order and the integration of a
finite number of orbits. The PME method, as used in
\citet{omurkanov14}, uses 10 Fourier terms and 1000 orbits. The ECB
method uses an expansion of the response DF and potential in a basis
of 15 rather contrived trial functions of the action variables. For
the FEM method, the disc is divided in $N$ rings in which the response
potential is expanded in $N_d$ basis functions. Here, $N=100$ and
$N_d=2$. The ECB results reported in \citet{jalali07} were obtained
using 10 Fourier terms and 15 trial functions.

Overall, the agreement between {\sc pyStab} and these other mode
analysis methods is very satisfactory. We list the frequencies of the
retrieved eigenmodes, together with their names as given by
\citet{jalali07}, in Table \ref{tbl-1}. The surface density
perturbation of the nine most prominent $m=2$ eigenmodes of the cored
exponential disc model, as named by \citet{jalali07}, is presented in
Fig. \ref{fig:modes.pdf}. These density plots can be directly compared
with Figs. 3 and 4 in \citet{omurkanov14}. Overall, the agreement is
satisfactory in terms of size, shape, and number of density
enhancements along the arms. This code comparison makes us confident
that {\sc pyStab} works properly and that the results are reliable.

\begin{figure*}
\includegraphics[trim=0 20 0 0,clip,width=0.48\textwidth]{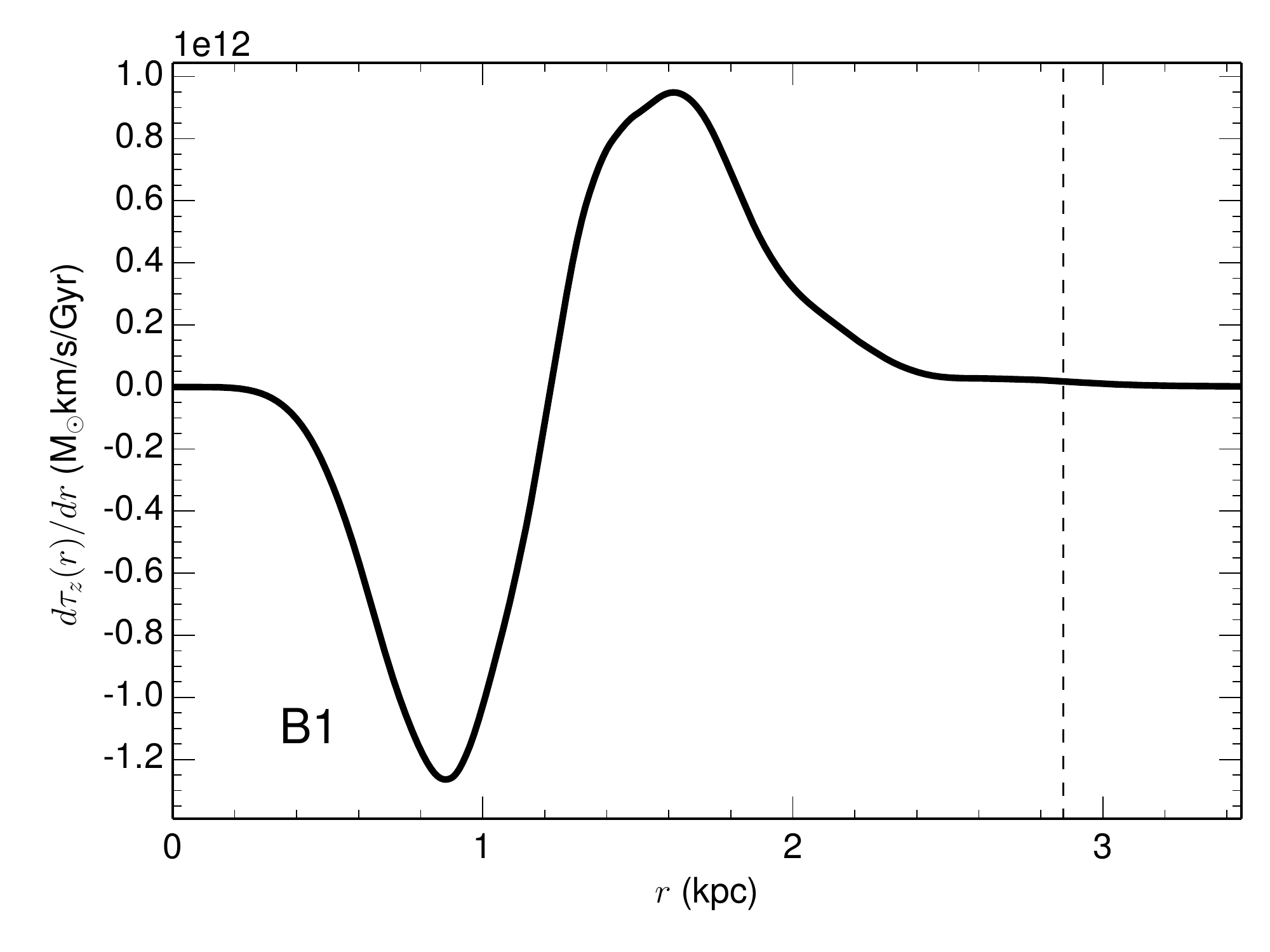}
\includegraphics[trim=0 20 0 0,clip,width=0.48\textwidth]{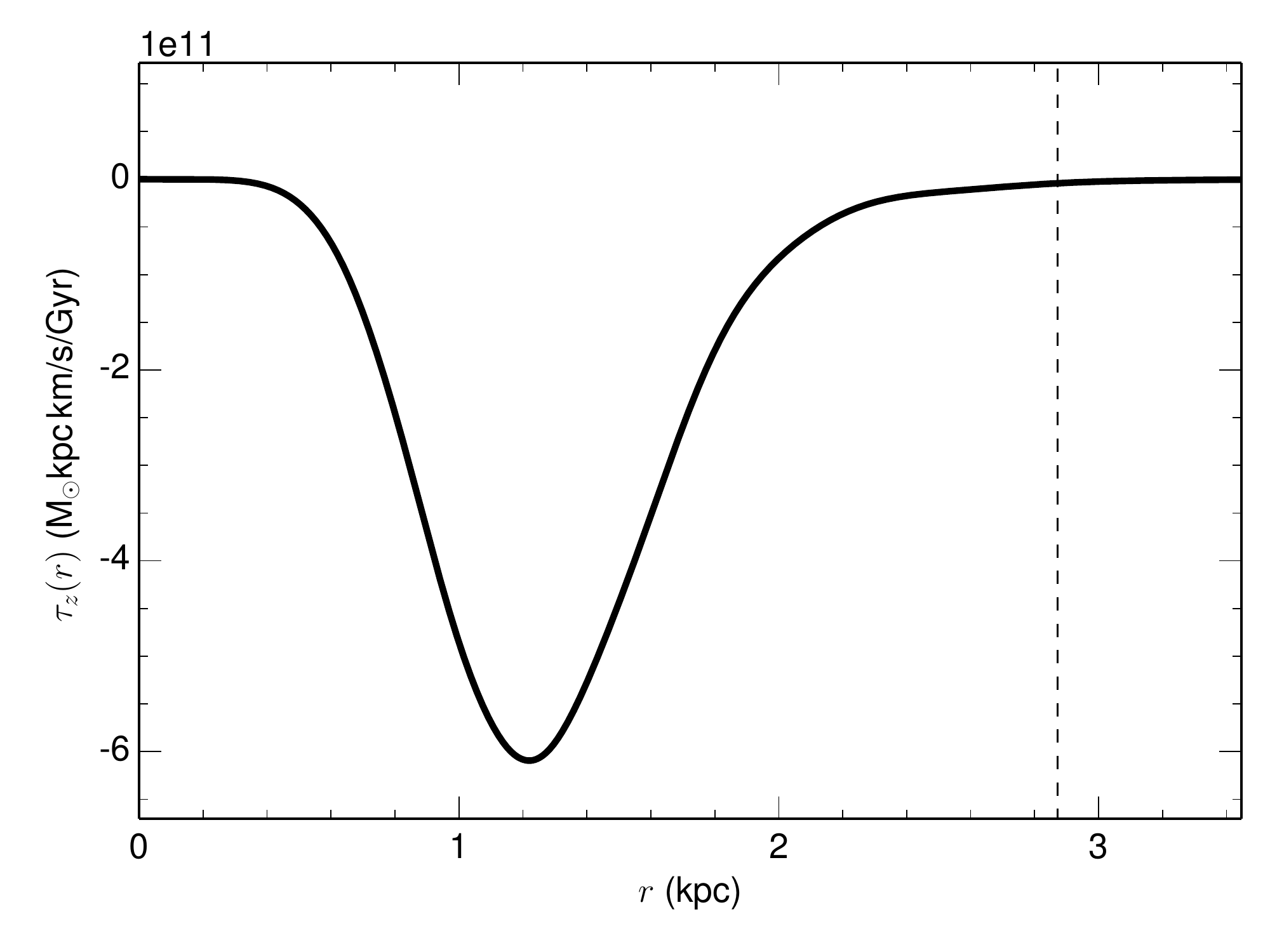}
\includegraphics[trim=0 20 0 0,clip,width=0.48\textwidth]{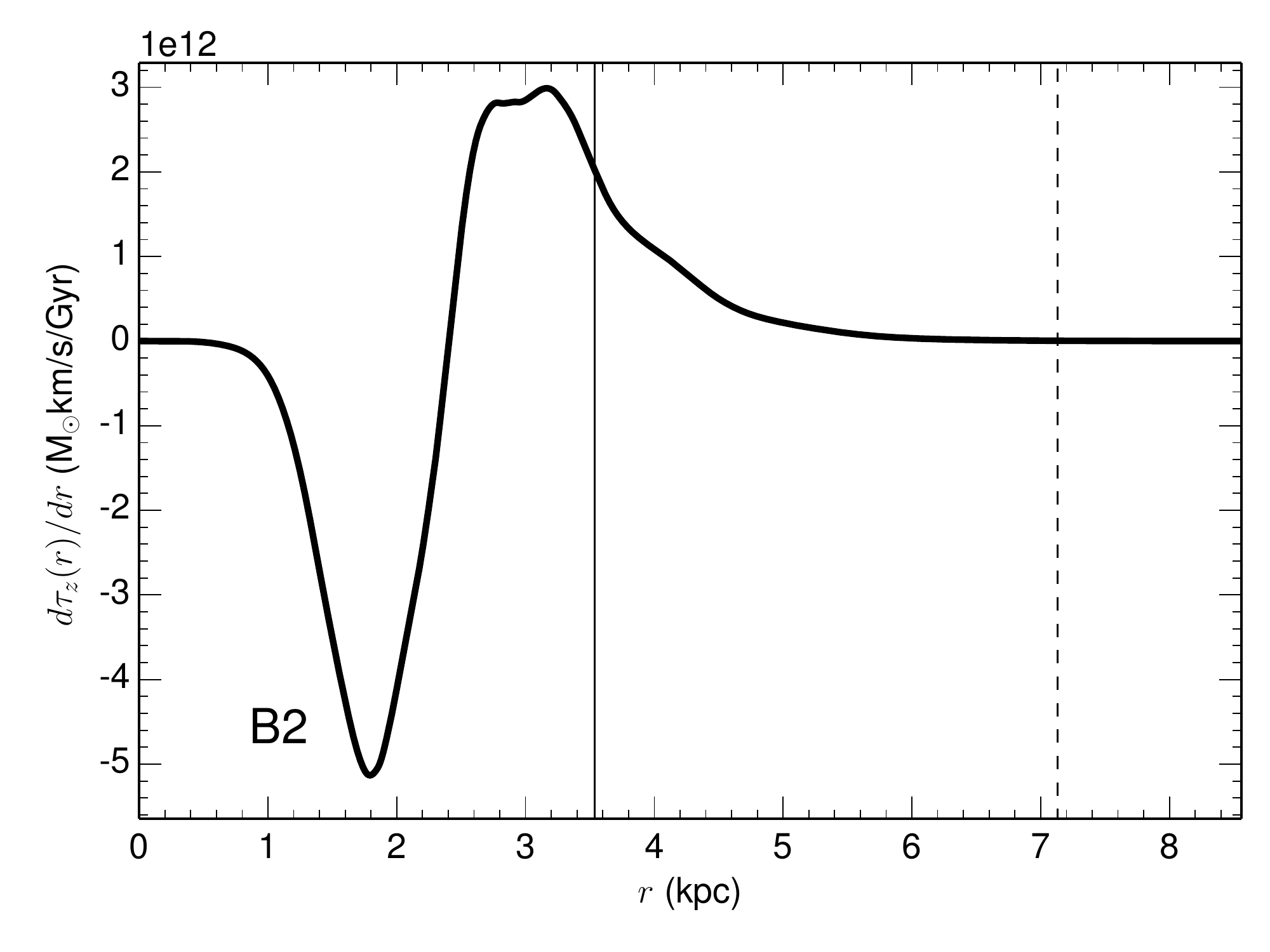}
\includegraphics[trim=0 20 0 0,clip,width=0.48\textwidth]{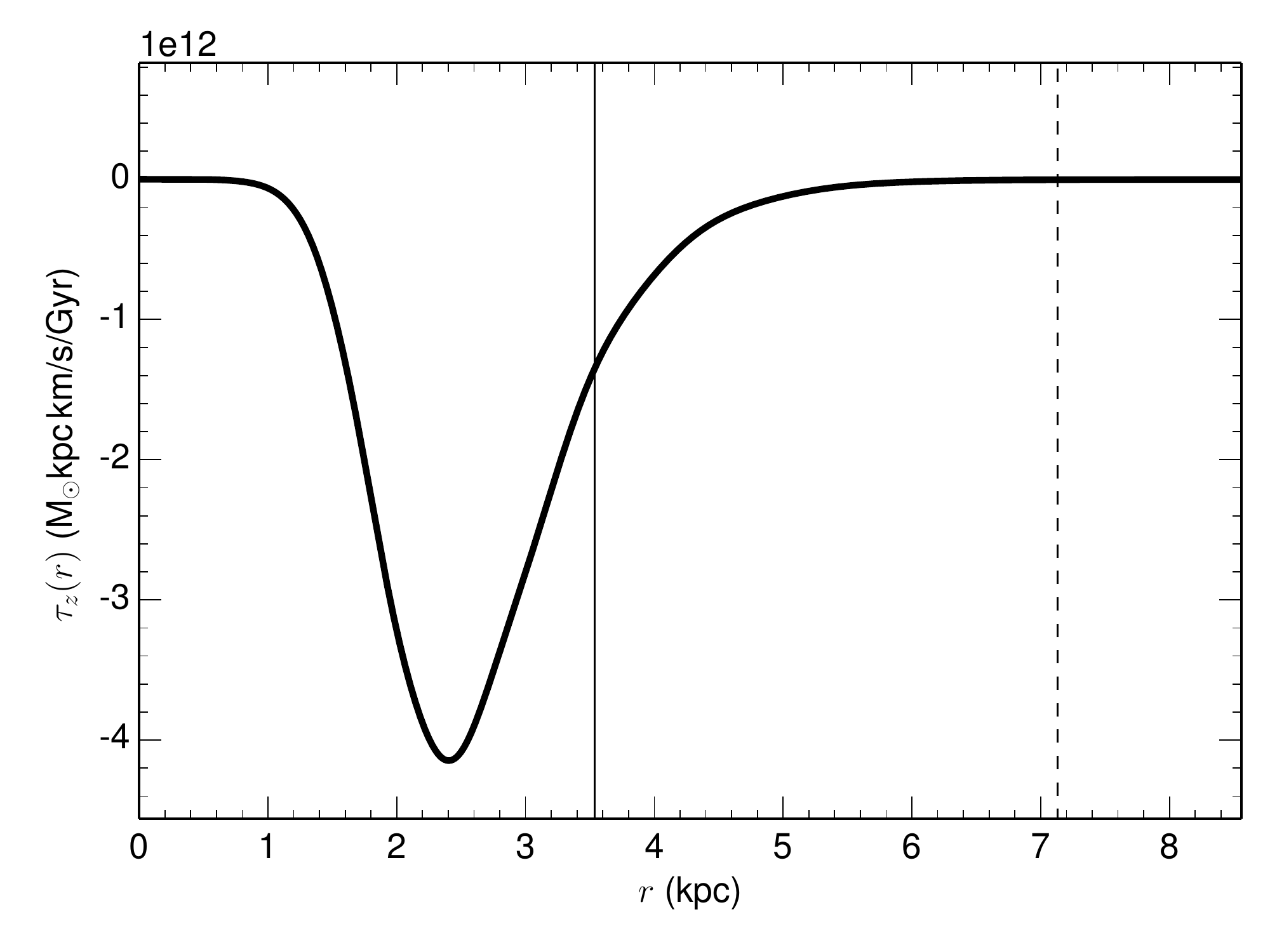}
\includegraphics[trim=0 20 0 0,clip,width=0.48\textwidth]{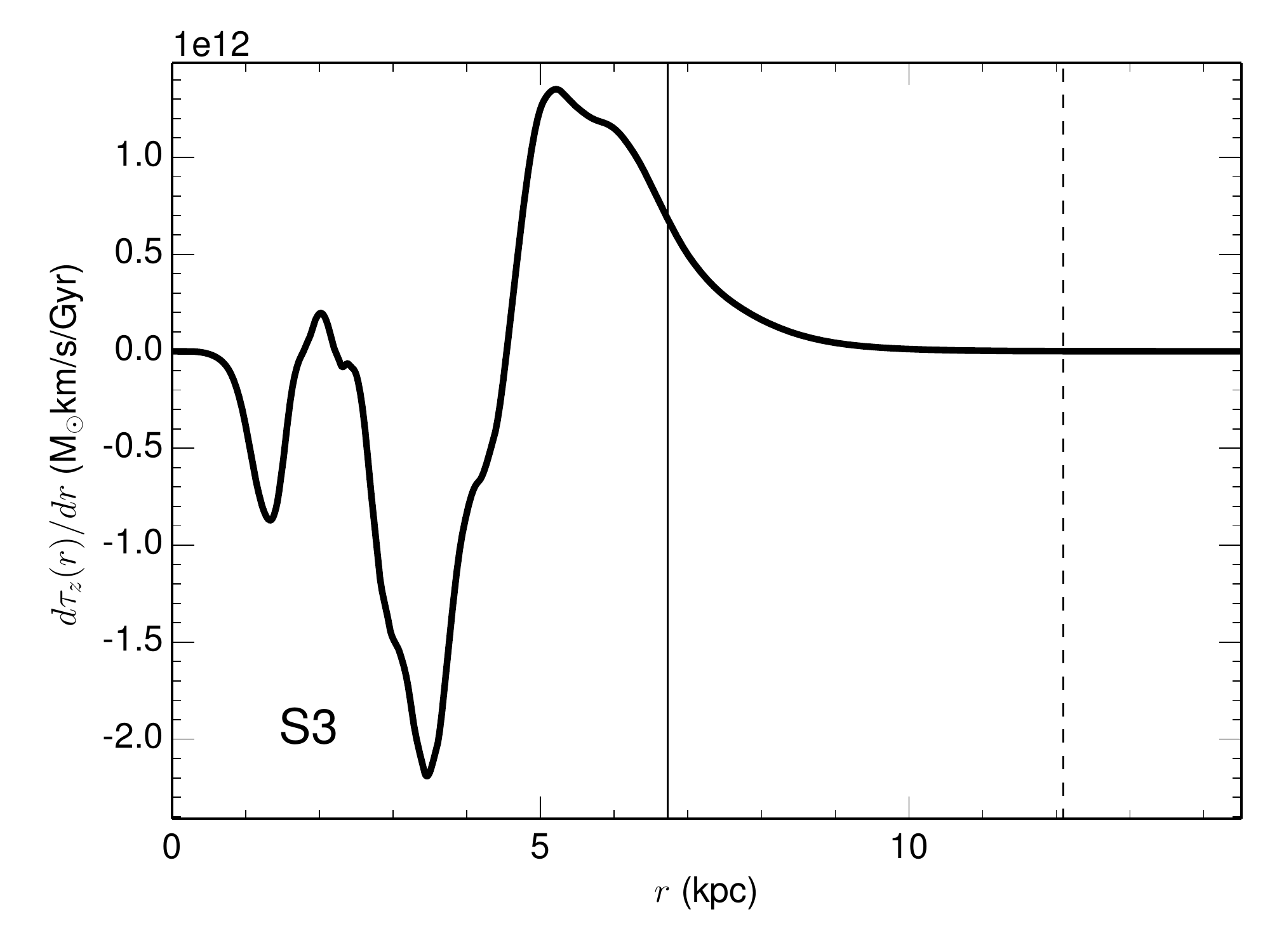}
\includegraphics[trim=0 20 0 0,clip,width=0.48\textwidth]{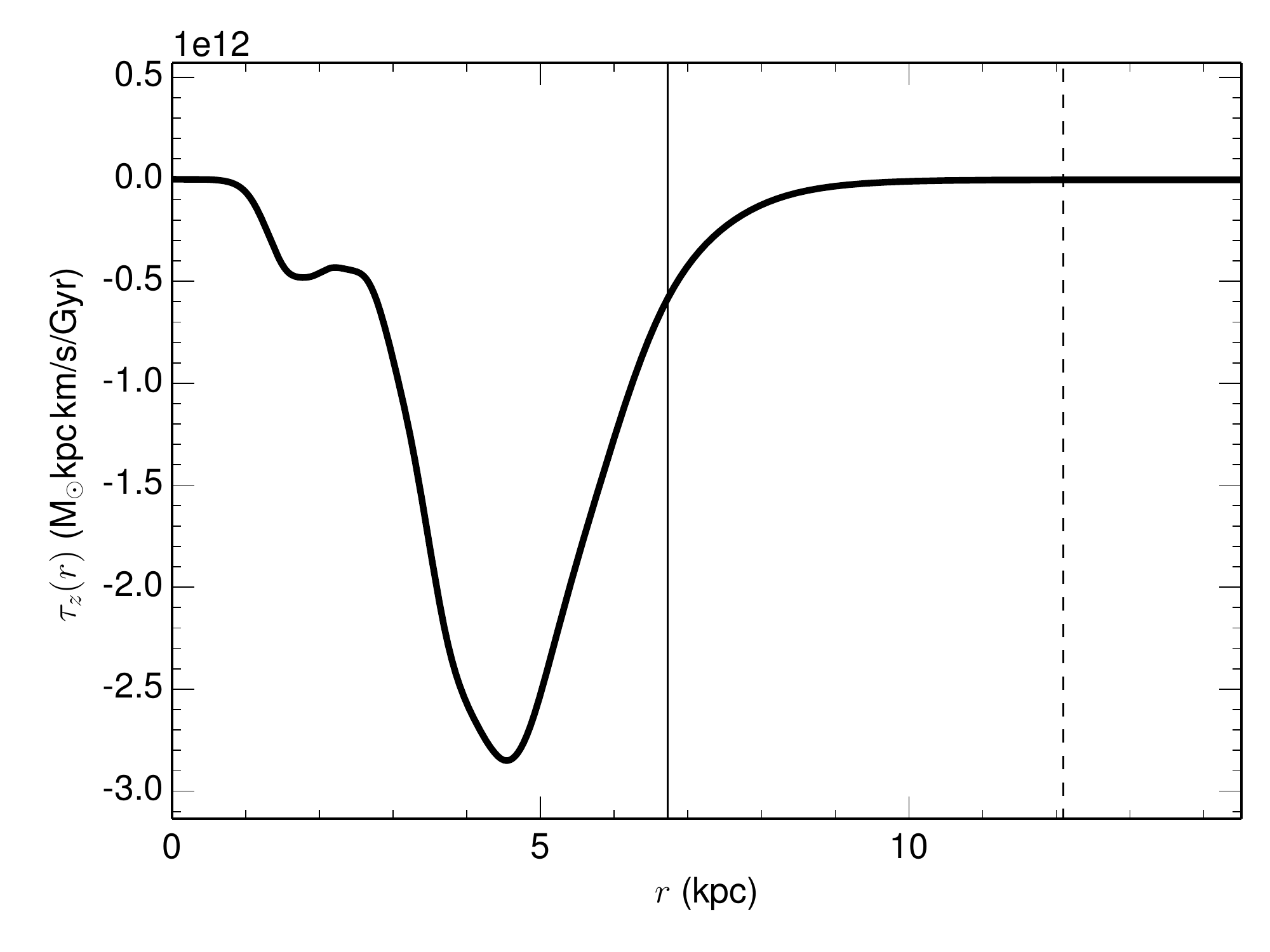}
\caption{Left column:~the torque $d\tau_z(r)$ exerted by the spiral
  pattern on an annular ring of stars with radius $r$ and width
  $dr$. Right column:~the torque $\tau_z(r)$ integrated out to radius
  $r$. Top to bottom:~the eigenmodes B1, B2, and S3 of the cored
  exponential disc model. Vertical full line:~the CR radius; vertical
  dashed line:~the OLR radius.
 \label{fig:tau.pdf}}
\end{figure*}
 As a final test of the formalism, we compute the torque exerted
  by the spiral pattern on the stellar disc. As shown by
  \citet{zhang96,zhang98}, the torque $d\tau_z(r)$ on an annular ring
  of stars with radius $r$ and width $dr$ is given by 
\begin{align}
d\tau_z(r) & = r dr \int_0^{2\pi} \rho_{\rm resp} \frac{\partial V_{\rm
      resp}}{\partial \theta} d\theta \nonumber \\
&= -\frac{m}{2} \tilde{\rho}_{\rm resp}(r) \tilde{V}_{\rm resp}(r) \sin( m \gamma_0(r))dr.
\end{align}
Here, $\tilde{\rho}_{\rm resp}$ and $\tilde{V}_{\rm resp}$ are the
real amplitudes of the spiral density and binding potential, and
$\gamma_0$ is the phase shift between the pattern potential and
density. If $d\tau_z$ is negative, the stars lose angular momentum to
the spiral pattern; otherwise, they gain angular momentum from the
pattern. For a pattern with a negligible radial amplitude variation,
the transition from angular momentum loss to angular momentum gain
should occur precisely at the CR radius although this is at best an
approximation. Since no external forces act on the stellar disc, the
total torque $\tau_z = \int_0^\infty d\tau_z(r)$ should be exactly
zero \citep{poly15}.

In Fig. \ref{fig:tau.pdf}, $d\tau_z(r)$ and $\tau_z(r) = \int_0^r
d\tau_z$ are shown for three representative eigenmodes of the cored
exponential disc model:~B1, B2, and S3. $d\tau_z(r)$ is negative at
small radius, changes sign at about two-thirds of the CR radius (if
there is a CR), and is positive at large radii. Hence, the cumulative
torque $\tau_z(r)$ first becomes zero and then rises again to zero at
large radii. Clearly, the formalism presented here conserves the total
angular momentum of the stellar disc with excellent precision:~the
asymptotic value of $\tau_z$ was found to be always smaller than
$\sim 10^{-4}$ times its extreme value.  

\section{Grooves and m=2 modes} \label{m2modes}

\begin{figure}
\includegraphics[trim=50 10 50 10,clip,width=0.49\textwidth]{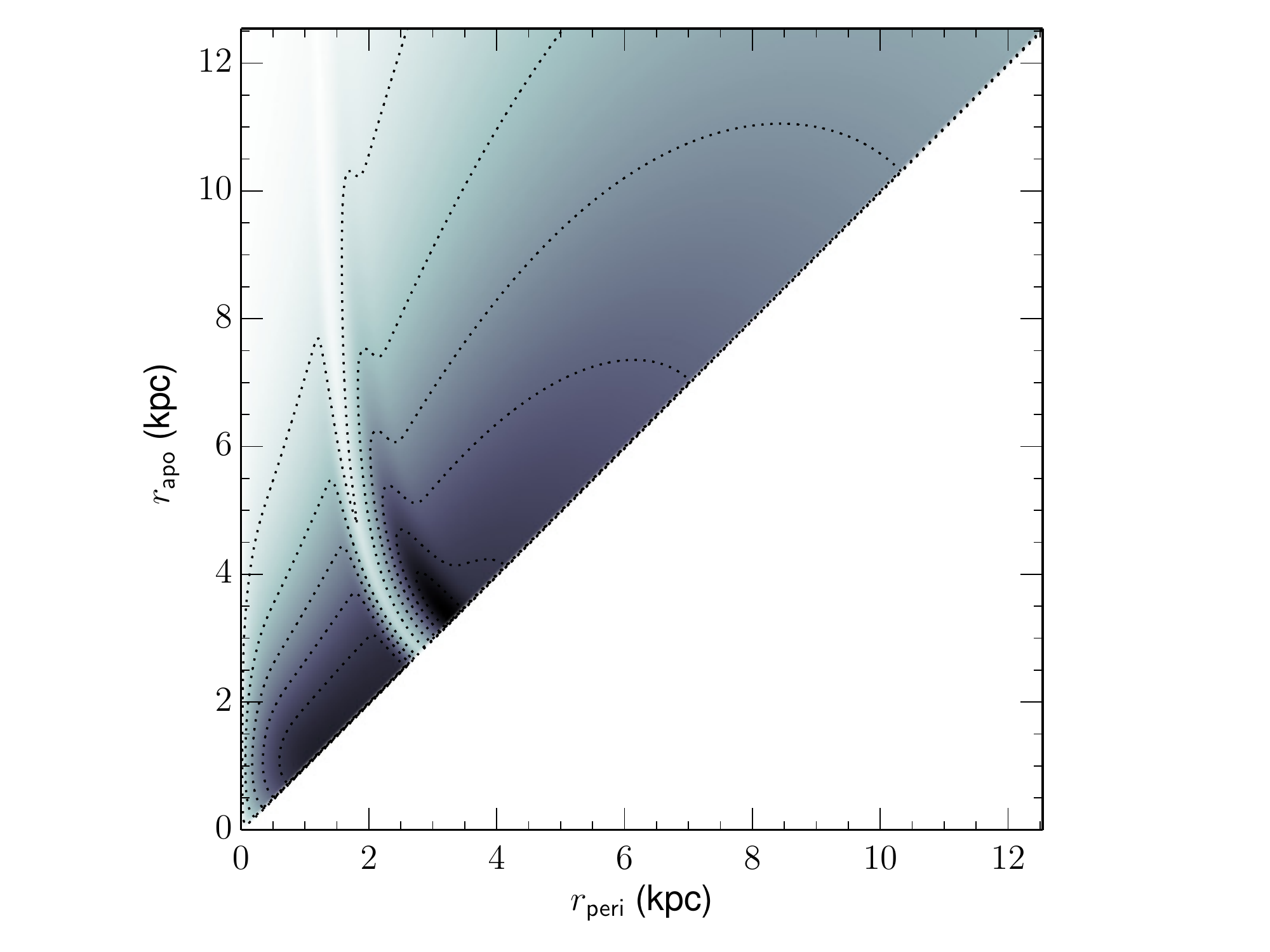}
\includegraphics[trim=60 0 60 1,clip,width=0.49\textwidth]{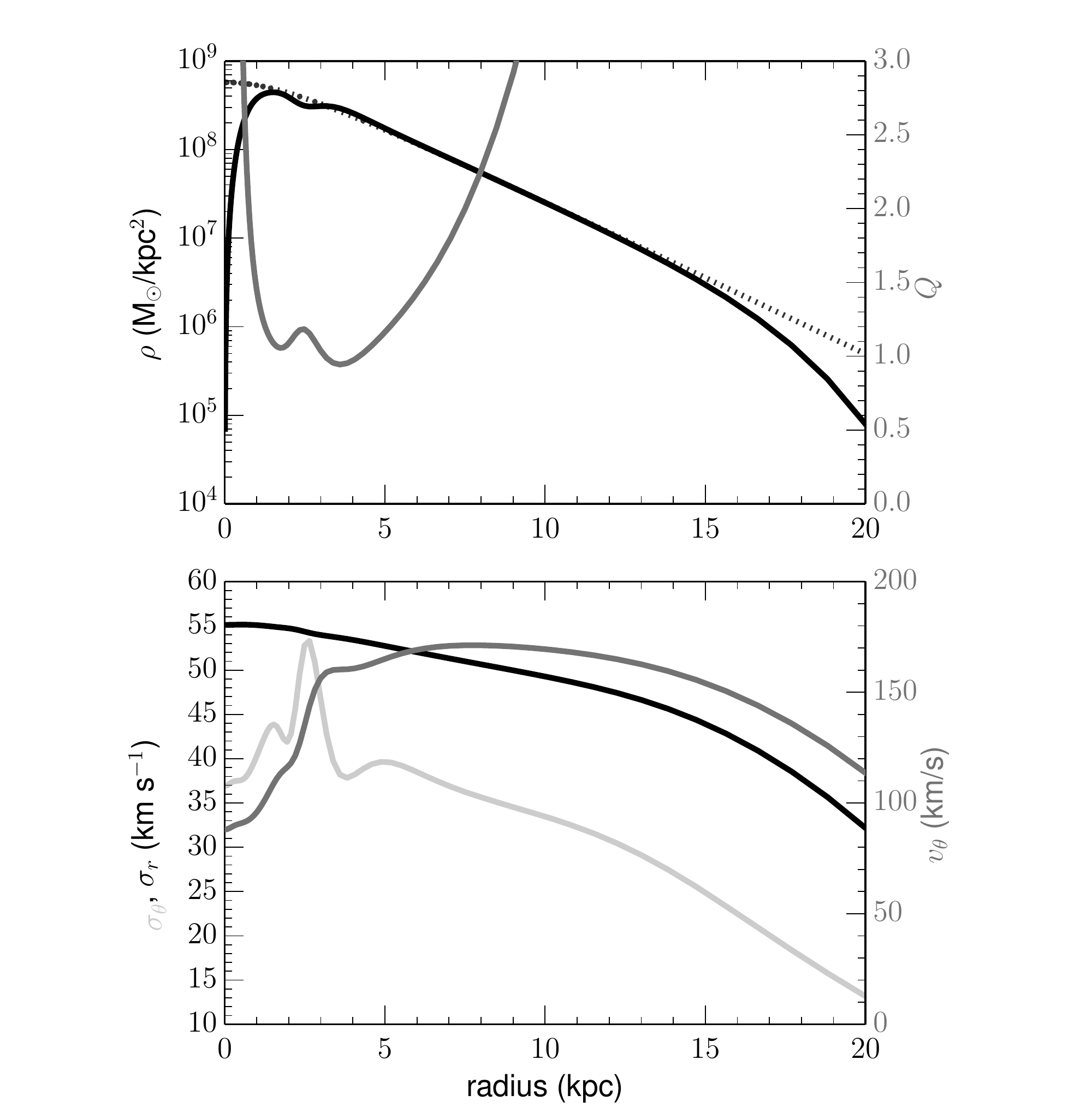}
\caption{Top panel:~the distribution function of the cored exponential
  disc model with a central cutout and with a groove at $J_{\rm
    groove}=433$~kpc~km~s$^{-1}$, shown in turning-point space. The
  colorbar indicates the phase-space density expressed in
  $M_\odot\,\text{kpc}^{-2}\,(\text{km}\,\text{s}^{-1})^{-2}$. Middle
  panel:~the stellar density, $\rho$ (black curve, left axis), and the
  Toomre $Q$-parameter (dark grey curve, right axis). The dotted black
  curve traces the stellar density (\ref{dens0}), without inner cutout
  and outer tapering. Bottom panel:~the radial velocity dispersion,
  $\sigma_r$ (black curve, left axis), tangential velocity dispersion,
  $\sigma_\theta$ (light grey curve, left axis), and the mean rotation
  velocity, $v_\theta$ (dark grey curve, right axis) of the cored
  exponential disc model with an inner cutout and with a groove at
  angular momentum $J_{\rm groove}=433$~kpc~km~s$^{-1}$, corresponding
  to a circular orbit with radius 2.87~kpc, the OLR of mode B1.
  \label{fig:kinematics_groove.pdf}}
\end{figure}

\citet{sellwood89} have reported on the occurrence of successive
generations of spiral patterns in numerical simulations of stellar
discs. As the inner disc gets steadily dynamically warmer, each
generation of patterns decays and a new one grows but with lower
pattern speeds and larger radial extent than the previous one. These
authors argue that the dominant member of one particular generation of
patterns is a true eigenmode of the stellar disc as it is at that time
but not of the original disc. Therefore, the dynamical changes wrought
by the previous generation of patterns are instrumental in triggering
the next one. A detailed analysis of the evolution of stars in phase
space has led \citet{sellwood89} to propose the following cyclical
mechanism for recurrent spiral modes:
\begin{itemize}
\item a depopulated narrow groove at a location in phase space to
  which the disc is very responsive causes a set of modes to grow
\item when each mode saturates and finally decays, it transports its
  angular momentum to its OLR, pushing stars there towards higher
  angular momenta, thus creating new phase-space grooves
\item the grooves that fall in a responsive part of phase space cause
  a new set of modes to grow
\end{itemize}
As long as the stellar disc can be cooled, e.g. by star formation,
this cycle can in principle continue. One caveat is that each
successive generation of spiral modes is spatially more extended than
the previous one and that the grooves can end up in unresponsive parts
of phase space and therefore produce no new modes, thus halting the
cycle. At least in this particular case, the disc is unresponsive
  to the high-$J_{\rm groove}$ grooves that are expected from later
  generations of spiral patterns.

In order to investigate how a narrow groove in phase space affects the
stability properties of a stellar disc, we adopt the cored exponential
model from the previous paragraph and remove stars from a narrow strip
around a fixed angular momentum $J_{\rm groove}$ by multiplying the
original DF, given by eqn. (\ref{DF0}), with a function $H_{\rm
  groove}(J)$ of the form \beqn H_{\rm groove}(J) = 1 -
e^{-\frac{x^6}{\sigma_1}} + A e^{-\frac{(x-1.5)^2}{\sigma_2}}
 \label{eq:Hgroove}\neqn with $x=(J-J_{\rm groove})/w_J$. Here, we choose
$w_J=60$~kpc~km~s$^{-1}$, $\sigma_1=2.0$, $\sigma_2=0.5$, and the
 forefactor $A$ such that the narrow positive bump cancels the broader
 negative groove. In other words:~stars are removed from the groove
 and deposited at the groove's high-$J$ edge.
As an example, the top panel of Fig. \ref{fig:kinematics_groove.pdf}
shows the DF of the model with a groove centered on $J_{\rm
  groove}=433$~kpc~km~s$^{-1}$. This is reflected in a narrow, curved
groove in the DF in turning-point space ending in the circular orbit
with radius $r_{\rm circle}=2.87$~kpc.

Since the ridge at the edge of the DF groove to a good approximation
conserves the number of stars, the epicyclic motions of the stars
cause the groove to have only a minor effect on the stellar surface
density. Only around the radius of the circular orbit with angular
momentum equal to $J_{\rm groove}$ is there a small wiggle in the
density, as can be seen in
Fig. \ref{fig:kinematics_groove.pdf}. Likewise, the radial velocity
dispersion remains virtually unaffected. The groove increases the
Toomre $Q$-parameter \citep{toomre64} by approximately 15\% around the
groove radius. The mean tangential velocity and, specifically, the
tangential velocity dispersion are significantly affected by the
groove. Since stars on circular orbits inside the groove have been
removed from the DF, it is mostly stars on eccentric orbits that
venture from the groove's high-$J$ bump towards the groove, thus
locally increasing the tangential velocity dispersion. Their epicyclic
motions locally contribute to the tangential velocity, leading to an
increase of the rotation velocity at the groove radius. The removal of
stars on eccentric orbits that move outside of the gap (see
Fig. \ref{fig:kinematics_groove.pdf}) explains the drop of the
dispersion at the edges of the groove.

\begin{figure*}
\includegraphics[trim=1 35 0 13,clip,width=0.448153\textwidth]{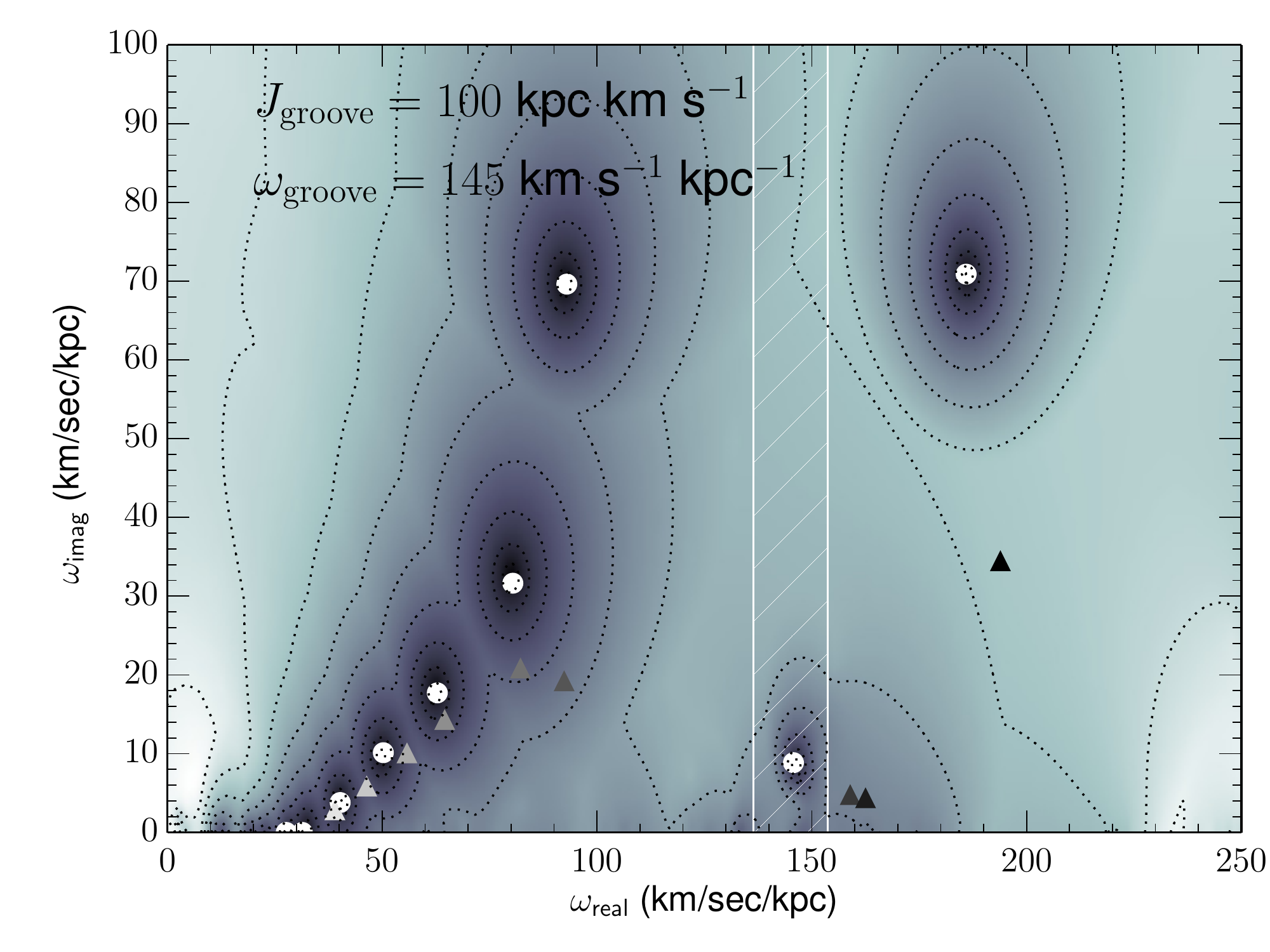} 
\includegraphics[trim=45 35 0 13,clip,width=0.420755\textwidth]{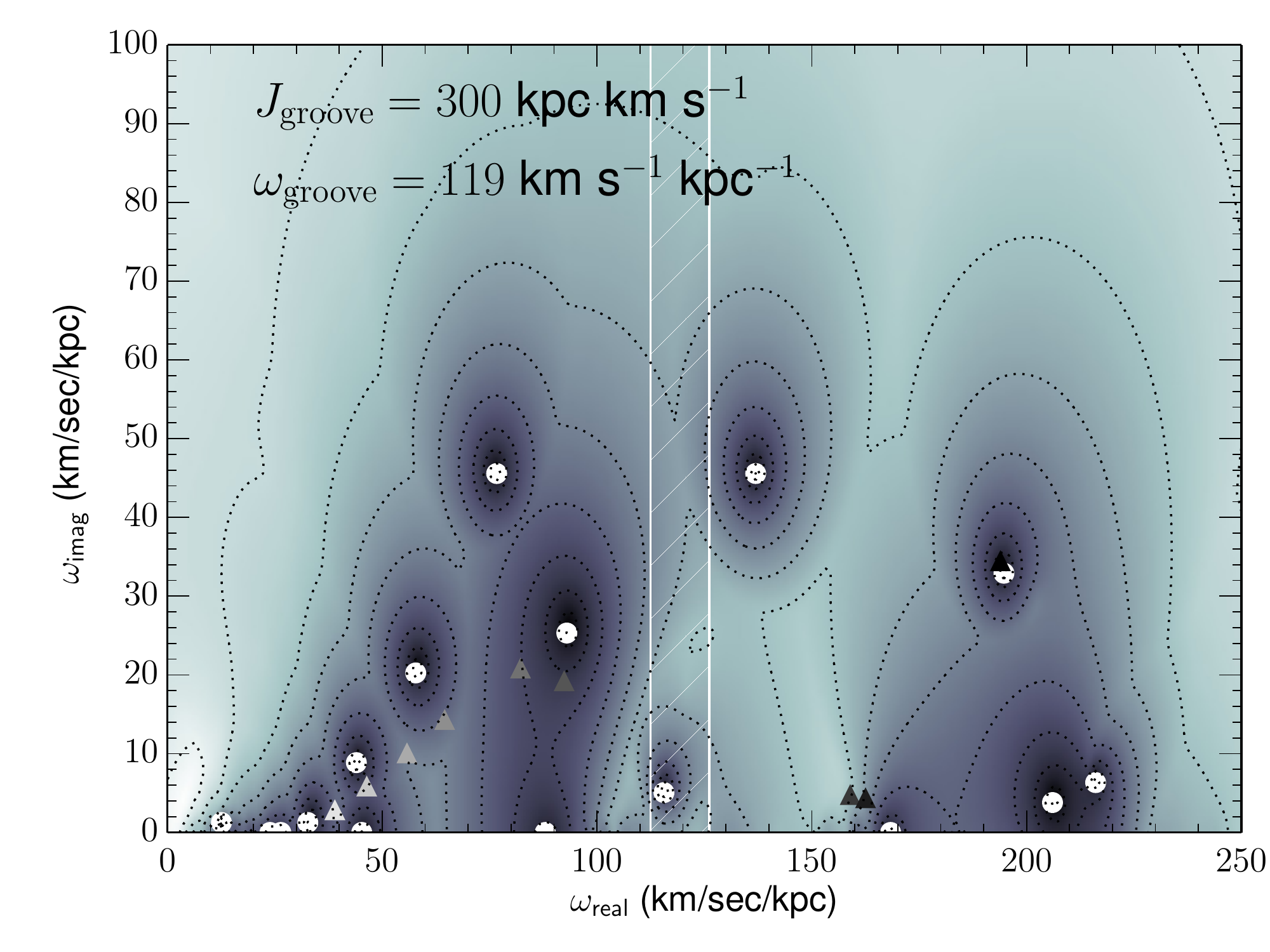} 
\includegraphics[trim=1 35 0 13,clip,width=0.448153\textwidth]{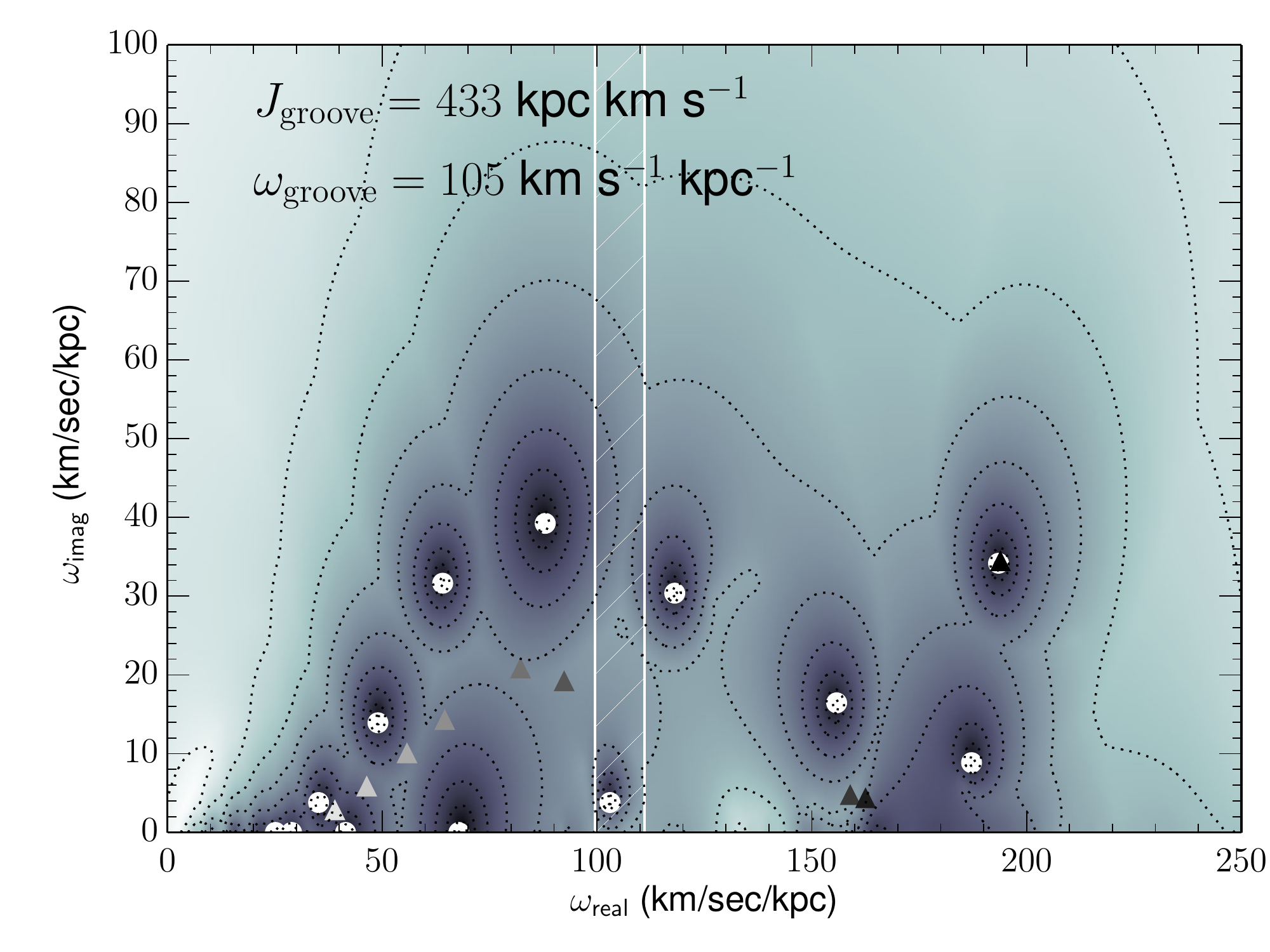} 
\includegraphics[trim=45 35 0 13,clip,width=0.420755\textwidth]{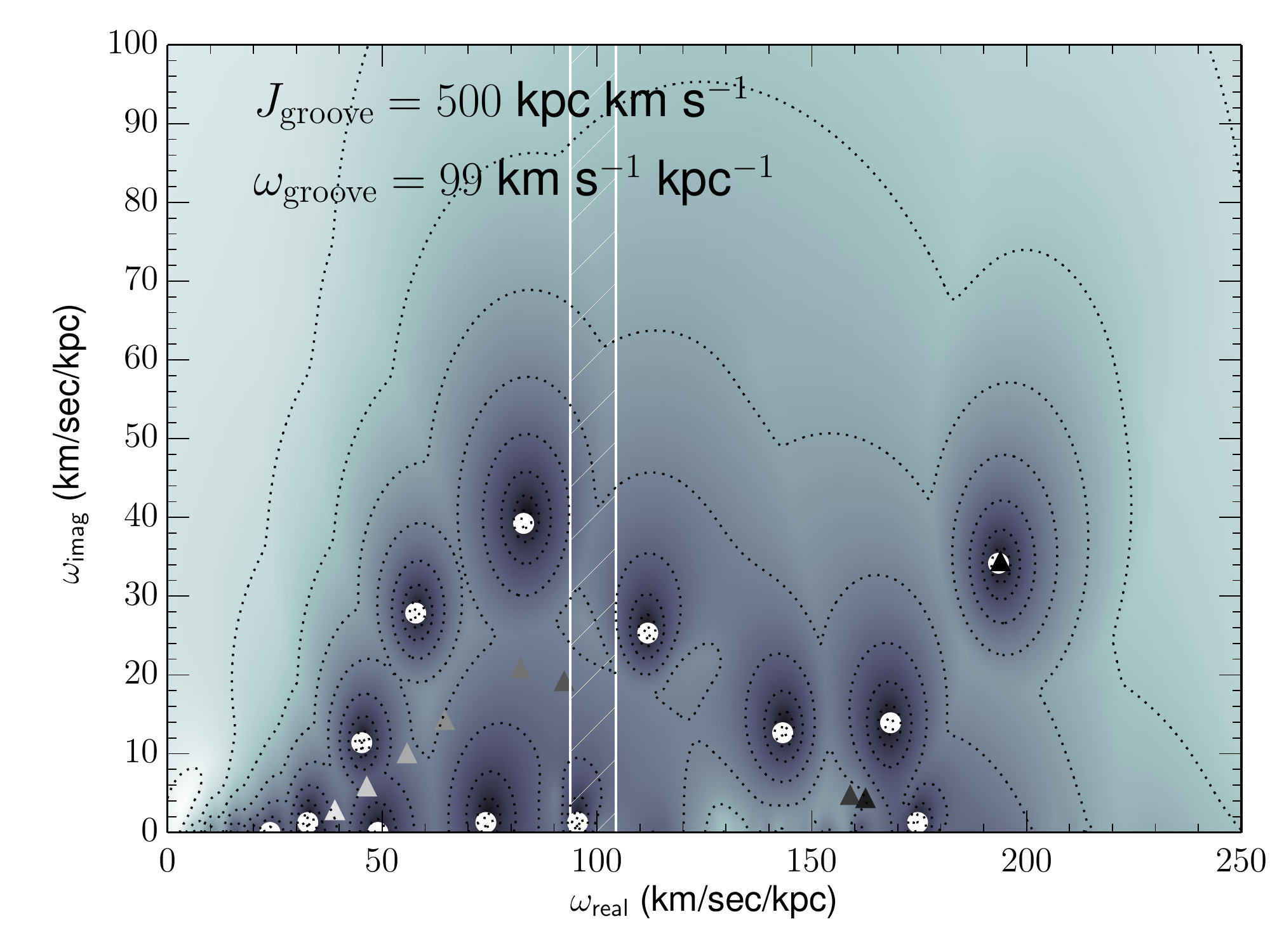} 
\includegraphics[trim=1 35 0 13,clip,width=0.448153\textwidth]{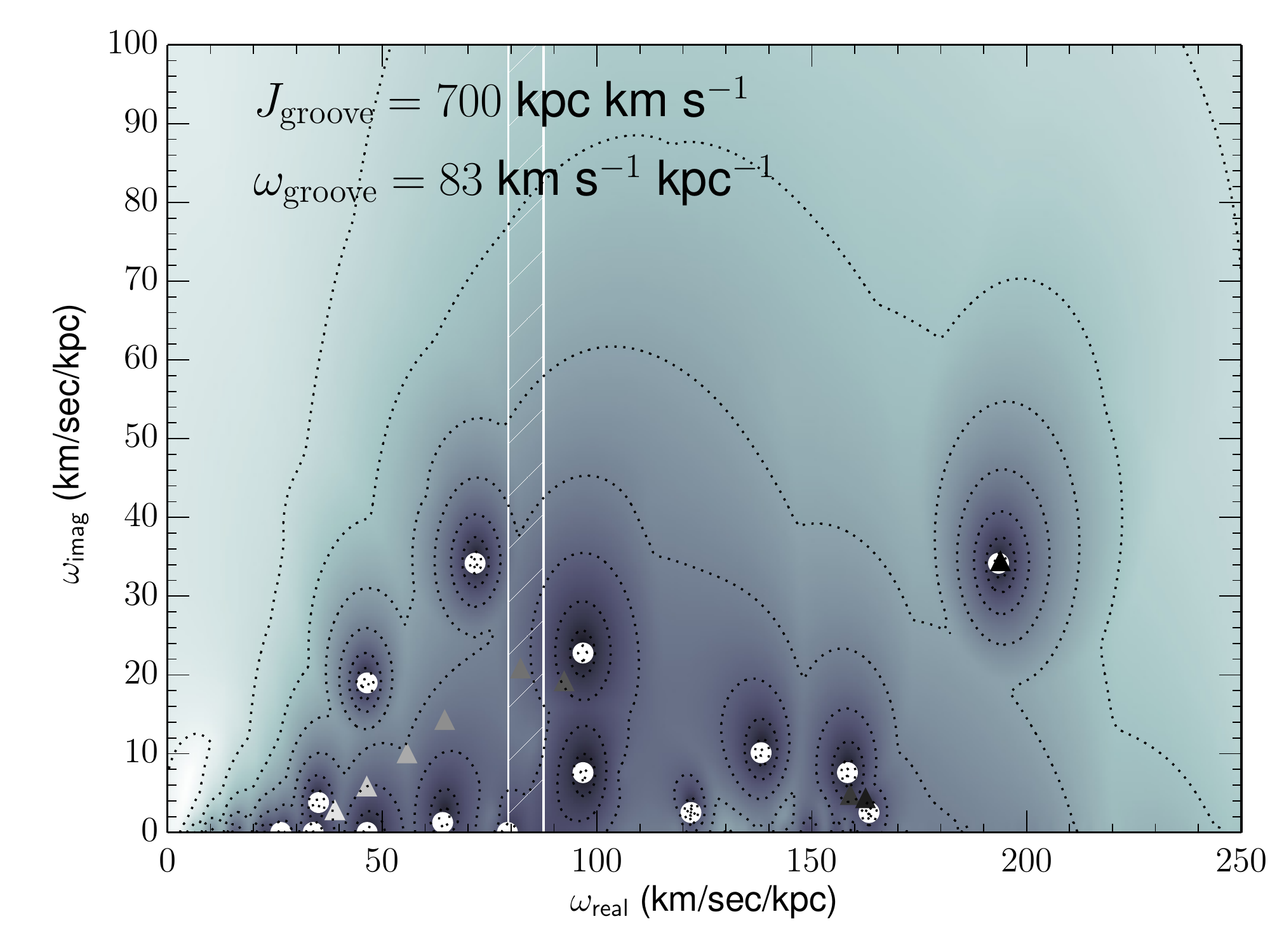} 
\includegraphics[trim=45 35 0 13,clip,width=0.420755\textwidth]{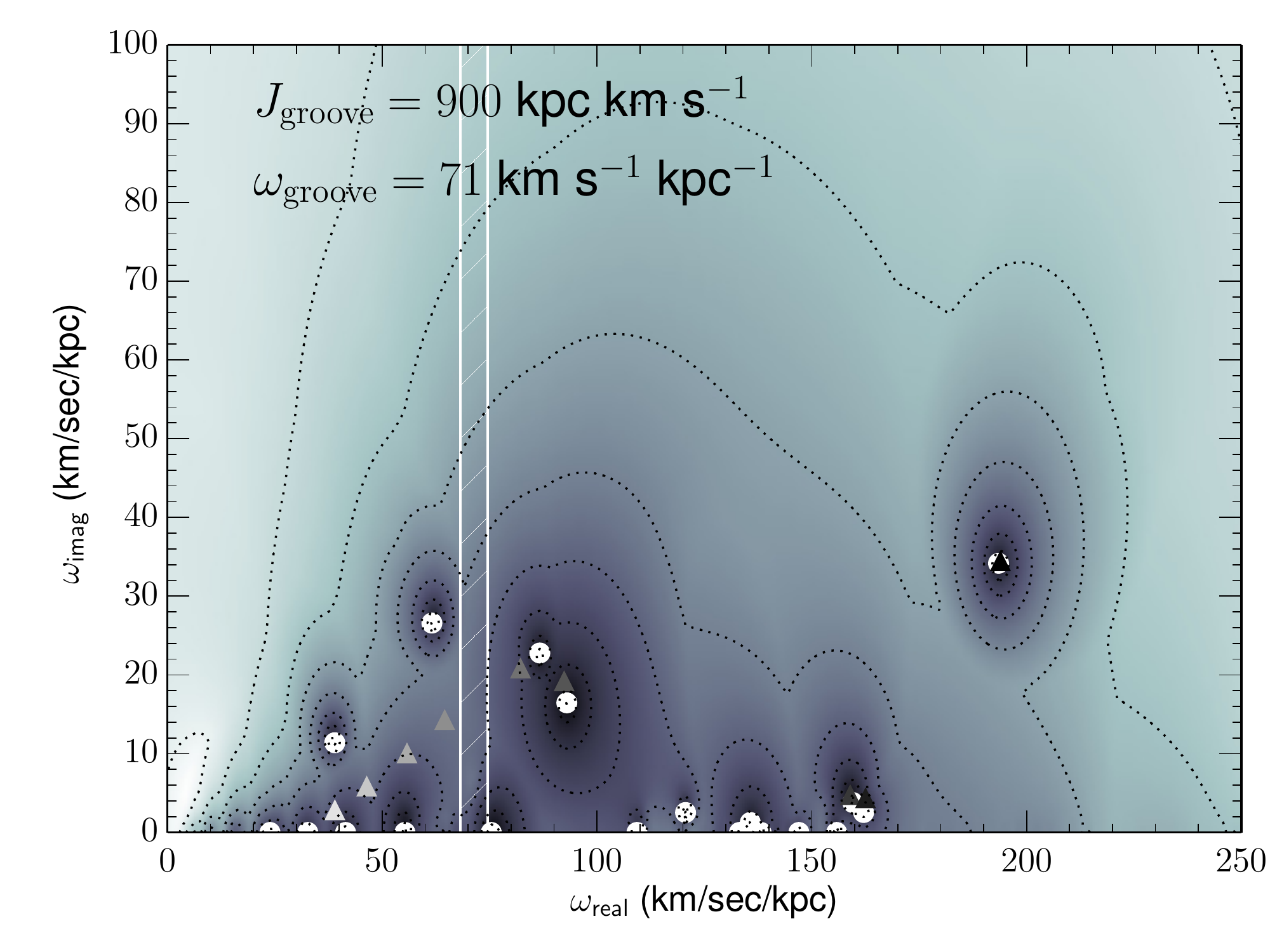} 
\includegraphics[trim=1 15 0 13,clip,width=0.448153\textwidth]{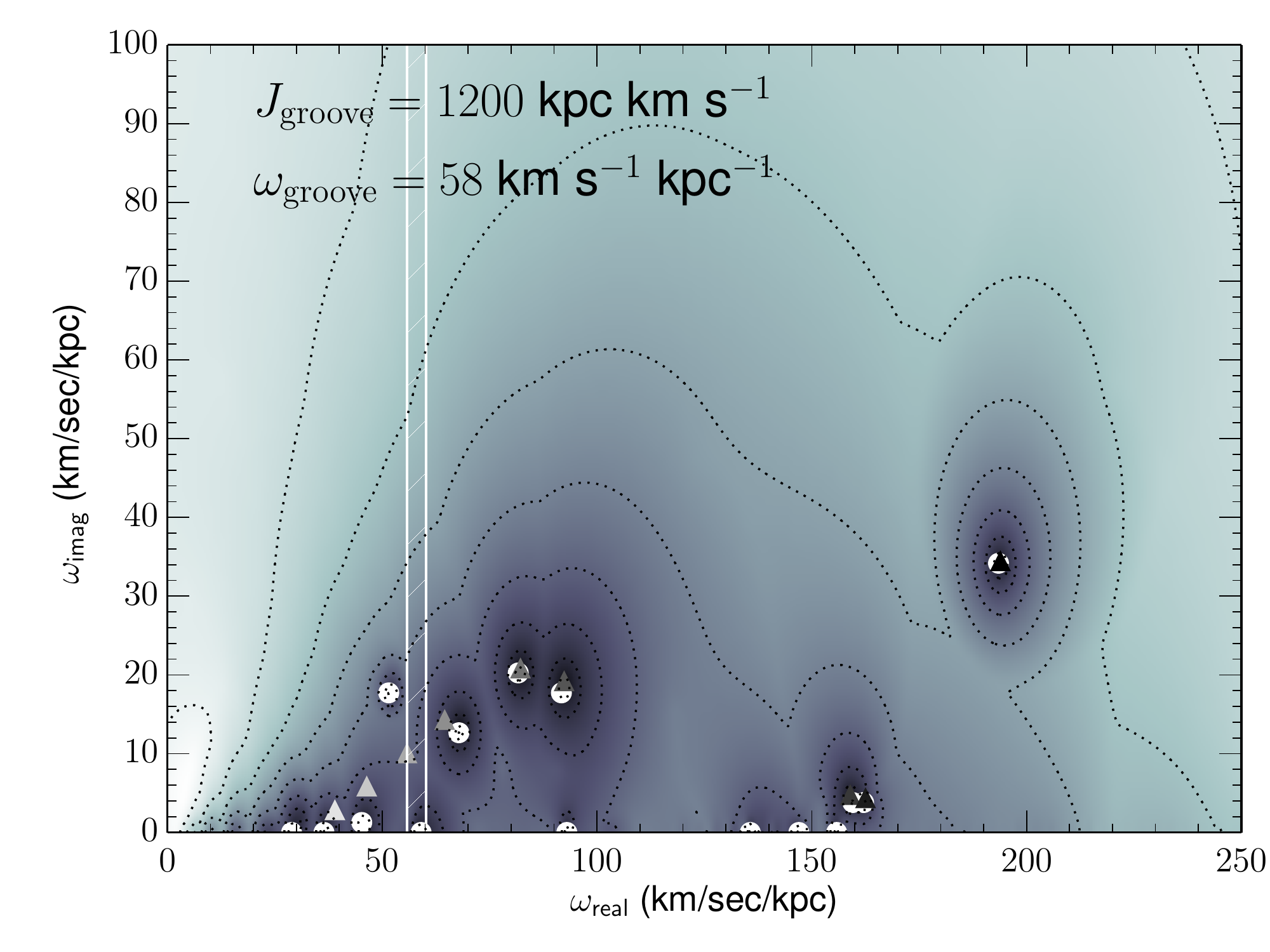} 
\includegraphics[trim=45 15 0 13,clip,width=0.420755\textwidth]{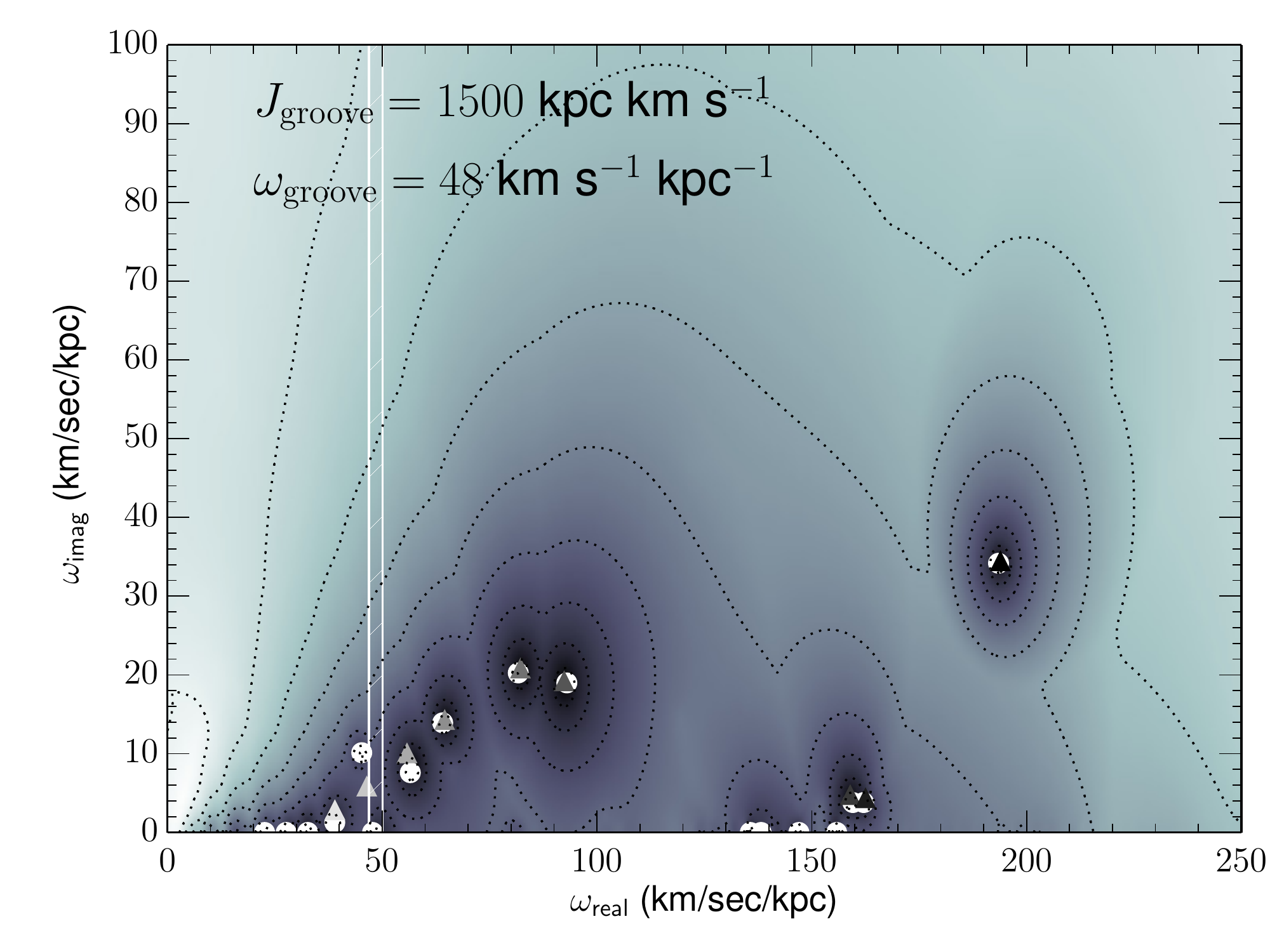} 
\caption{The $m=2$ mode spectrum in the complex frequency plane of
  cored exponential disc models with different narrow phase space
  grooves, centered on the angular momentum $J_{\rm groove}$ indicated
  in each panel. The white hatched region in each panel, centered on
  the frequency $\omega_{\rm groove}$, indicates the locus of the
  modes that corotate with stars on circular orbits inside the
  groove. The colored triangles indicate the position of the $m=2$
  eigenmodes of the original cored exponential disc model, as listed
  in Table \ref{tbl-1}.
 \label{fig:freqs.pdf}}
\end{figure*}

In Fig. \ref{fig:freqs.pdf}, the mode spectra of several cored
exponential disc models with different narrow phase space grooves,
centered on the angular momentum $J_{\rm groove}$ indicated in each
panel, are shown in the complex frequency plane. These grooved models
are listed in Table \ref{tbl-2}. From left to right and from top to
bottom in Fig. \ref{fig:freqs.pdf}, the angular momentum $J_{\rm
  groove}$ of the groove increases while the corotation frequency
$\omega_{\rm groove}$ decreases. The white hatched region in each
panel, centered on the frequency $\omega_{\rm groove}$, indicates the
locus of the modes that corotate with stars on circular orbits inside
the groove. The colored triangles indicate the position of the growing
eigenmodes of the original cored exponential disc model, as listed in
Table \ref{tbl-1}. For any choice of $J_{\rm groove}$, there are modes
present that grow faster than in the ungrooved model. Only for very
high $J_{\rm groove}$-values does the grooved model's eigenmode
spectrum approach that of the ungrooved model.

A groove in phase space clearly can have an impressive and
destabilizing effect on the eigenmode spectrum of a disc galaxy model,
dramatically affecting the number and the frequencies of the
modes. Below, we discuss the modes associated with the groove in more
detail.

\begin{table}
\begin{center}
\caption{Models with a groove around angular momentum $J_{\rm
    groove}$. $r_{\rm circle}$ is the radius of the circular orbit
  with angular momentum $J_{\rm groove}$. $\omega_{\rm groove}$ is the
  frequency of a mode with corotation at $r_{\rm circle}$.
\label{tbl-2}}
\begin{tabular}{ccc} \hline
$J_{\rm groove}$ (kpc~km~s$^{-1}$) & $\omega_{\rm groove}$ (km~s$^{-1}$~kpc$^{-1}$) & $r_{\rm circle}$ (kpc) \\ \cline{1-3}
100 & 144.8 & 1.18 \\
200 & 131.2 & 1.75 \\
300 & 119.0 & 2.25 \\
433 & 105.1 & 2.87 \\
500 &  98.9 & 3.18 \\
600 &  90.6 & 3.64 \\
700 &  93.3 & 4.10 \\
800 &  76.9 & 4.56 \\
900 &  71.3 & 5.03 \\
1200 & 57.9 & 6.44 \\
1500 & 48.4 & 7.87 \\ \cline{1-3}
\end{tabular}
\end{center}
\end{table}

\begin{figure*}
\includegraphics[trim=55 35 92 10,clip,width=0.325\textwidth]{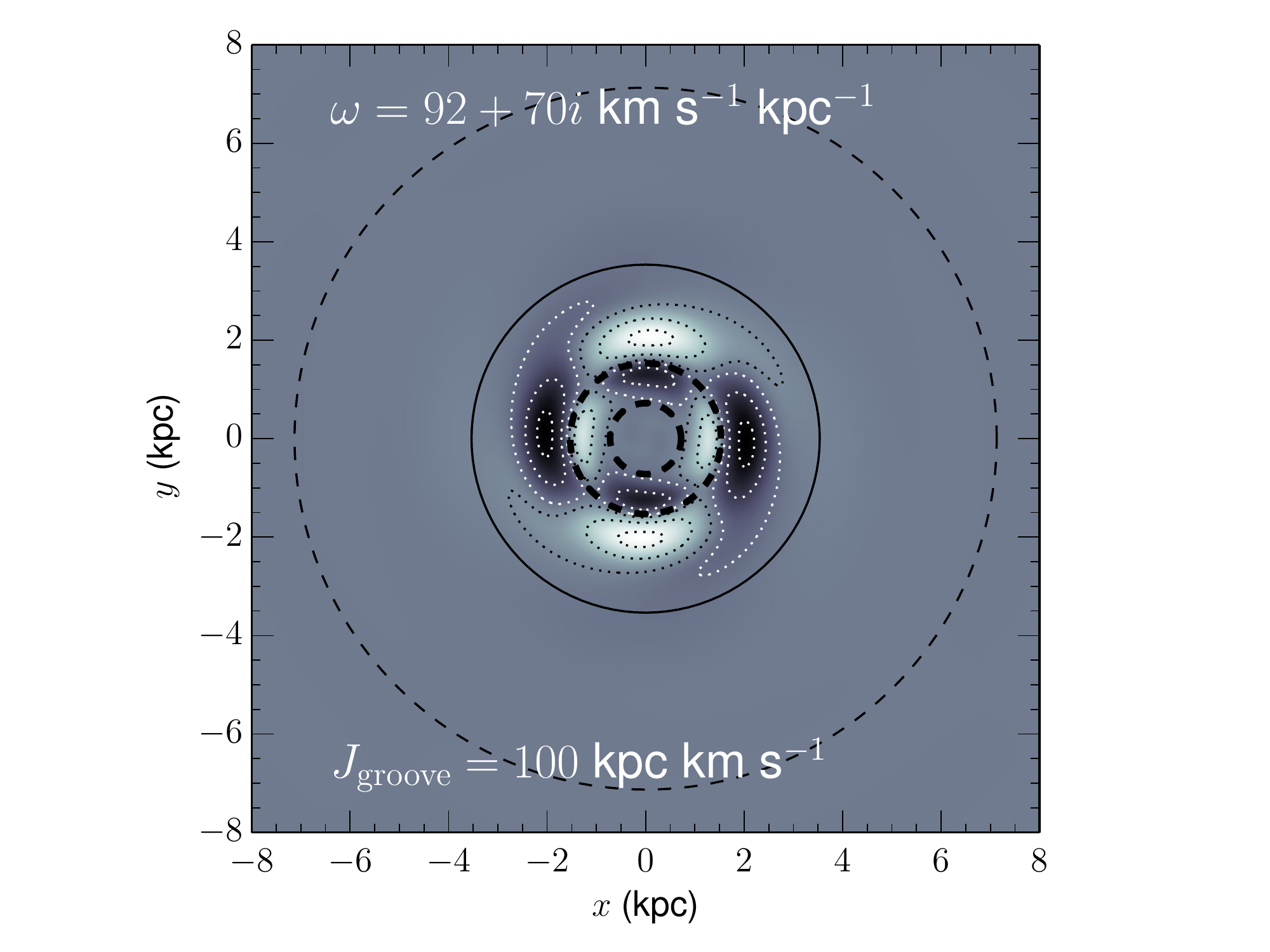}
\includegraphics[trim=55 35 92 10,clip,width=0.325\textwidth]{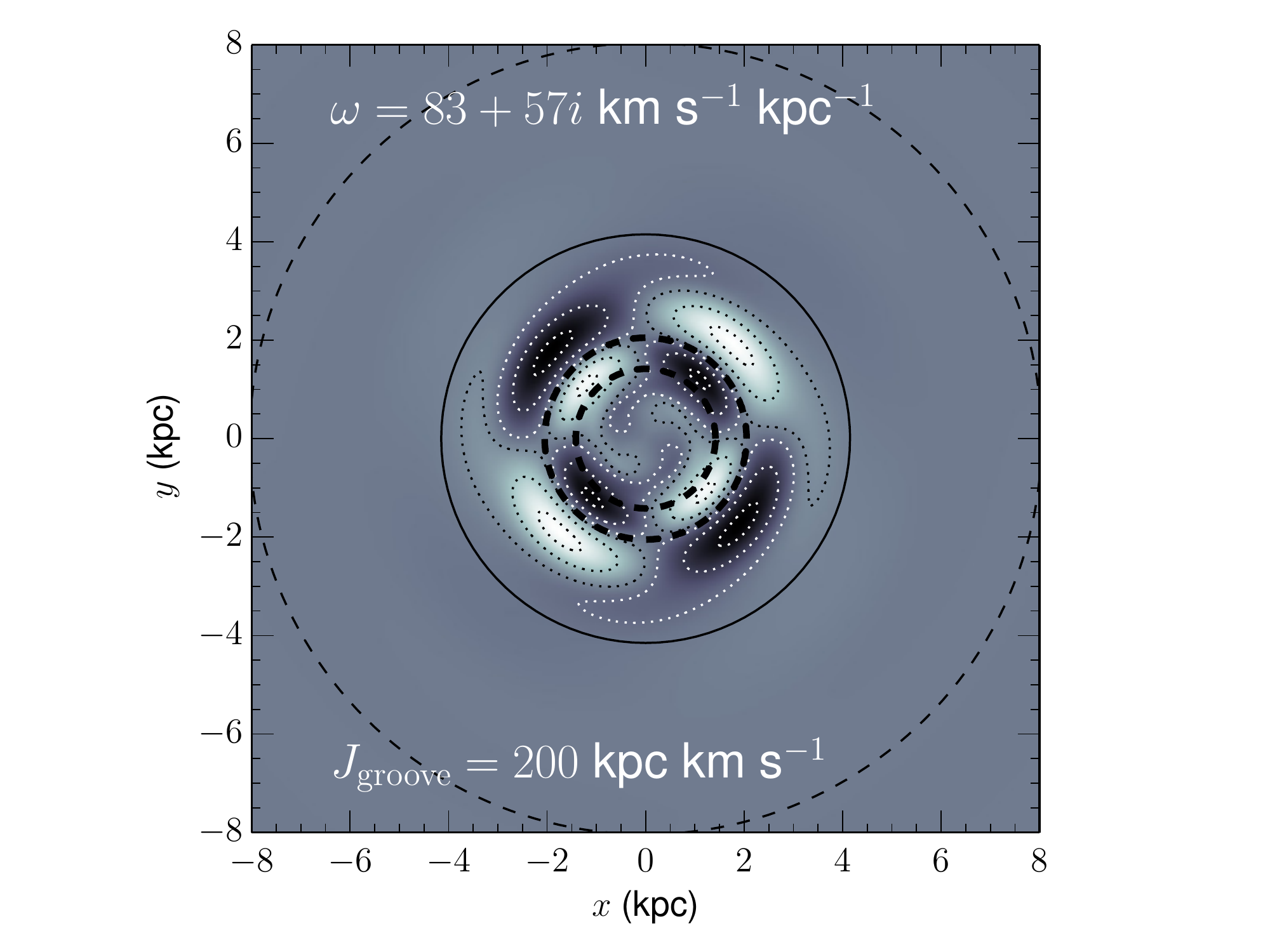}
\includegraphics[trim=55 35 92 10,clip,width=0.325\textwidth]{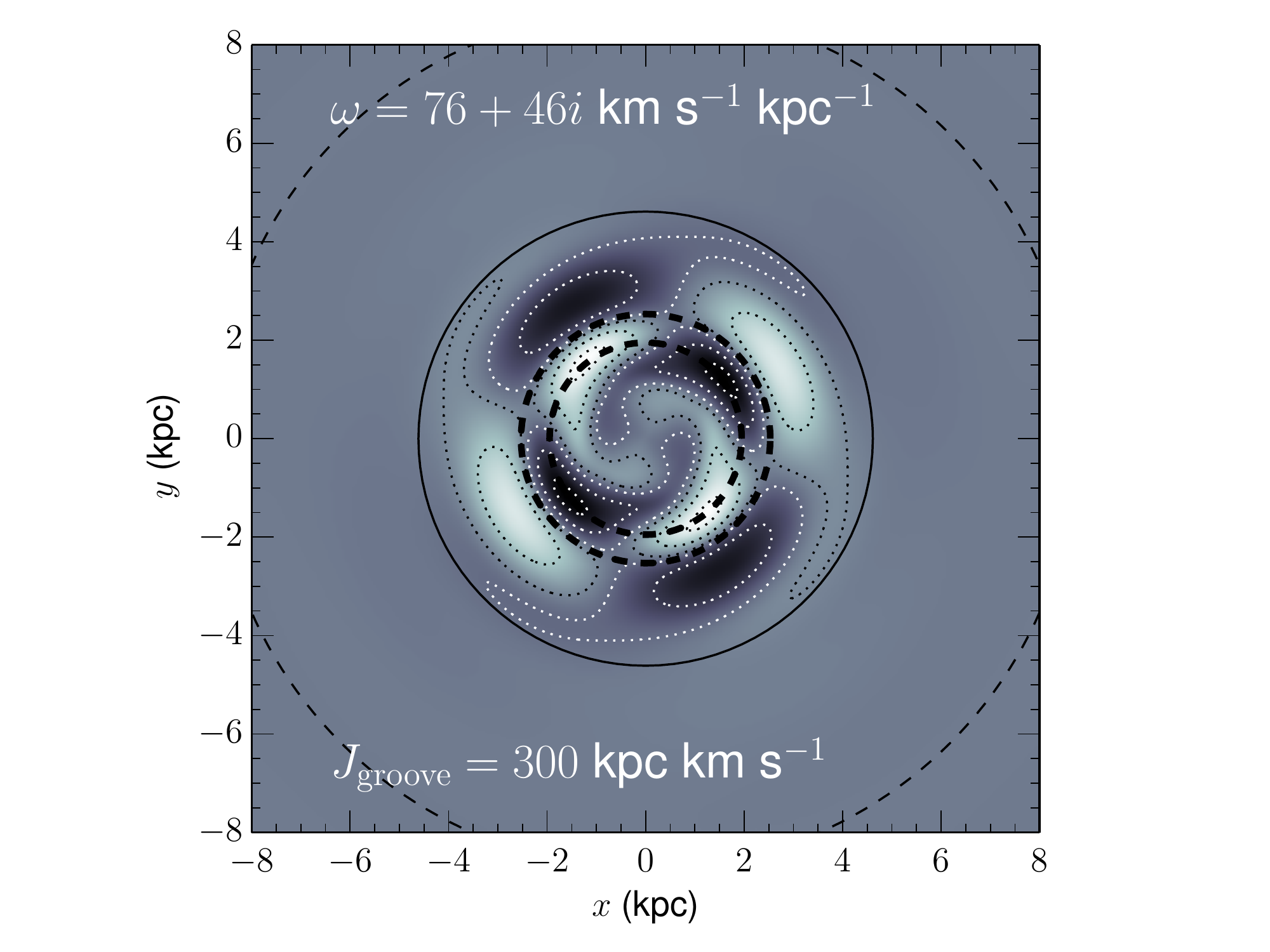}
\includegraphics[trim=55 35 92 10,clip,width=0.325\textwidth]{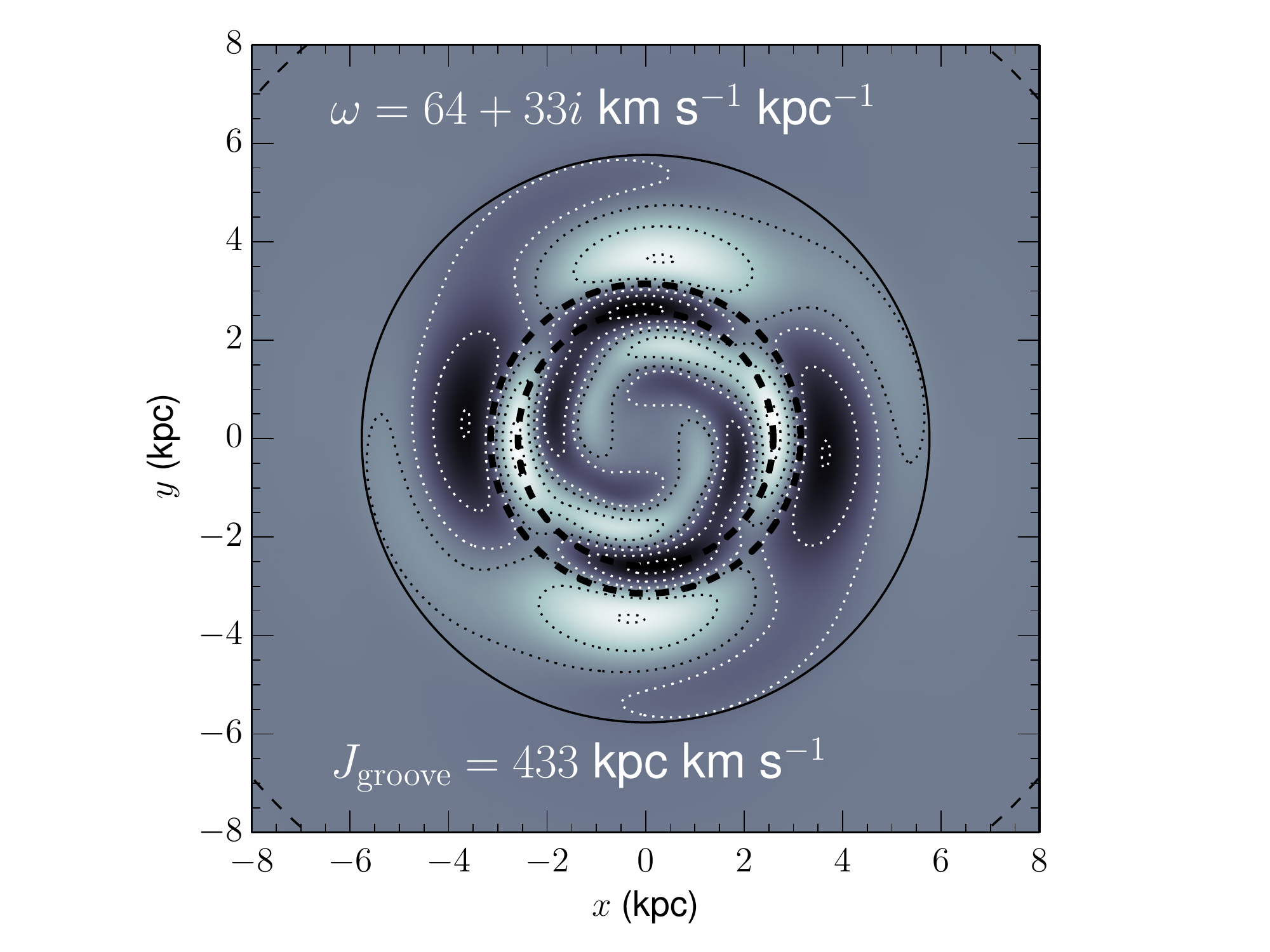}
\includegraphics[trim=55 35 92 10,clip,width=0.325\textwidth]{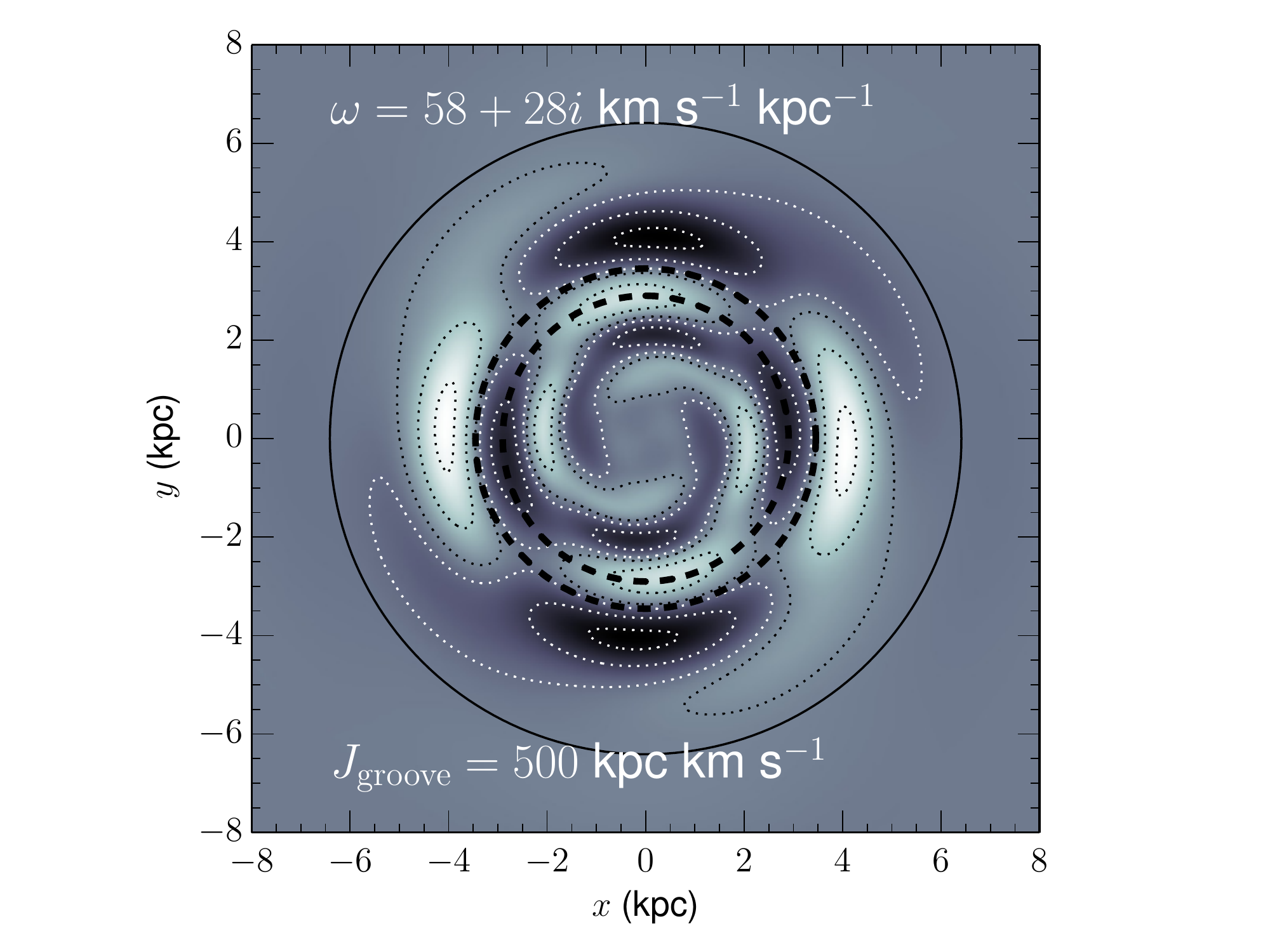}
\includegraphics[trim=55 35 92 10,clip,width=0.325\textwidth]{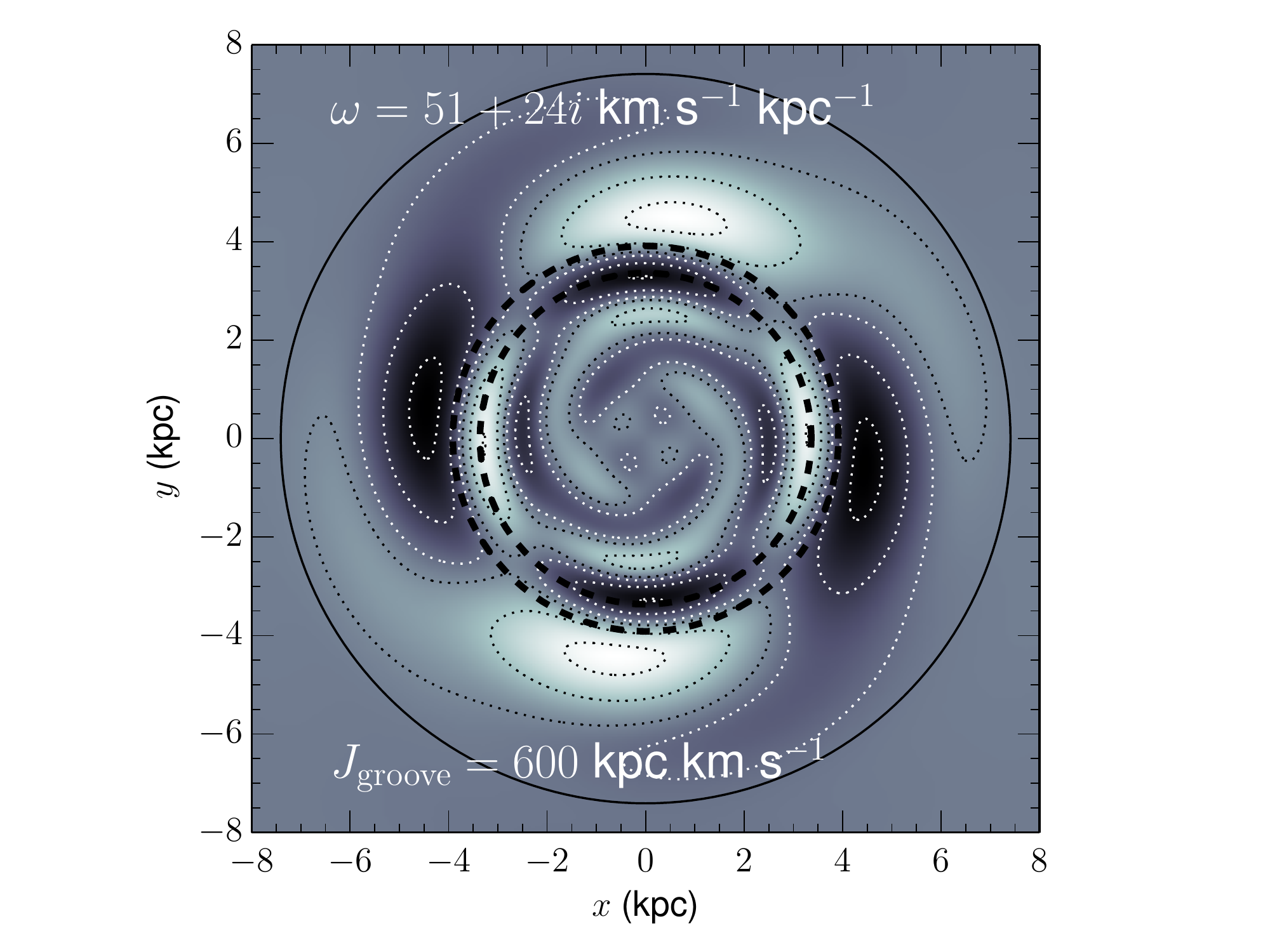}
\includegraphics[trim=55 10 92 10,clip,width=0.325\textwidth]{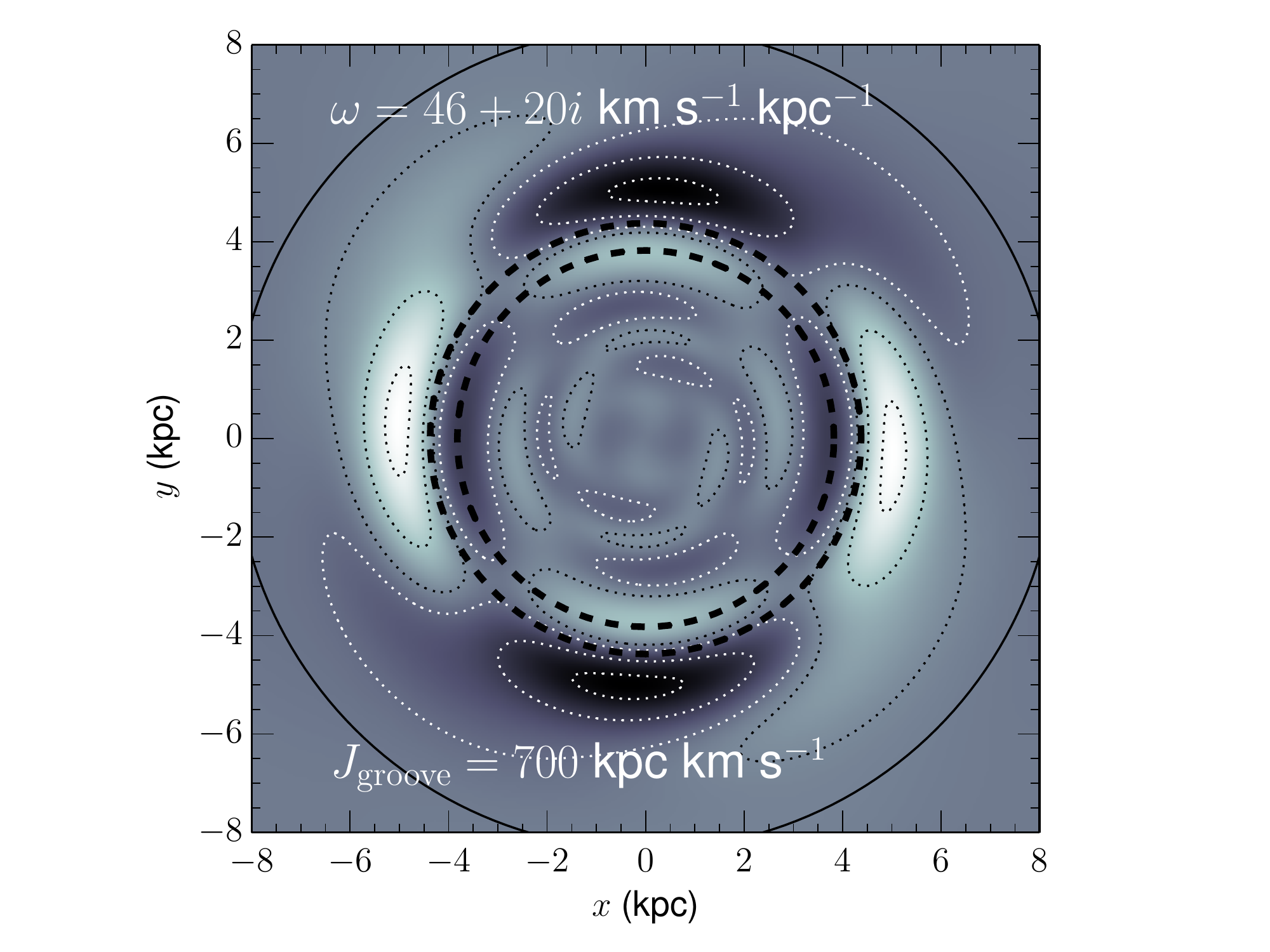}
\includegraphics[trim=55 10 92 10,clip,width=0.325\textwidth]{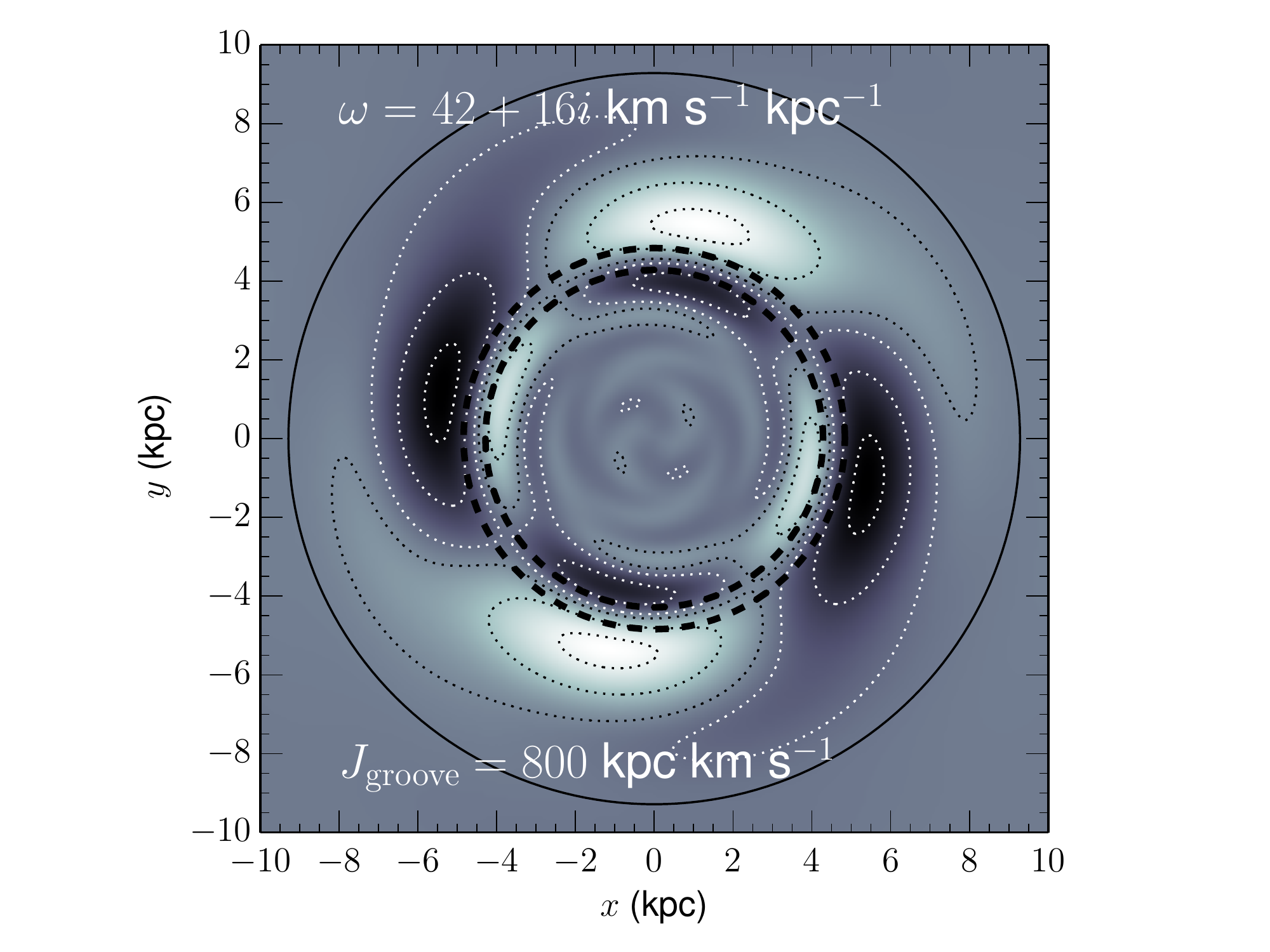}
\includegraphics[trim=55 10 92 10,clip,width=0.325\textwidth]{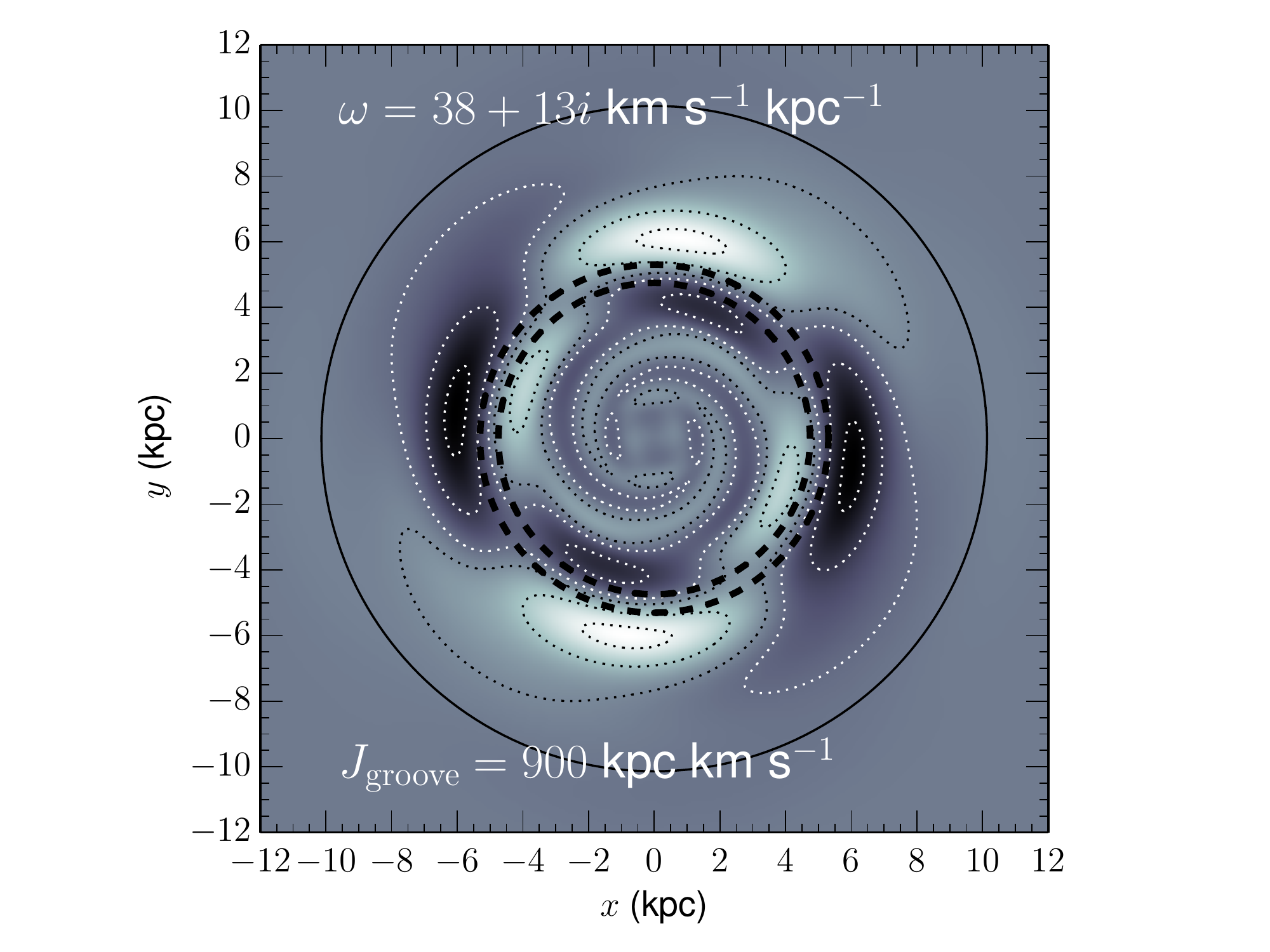}
\caption{Surface density of the $m=2$ ``low-frequency'' modes in the
  cored exponential disc model with a groove at $J_{\rm groove}$ as
  indicated in each panel. Each panel is labeled with the complex
  frequency $\omega$ of the mode in question. The groove edges are
  indicated in thick dashed lines, the corotation radius in a thin
  full line, and the outer Lindblad resonance in a thin dashed line.
  \label{fig:left.pdf}}
\end{figure*}

\begin{figure*}
\includegraphics[trim=55 35 92 10,clip,width=0.325\textwidth]{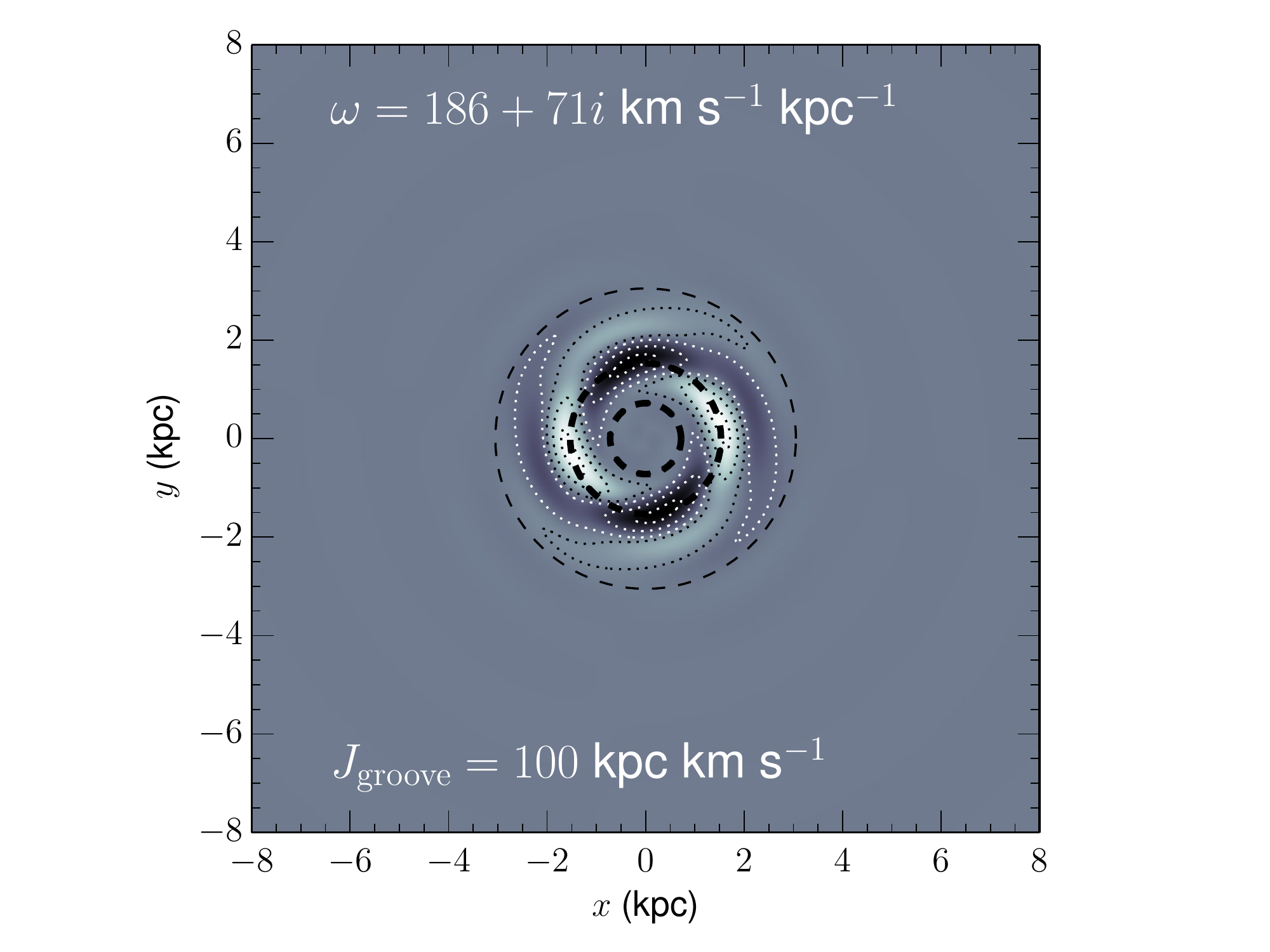}
\includegraphics[trim=55 35 92 10,clip,width=0.325\textwidth]{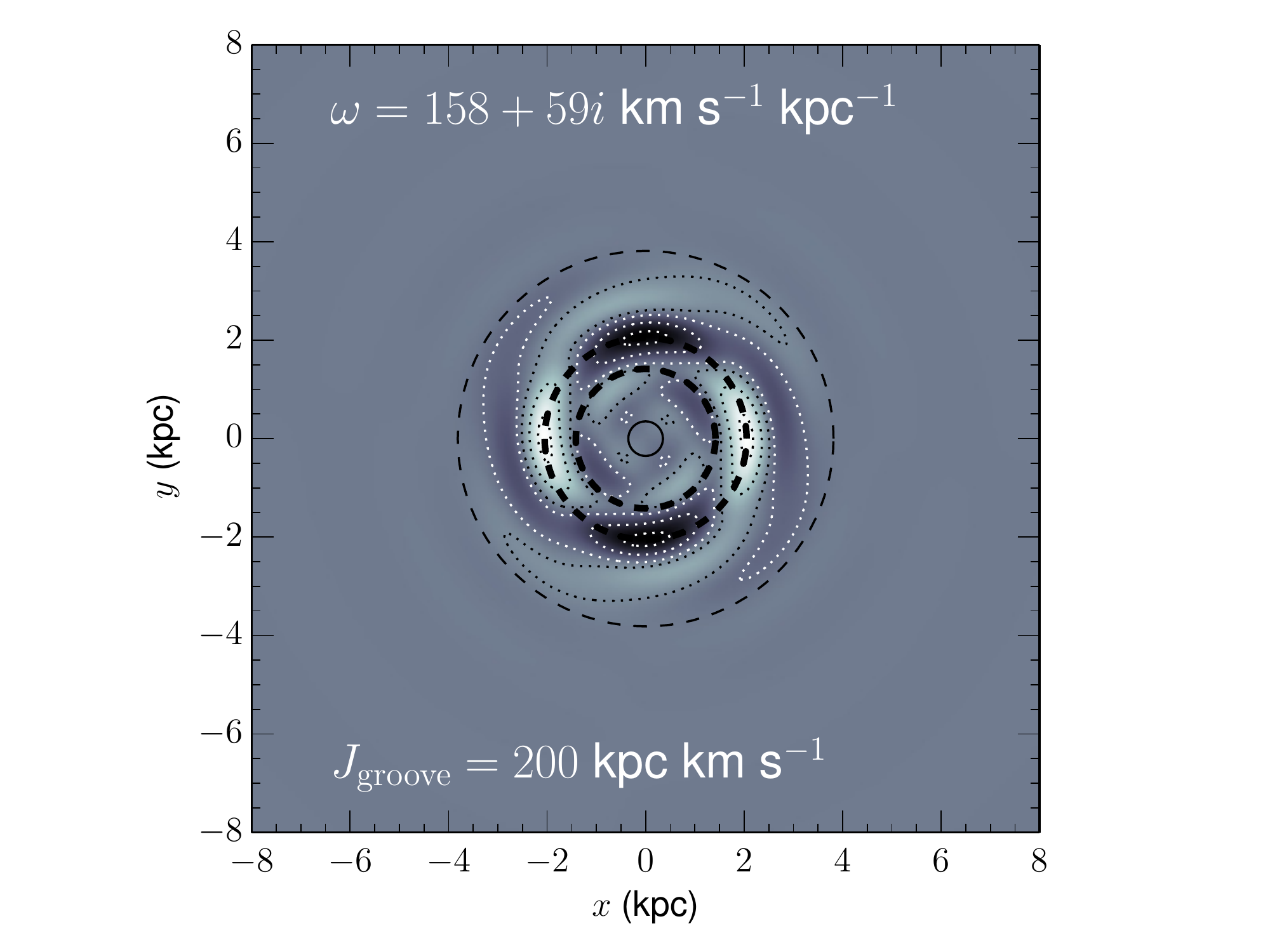}
\includegraphics[trim=55 35 92 10,clip,width=0.325\textwidth]{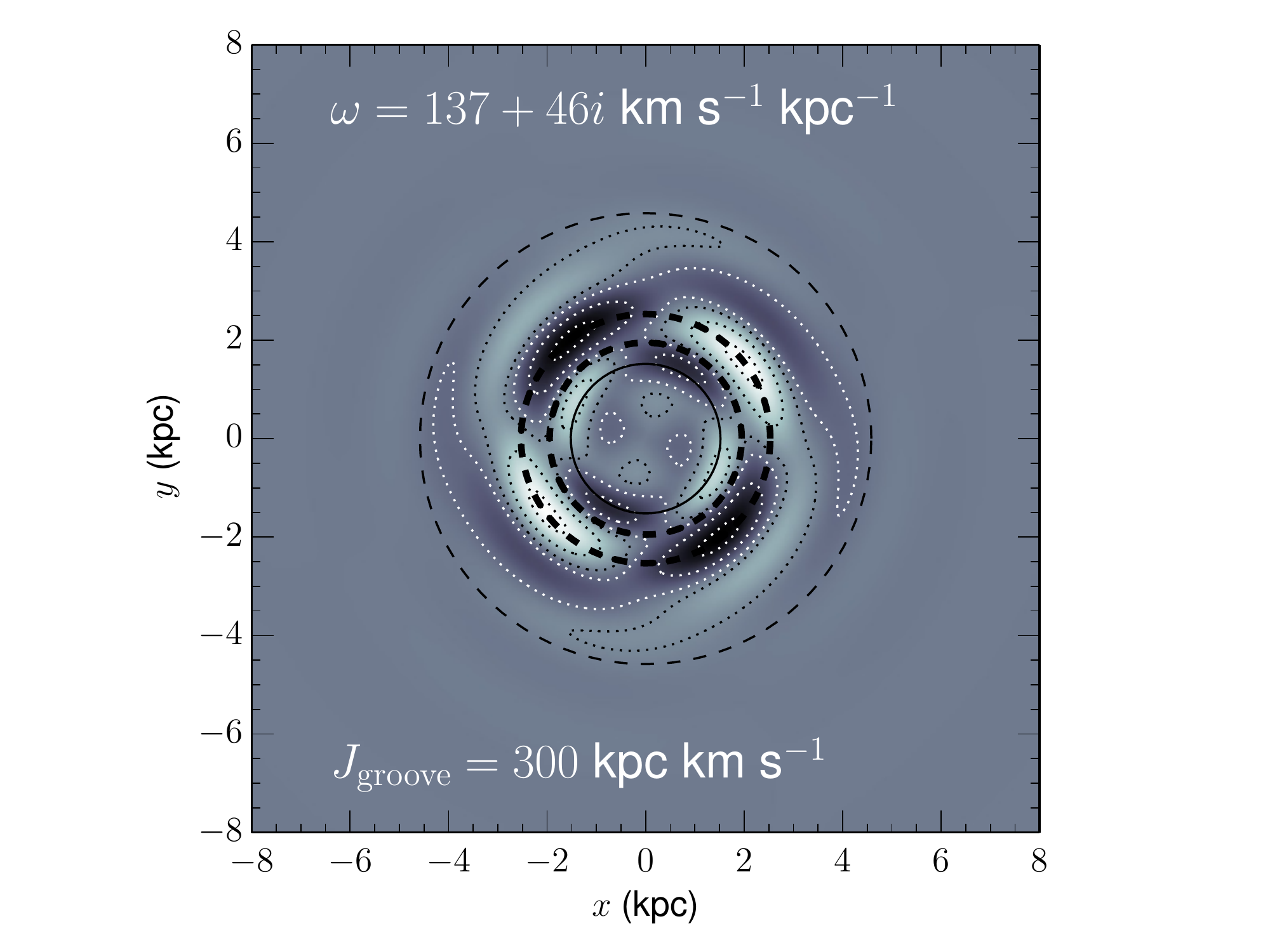}
\includegraphics[trim=55 35 92 10,clip,width=0.325\textwidth]{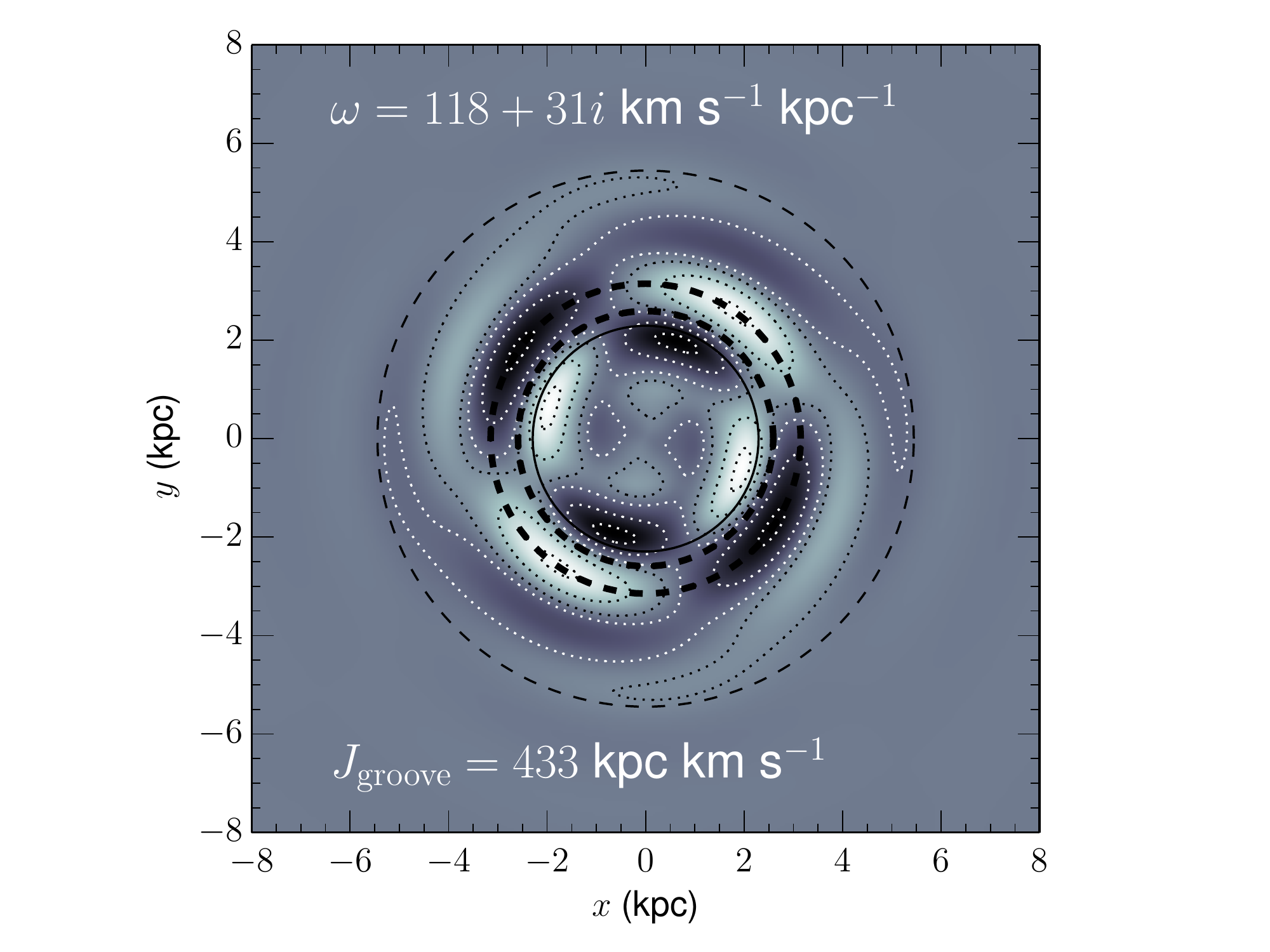}
\includegraphics[trim=55 35 92 10,clip,width=0.325\textwidth]{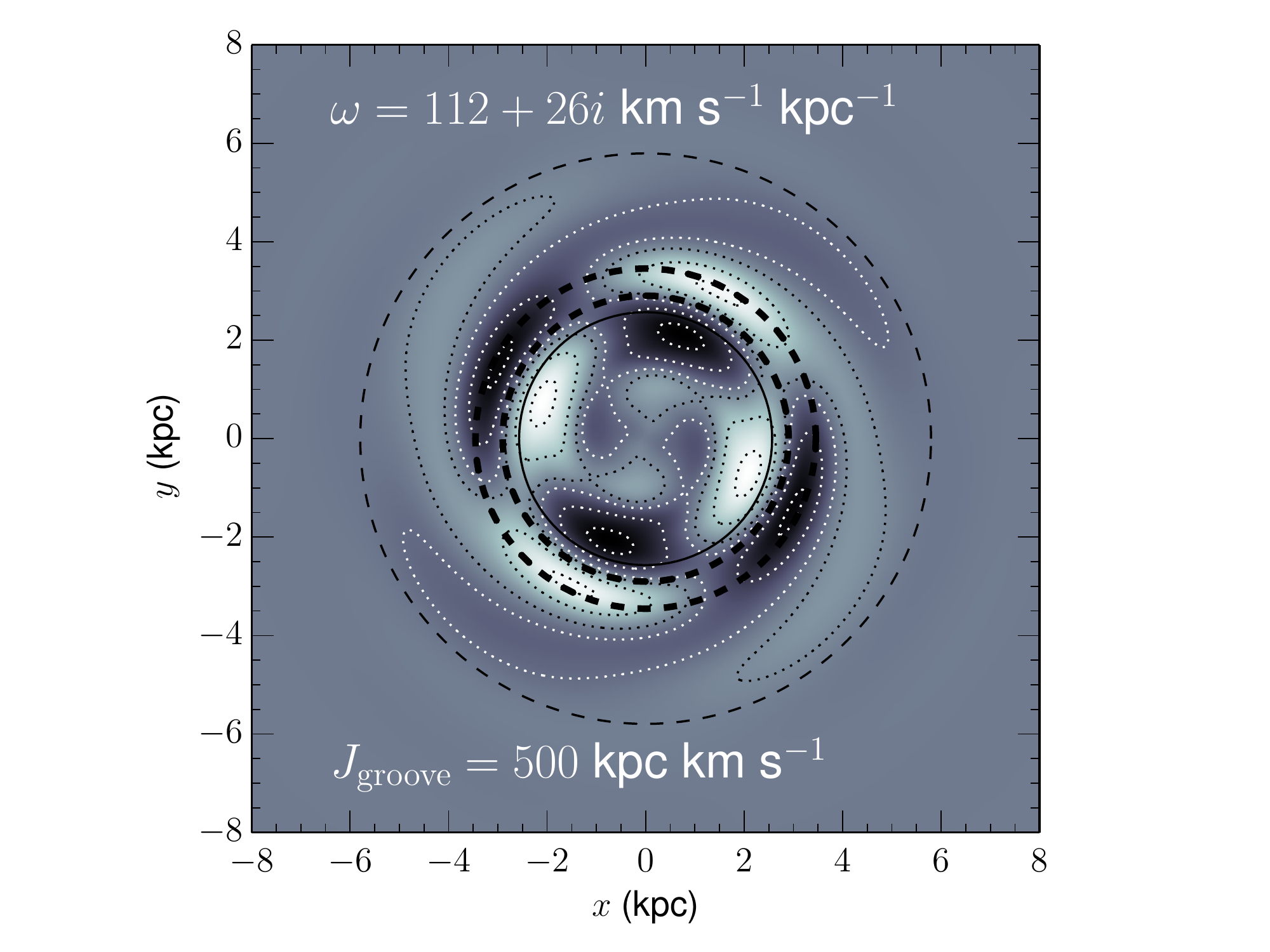}
\includegraphics[trim=55 35 92 10,clip,width=0.325\textwidth]{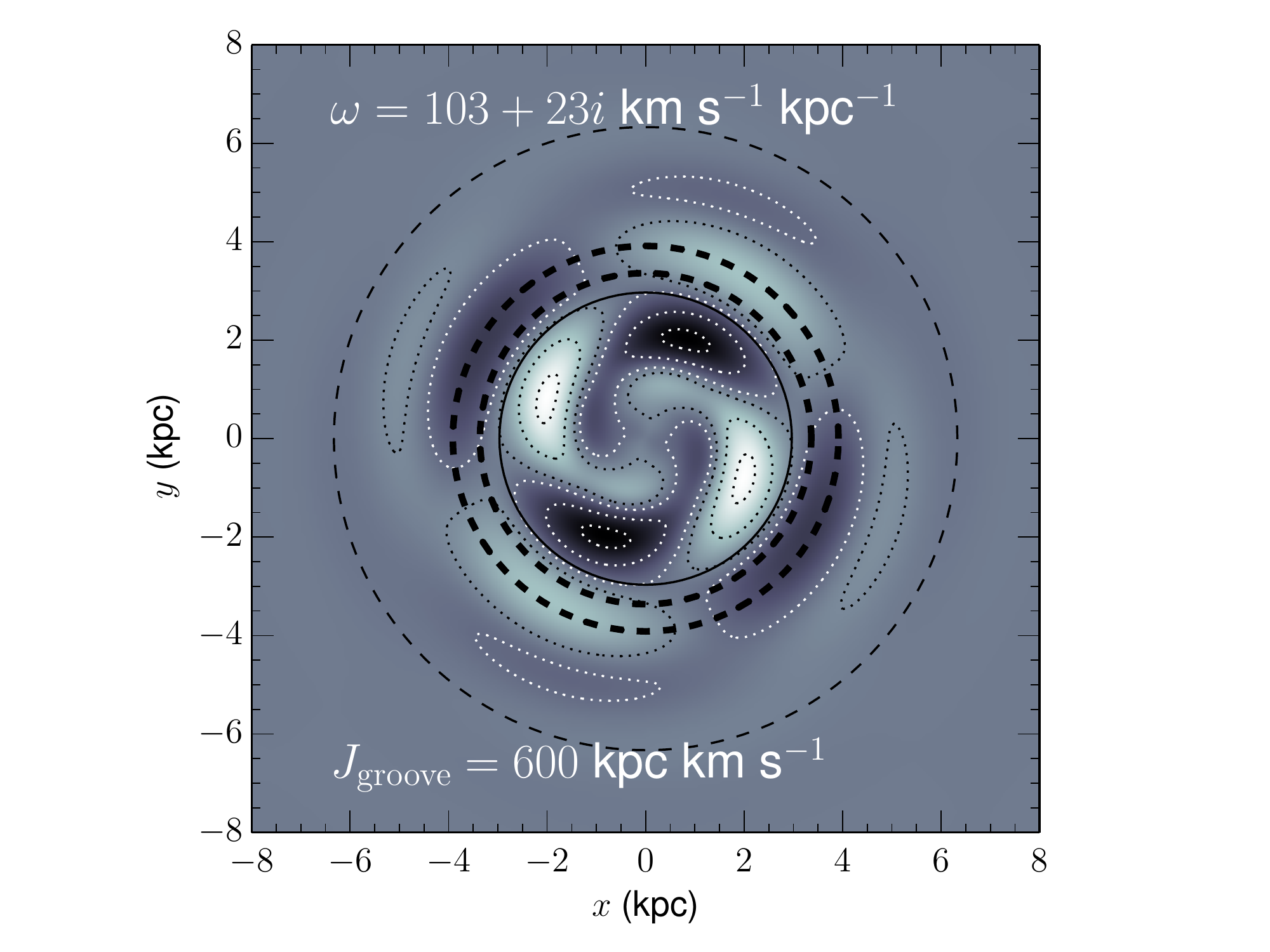}
\includegraphics[trim=55 10 92 10,clip,width=0.325\textwidth]{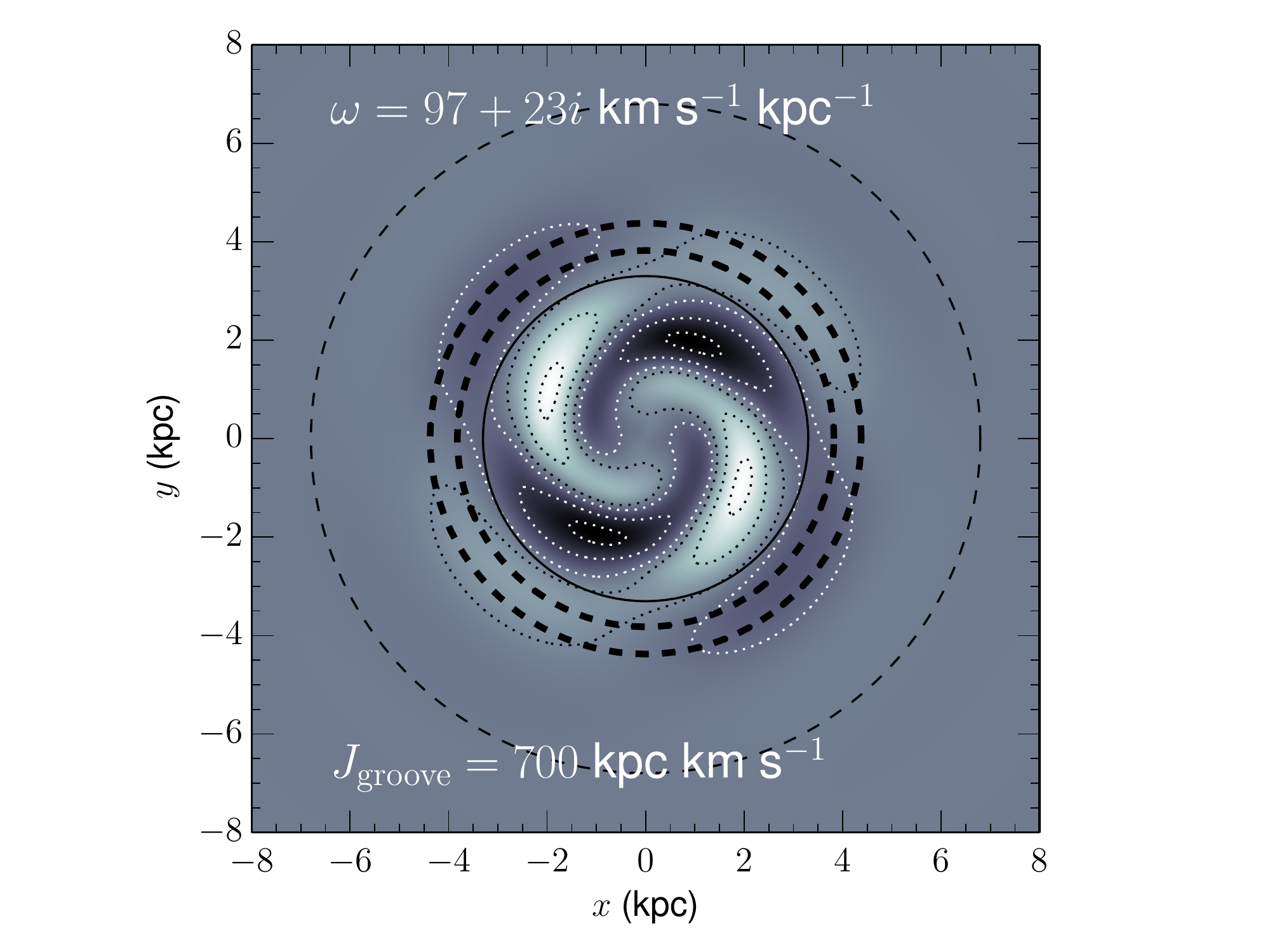}
\includegraphics[trim=55 10 92 10,clip,width=0.325\textwidth]{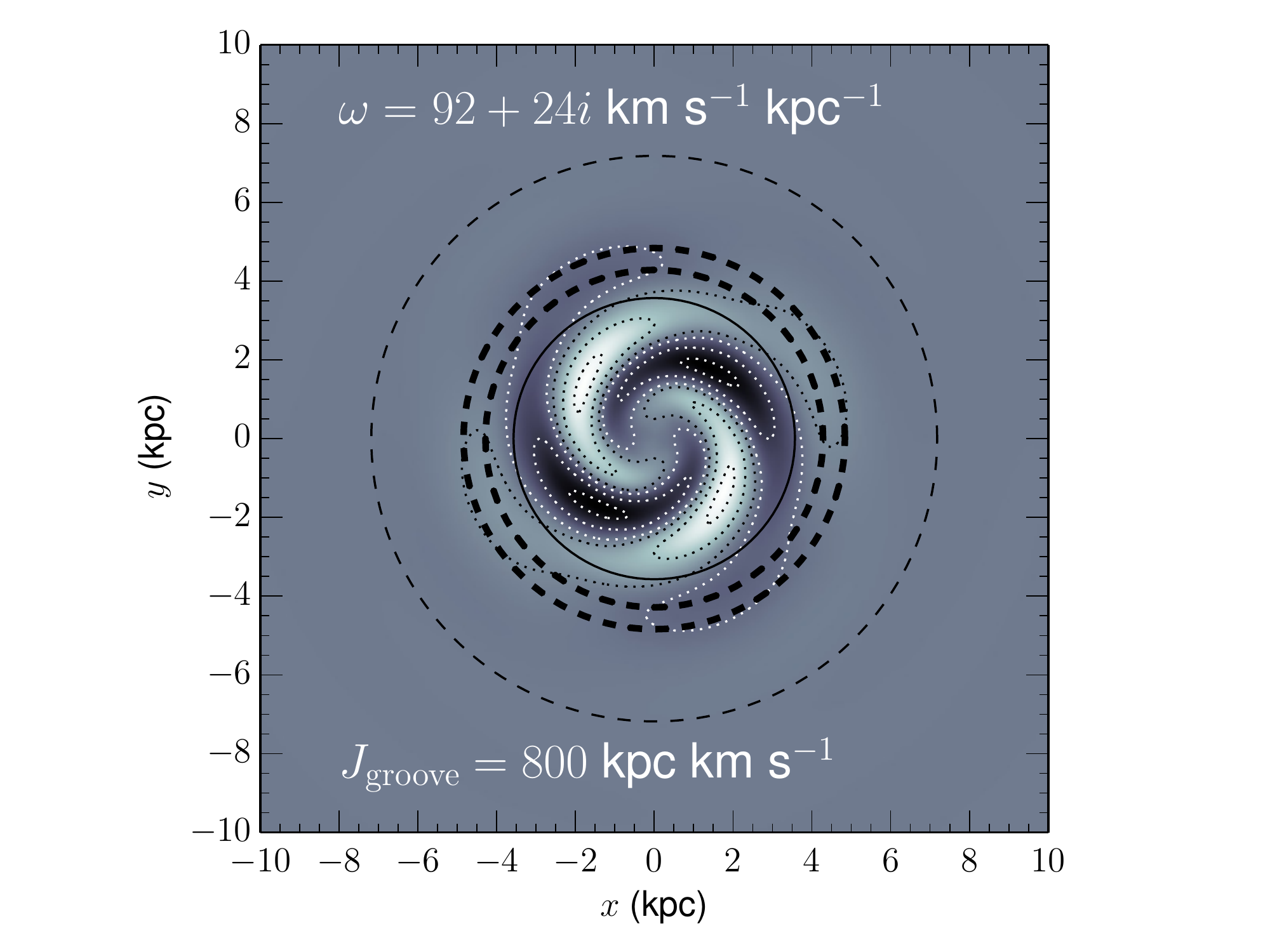}
\includegraphics[trim=55 10 92 10,clip,width=0.325\textwidth]{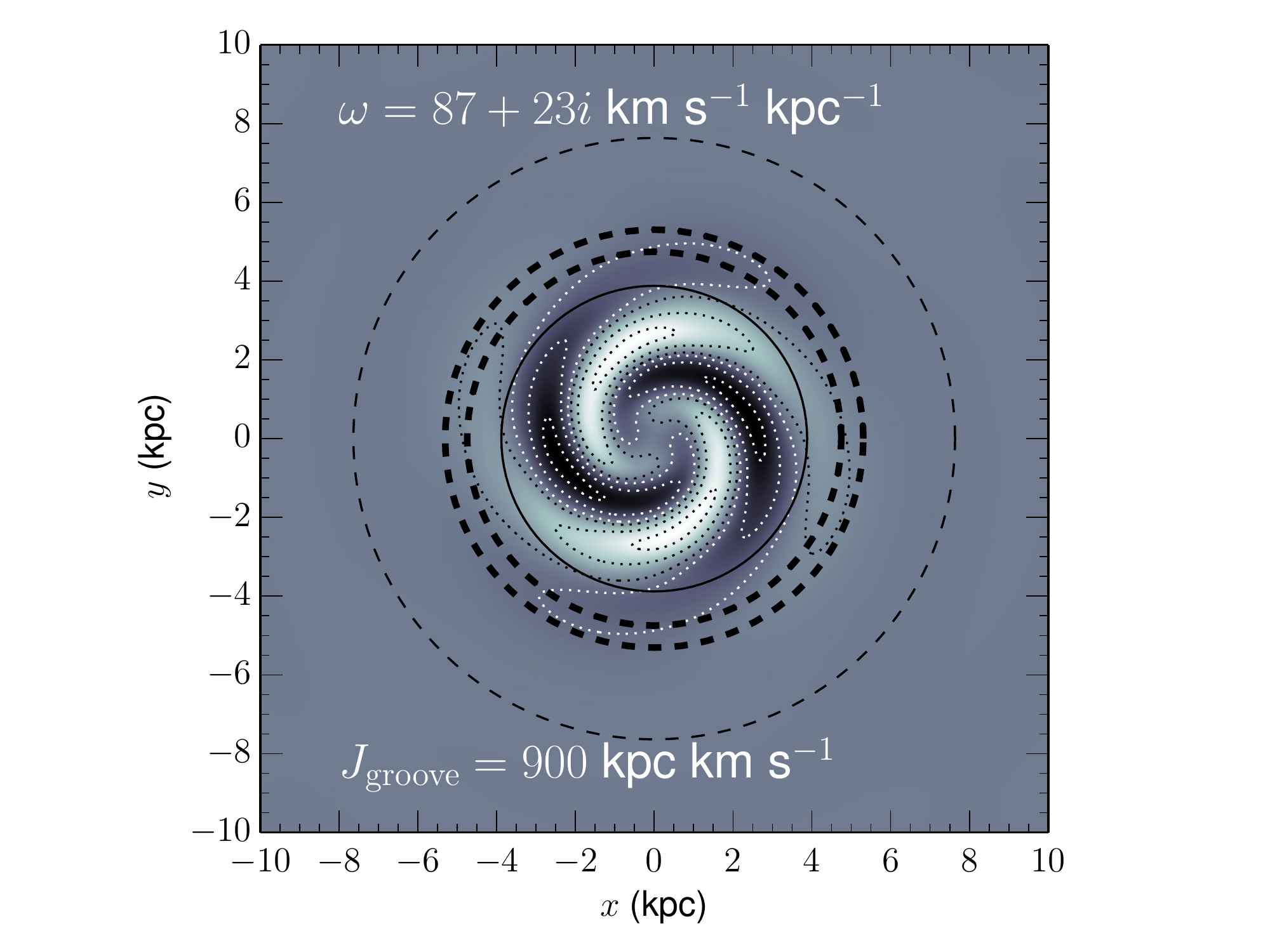}
\caption{Surface density of the $m=2$ ``high-frequency'' modes in the
  cored exponential disc model with a groove at $J_{\rm groove}$ as
  indicated in each panel. Each panel is labeled with the complex
  frequency $\omega$ of the mode in question.  The groove edges are
  indicated in thick dashed lines, the corotation radius in a thin
  full line, and the outer Lindblad resonance in a thin dashed line.
  \label{fig:right.pdf}}
\end{figure*}

\subsection{High and low frequency modes}

The grooved exponential disc supports couples of modes that straddle
the groove in frequency space. In other words:~these modes have
corotation radii either inside (the so-called ``high-frequency''
modes) or outside (the so-called ``low-frequency'' modes) the
groove. Their density distributions are shown in
Figs. \ref{fig:left.pdf} and \ref{fig:right.pdf}. Moreover, the two
modes of each couple have virtually identical growth rates, given by
$\omega_{\rm imag}$, and will co-evolve to non-linearity. Remarkably,
the OLR of the fastest rotating mode often lies very close to the CR
of the slowest rotating mode. Given this close resonance proximity,
they are likely to interact with each other, producing $m=0$ and $m=4$
beat waves \citep{sygnet88, masset97}.

Especially for grooves at small angular momentum, in this case this is
for $J_{\rm groove} \lesssim 400$~kpc~km~s$^{-1}$, these are the
fastest growing members of the grooved disc's eigenmode
spectrum. There are no $m=2$ modes in the ungrooved exponential disc
that could carve a groove at these small $J_{\rm groove}$-values
although the fastest rotating $m=4$ mode we found (cf. section
\ref{m4modes}) has its OLR at 2.36~kpc, which corresponds to a groove
angular momentum of $J_{\rm groove} = 323$~kpc~km~s$^{-1}$. For higher
$J_{\rm groove}$-values, these modes shift gradually towards lower
pattern speeds and smaller growth rates and, finally, they disappear
among the modes of the ungrooved model.

The low-frequency modes have a radial node at the groove's outer edge
while the high-frequency modes have a radial node at the groove's
inner edge (although the strength of the part of the mode outside the
groove diminishes with increasing $J_{\rm groove}$). This is likely to
be a strong clue regarding the origin of these modes as standing wave
patterns formed by traveling waves reflecting off the groove's inner
and outer edge.

\subsection{Medium frequency mode}

\begin{figure*}
\includegraphics[trim=55 35 92 10,clip,width=0.325\textwidth]{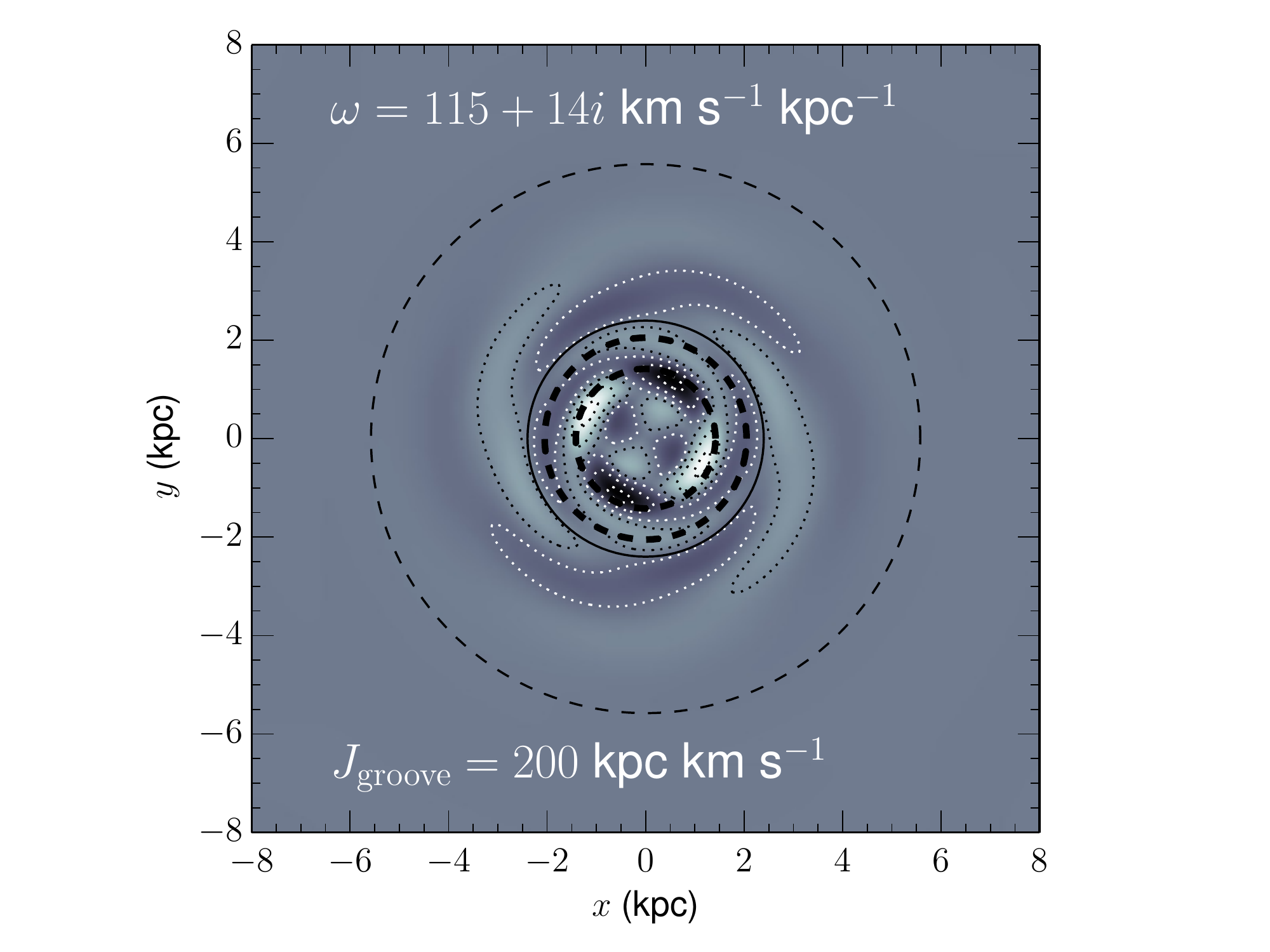}
\includegraphics[trim=55 35 92 10,clip,width=0.325\textwidth]{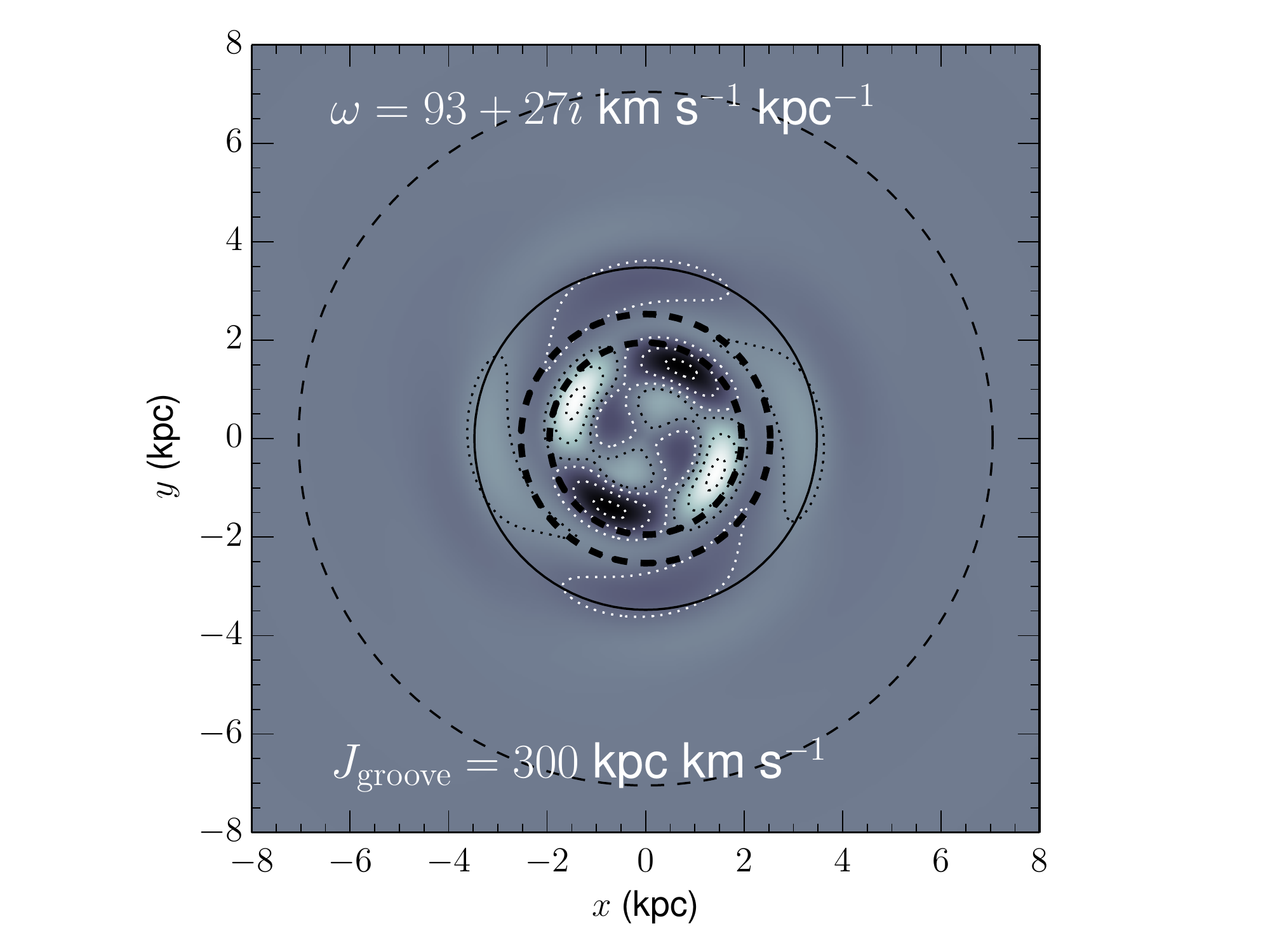}
\includegraphics[trim=55 35 92 10,clip,width=0.325\textwidth]{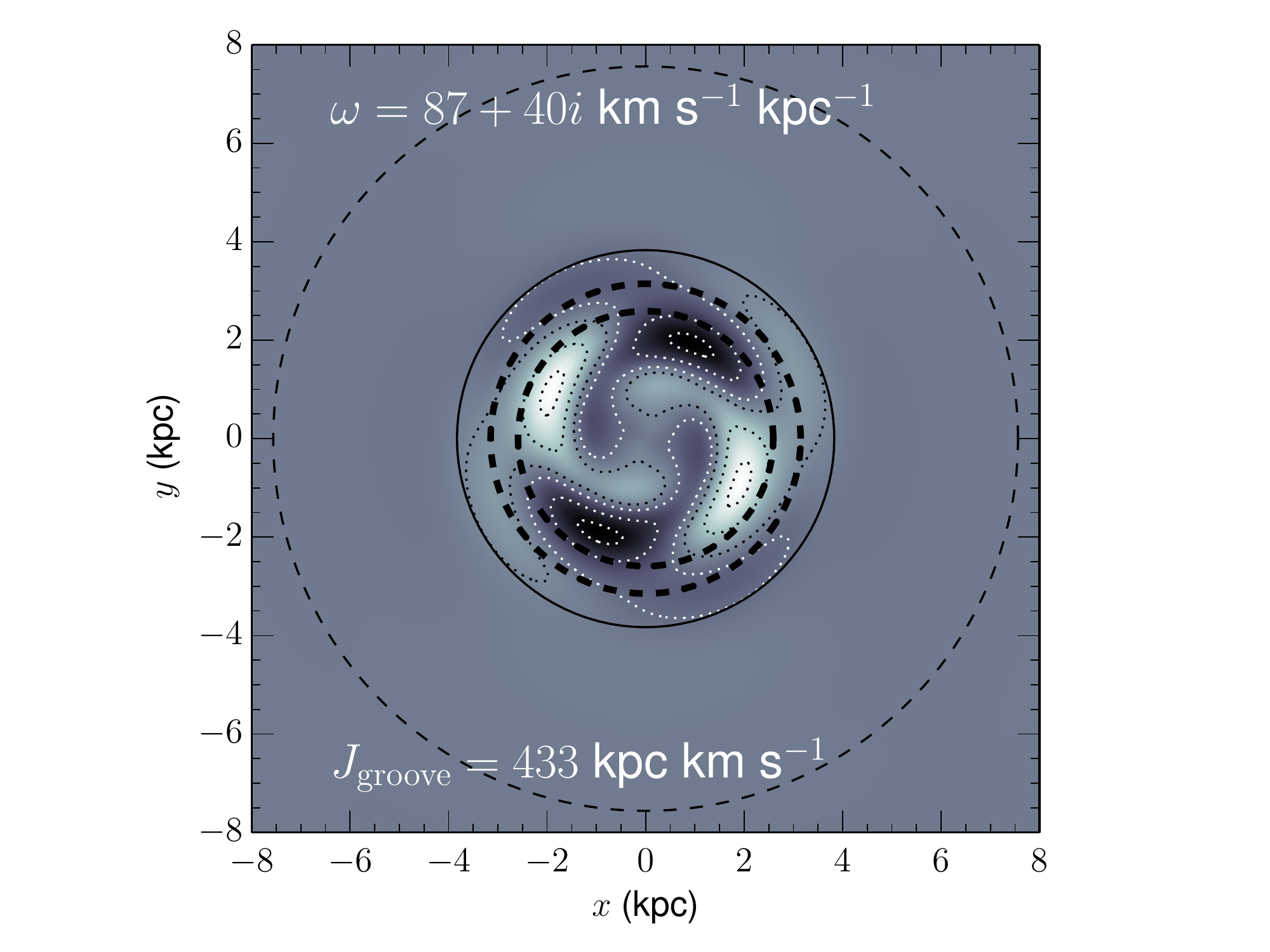}
\includegraphics[trim=55 35 92 10,clip,width=0.325\textwidth]{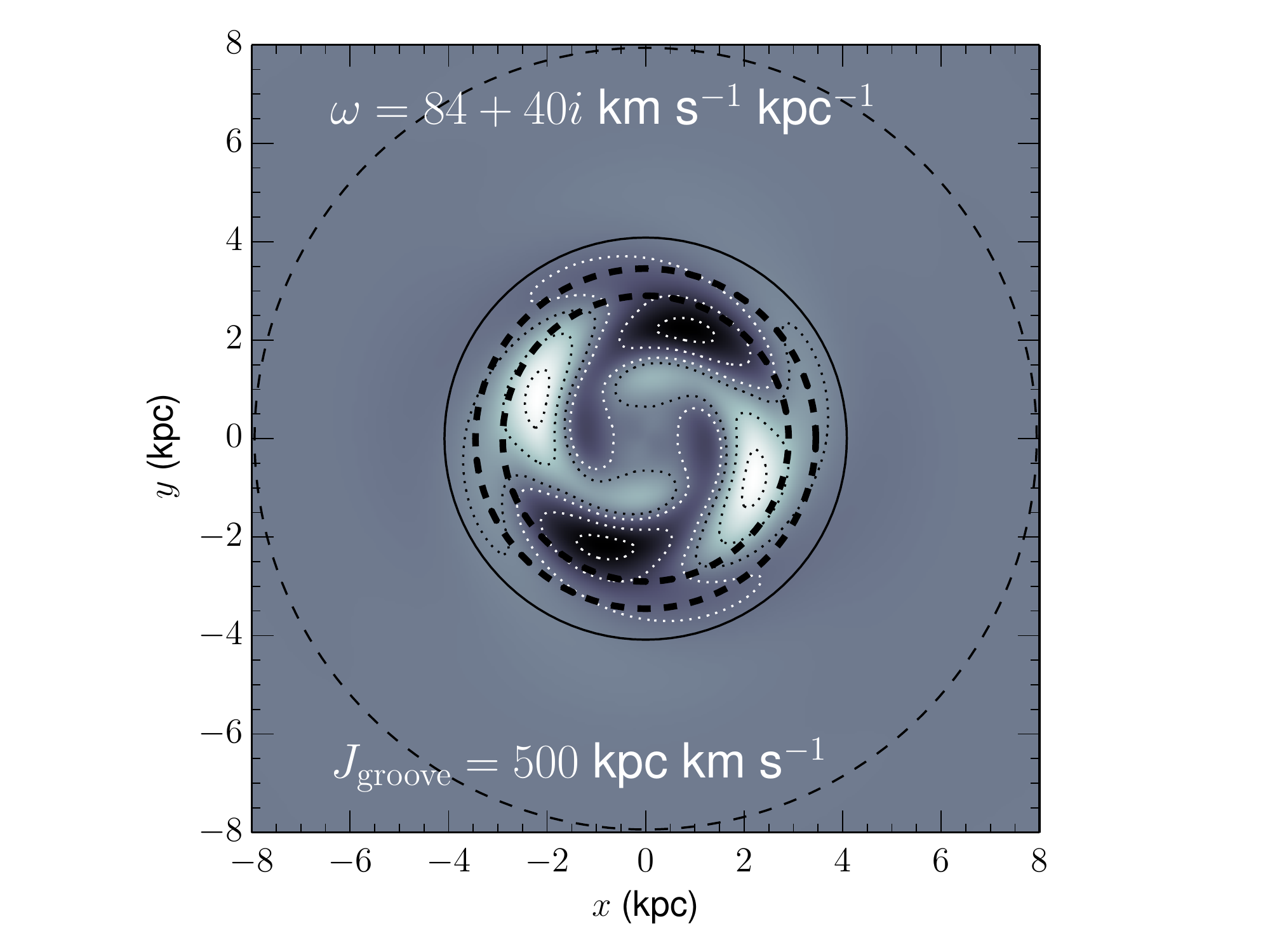}
\includegraphics[trim=55 35 92 10,clip,width=0.325\textwidth]{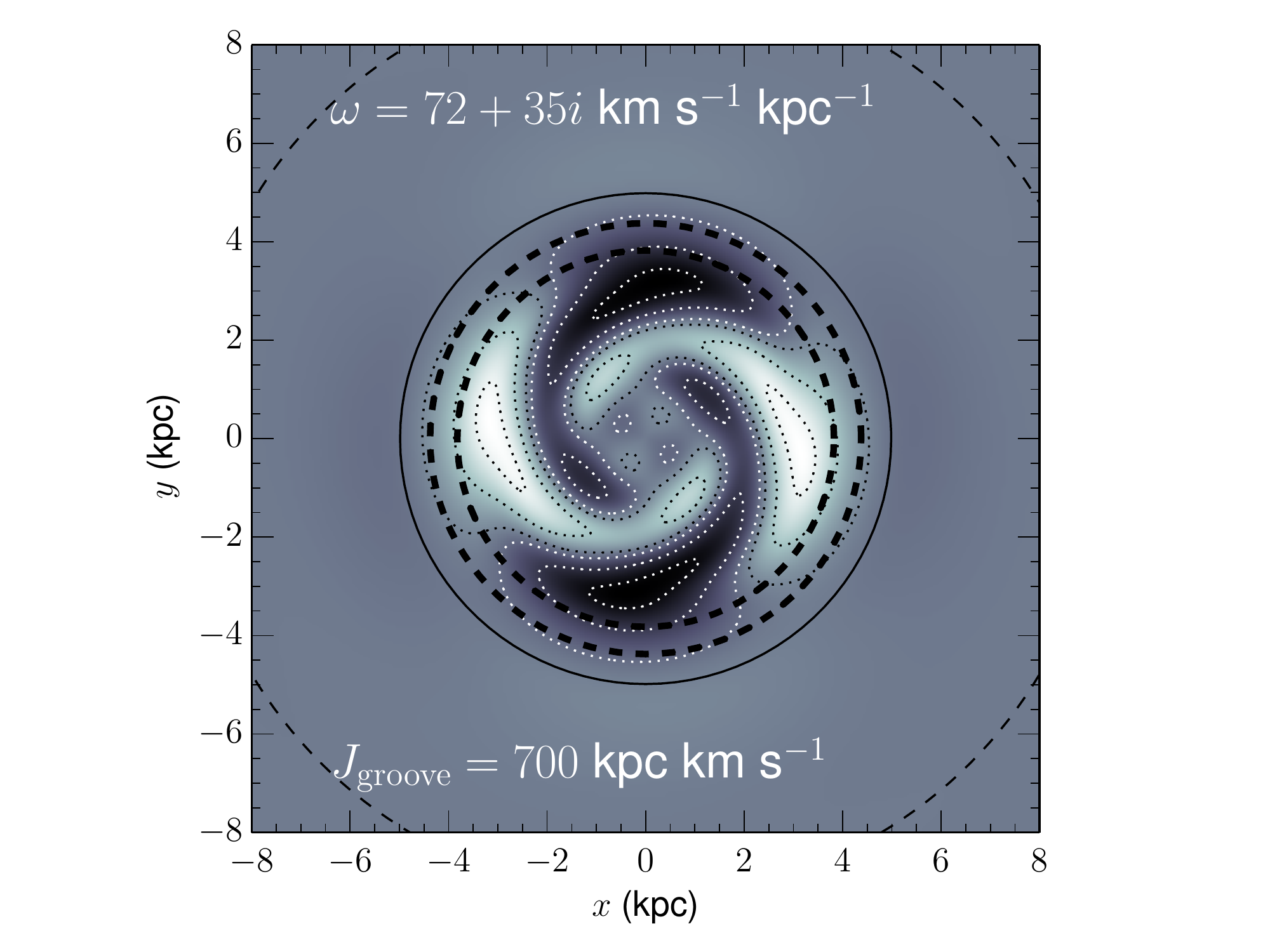}
\includegraphics[trim=55 35 92 10,clip,width=0.325\textwidth]{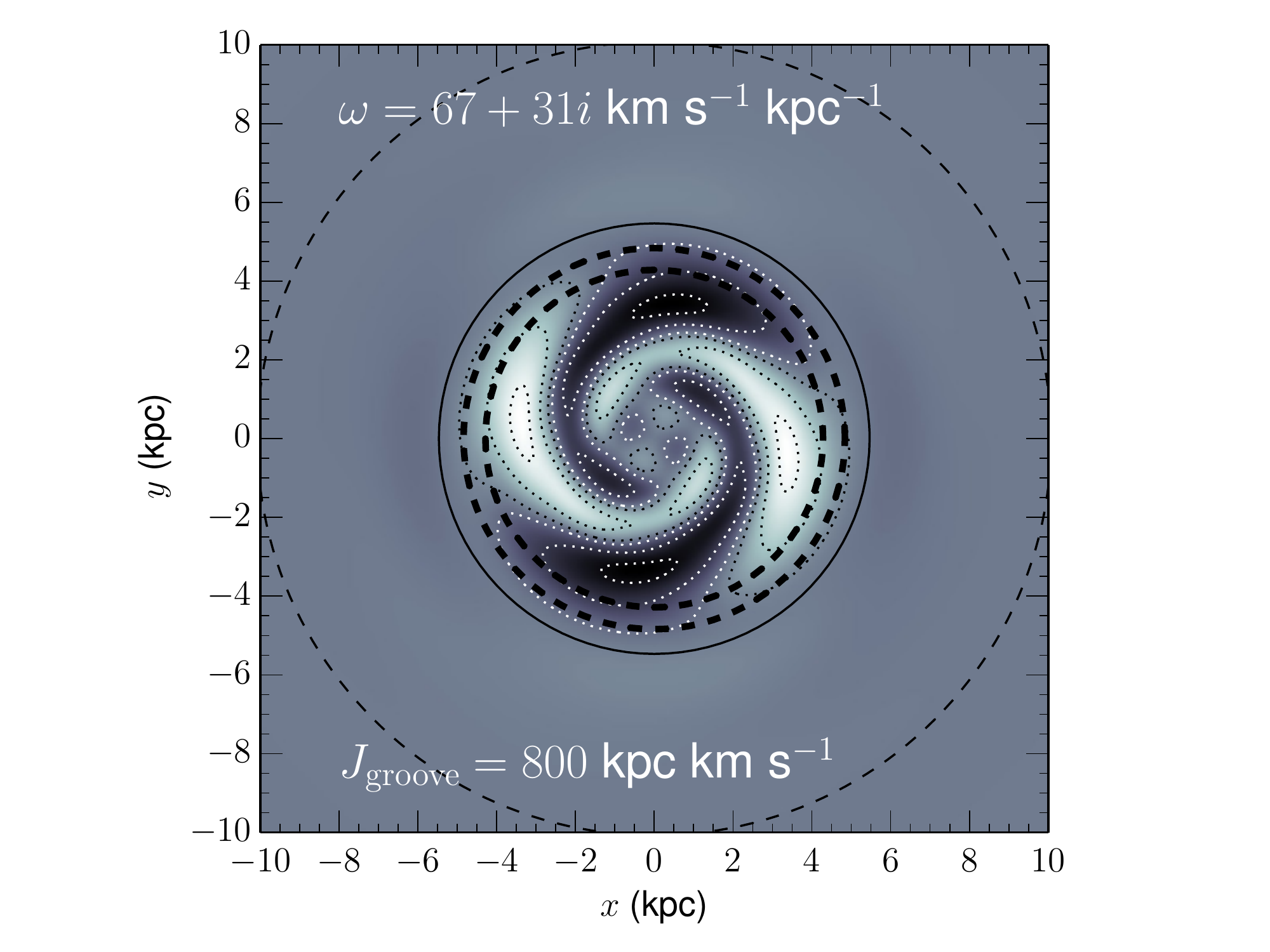}
\includegraphics[trim=55 10 92 10,clip,width=0.325\textwidth]{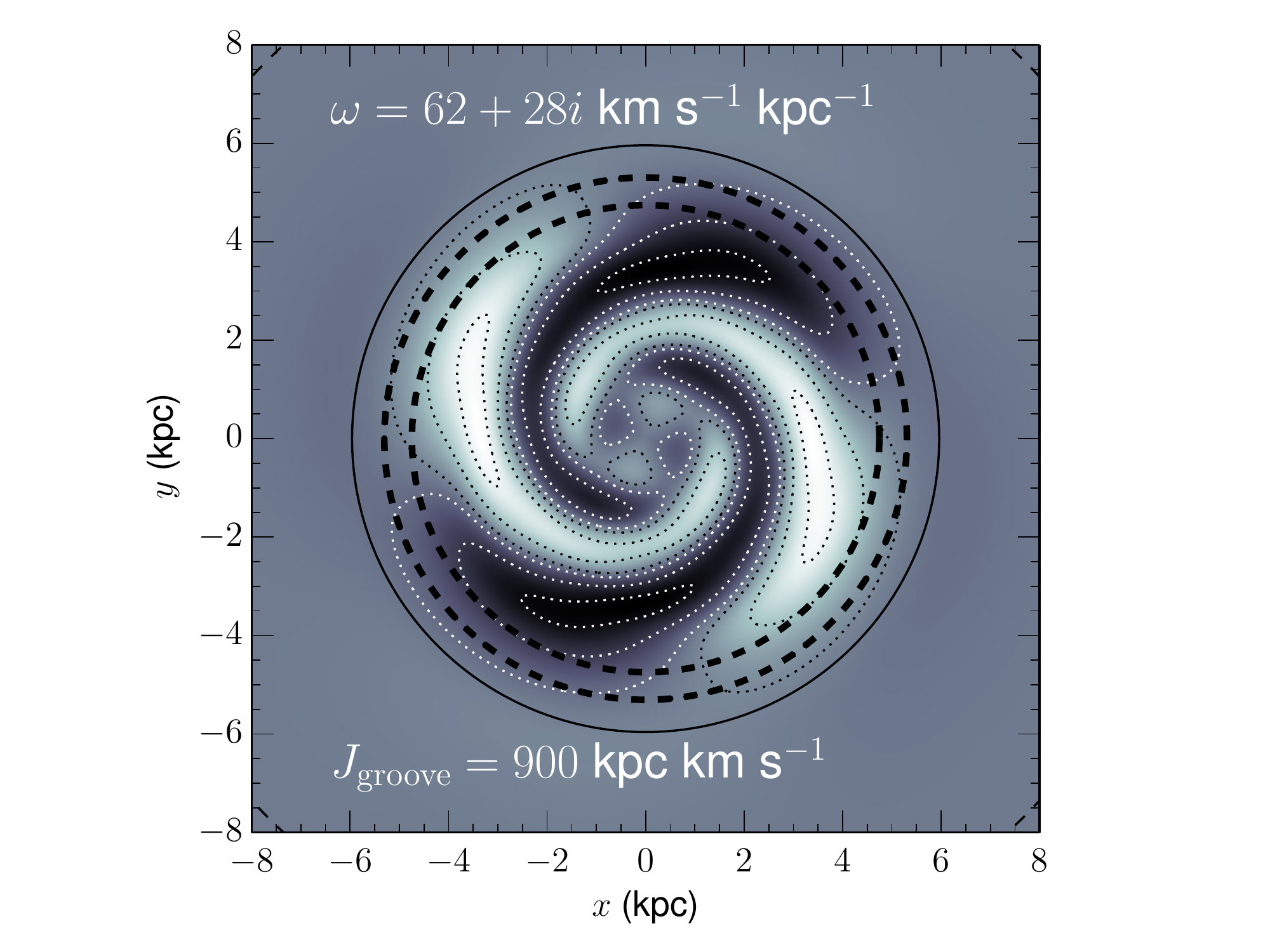}
\includegraphics[trim=55 10 92 10,clip,width=0.325\textwidth]{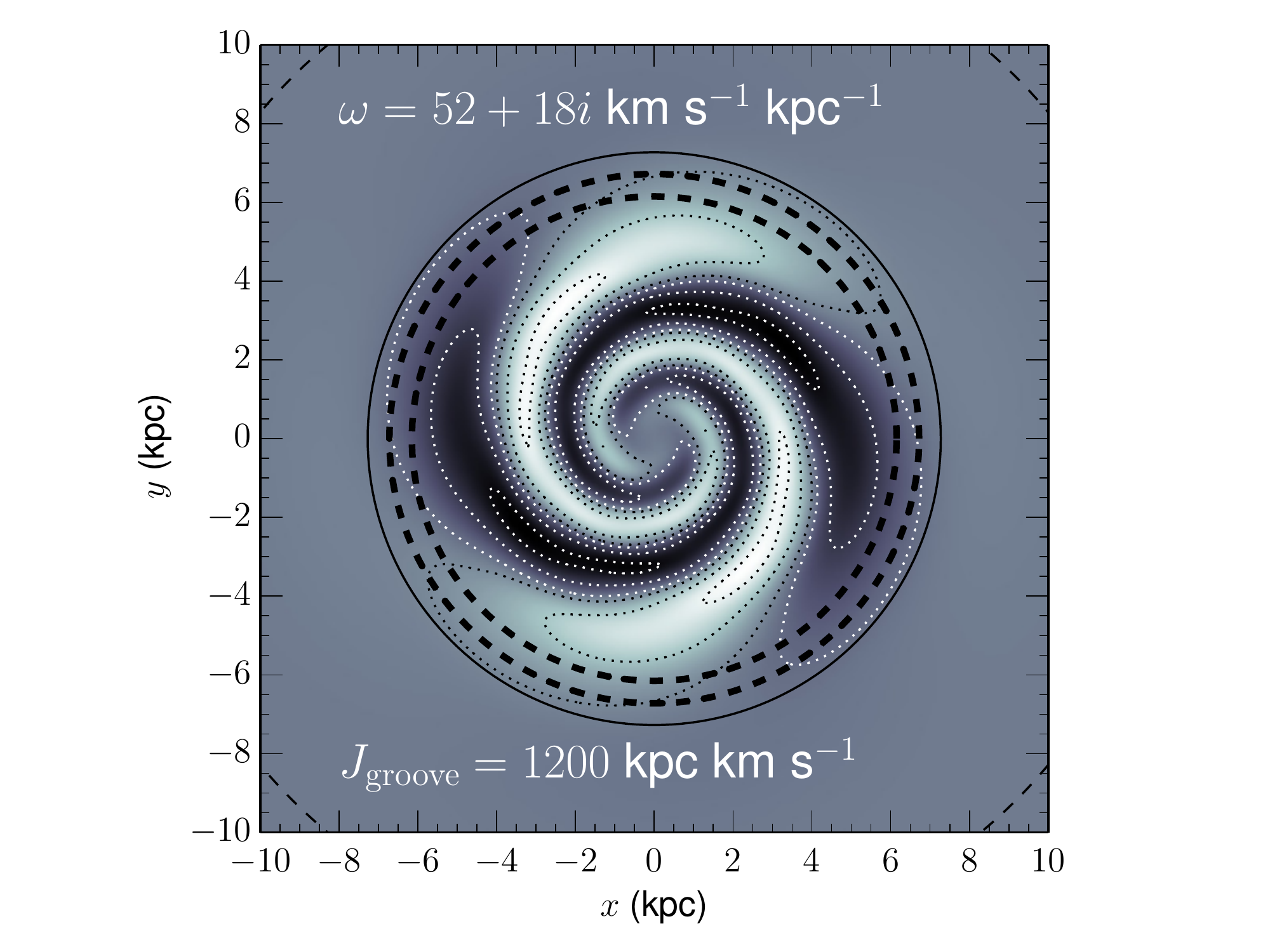}
\includegraphics[trim=55 10 92 10,clip,width=0.325\textwidth]{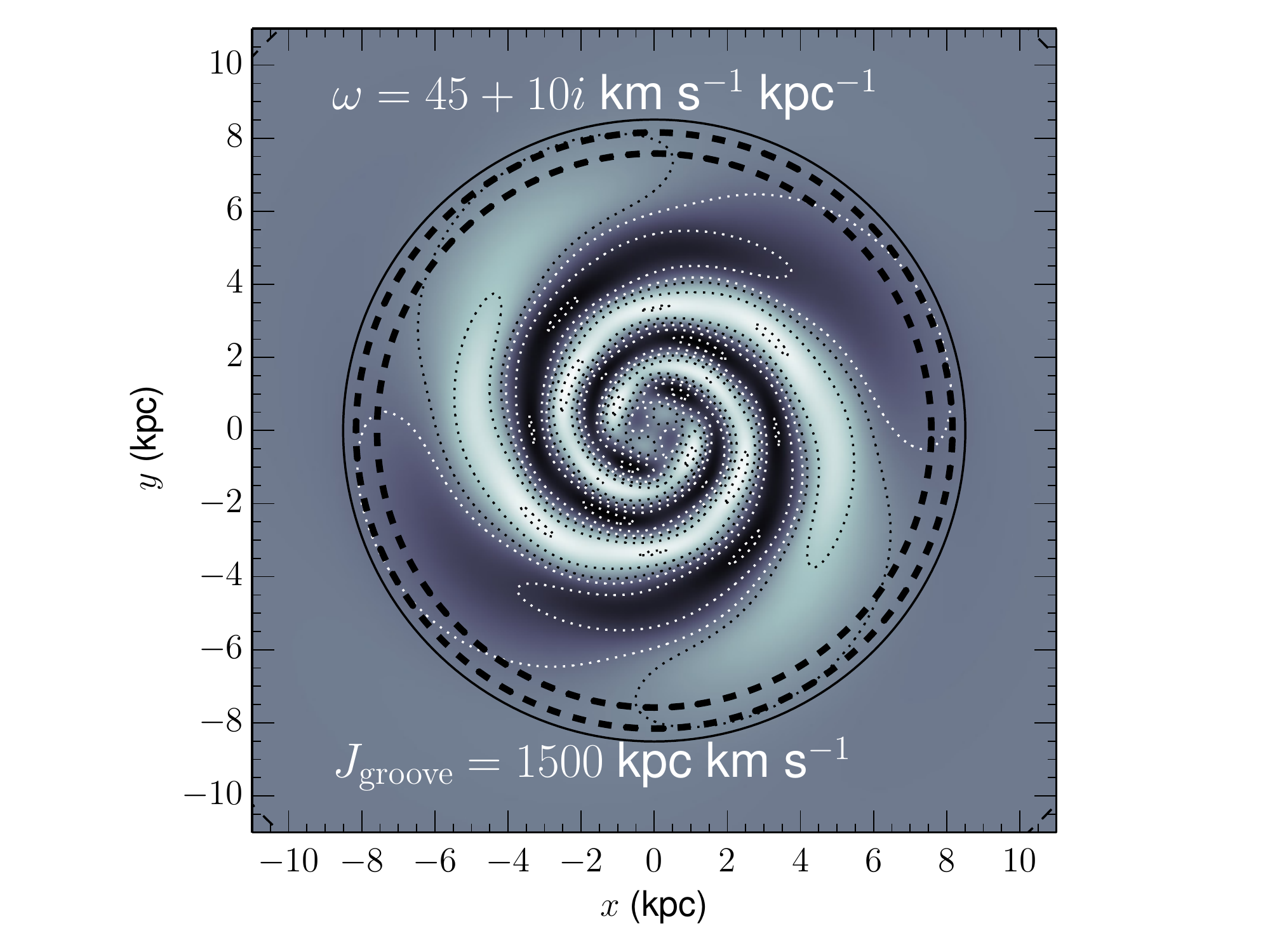}
\caption{Surface density of the $m=2$ ``medium-frequency'' modes in
  the cored exponential disc model with a groove at $J_{\rm groove}$,
  as indicated in each panel. Each panel is labeled with the complex
  frequency $\omega$ of the mode in question.  The groove edges are
  indicated in thick dashed lines, the corotation radius in a thin
  full line, and the outer Lindblad resonance in a thin dashed line.
  \label{fig:middle.pdf}}
\end{figure*}

Around $J_{\rm groove} \approx 200$~kpc~km~s$^{-1}$, a
``medium-frequency'' mode is destabilized with a frequency in between
that of the high- and low-frequency modes. It sits just at the
low-frequency side of the groove in frequency space (see
Fig. \ref{fig:freqs.pdf}). Hence, the mode's CR radius sits just
outside the groove (see Fig. \ref{fig:middle.pdf}). Although absent
for the very lowest $J_{\rm groove}$-values, it overtakes the high-
and low-frequency modes in growth-rate around $J_{\rm groove} \sim
400$~kpc~km~s$^{-1}$. While the eigenmode spectrum approaches that of
the ungrooved model for very high $J_{\rm groove}$-values, this
``medium-frequency'' mode stays present.

\begin{figure*}
\includegraphics[trim=0 20 0 0,clip,width=0.48\textwidth]{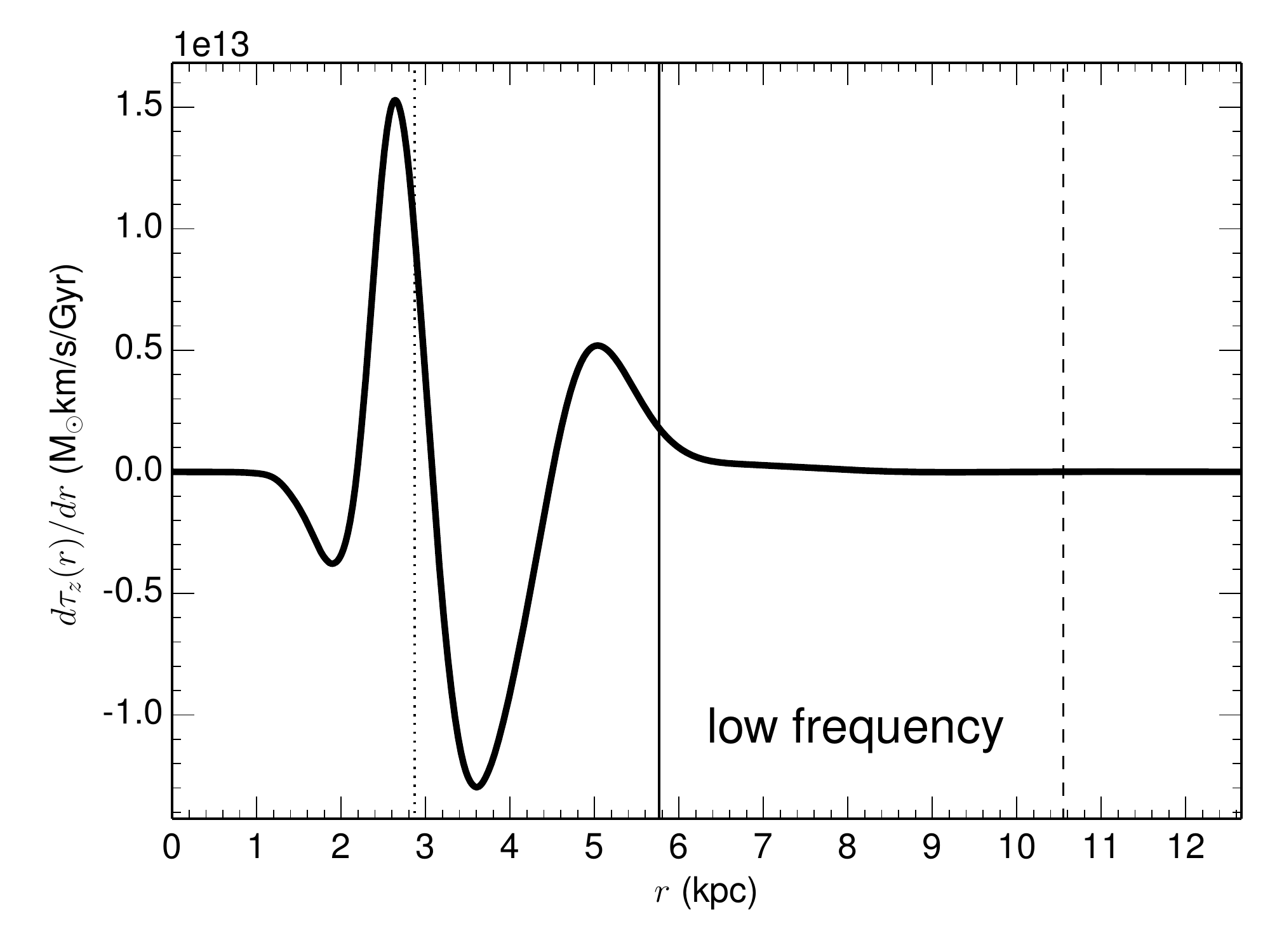}
\includegraphics[trim=0 20 0 0,clip,width=0.48\textwidth]{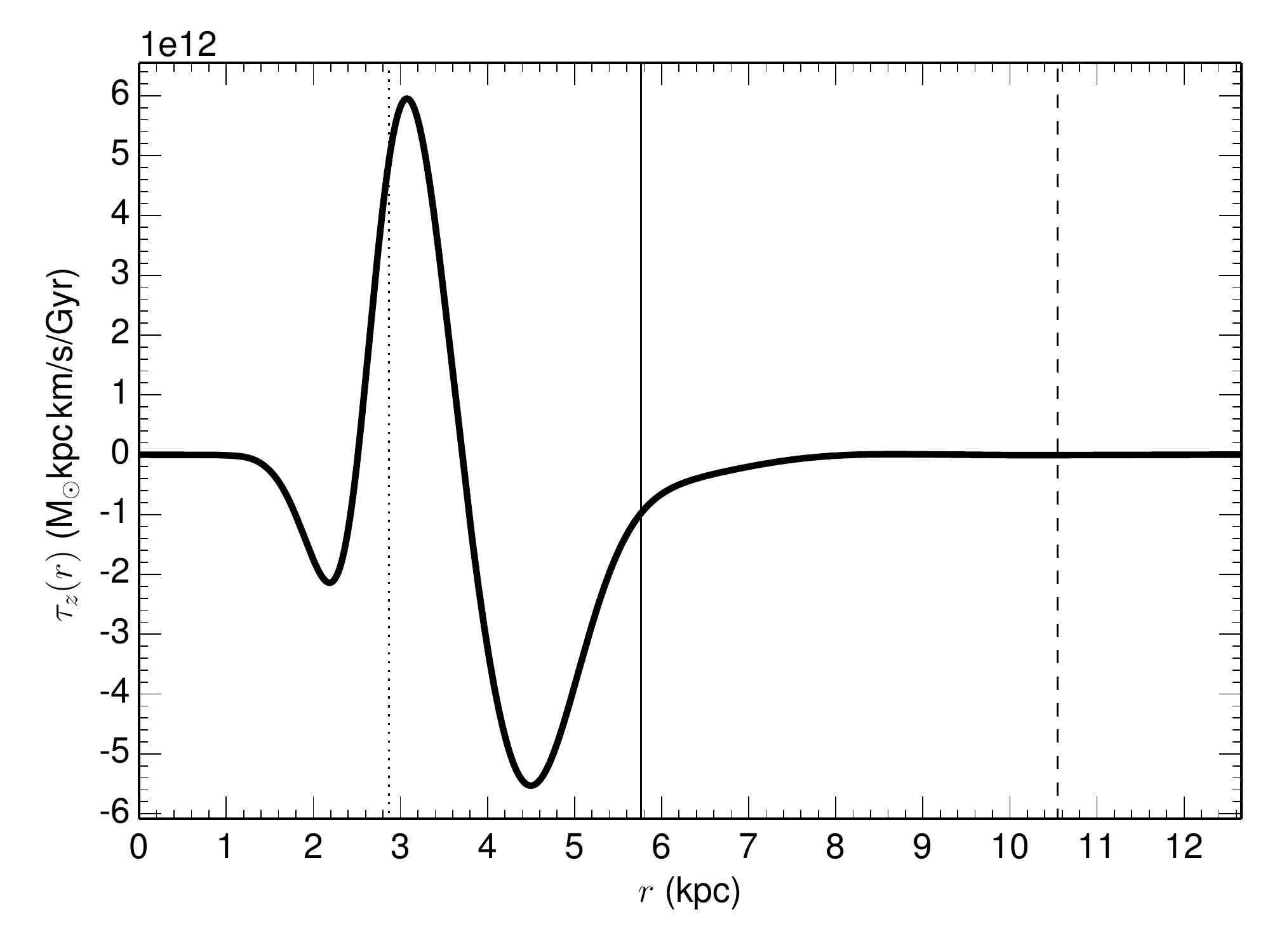}
\includegraphics[trim=0 20 0 0,clip,width=0.48\textwidth]{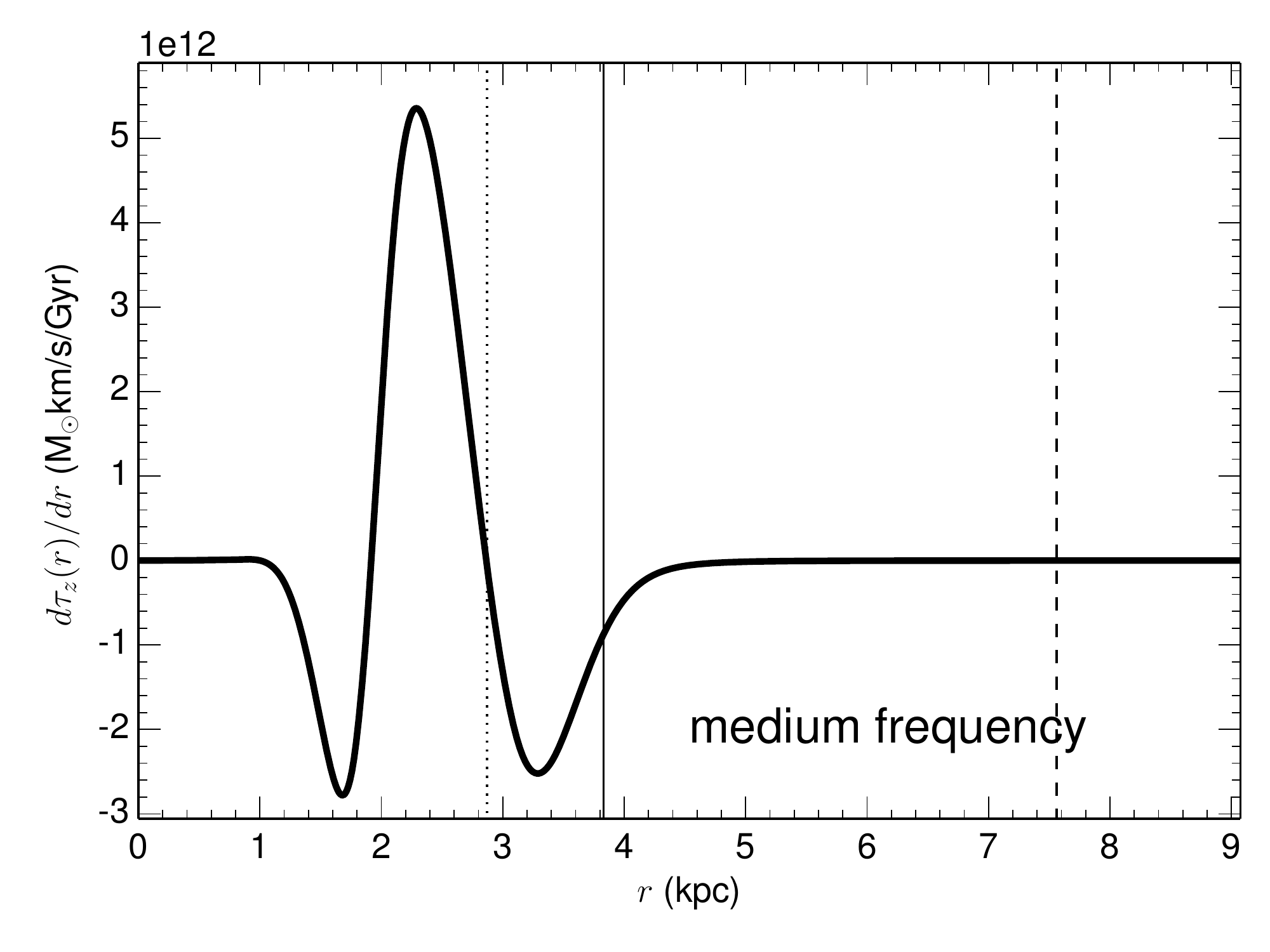}
\includegraphics[trim=0 20 0 0,clip,width=0.48\textwidth]{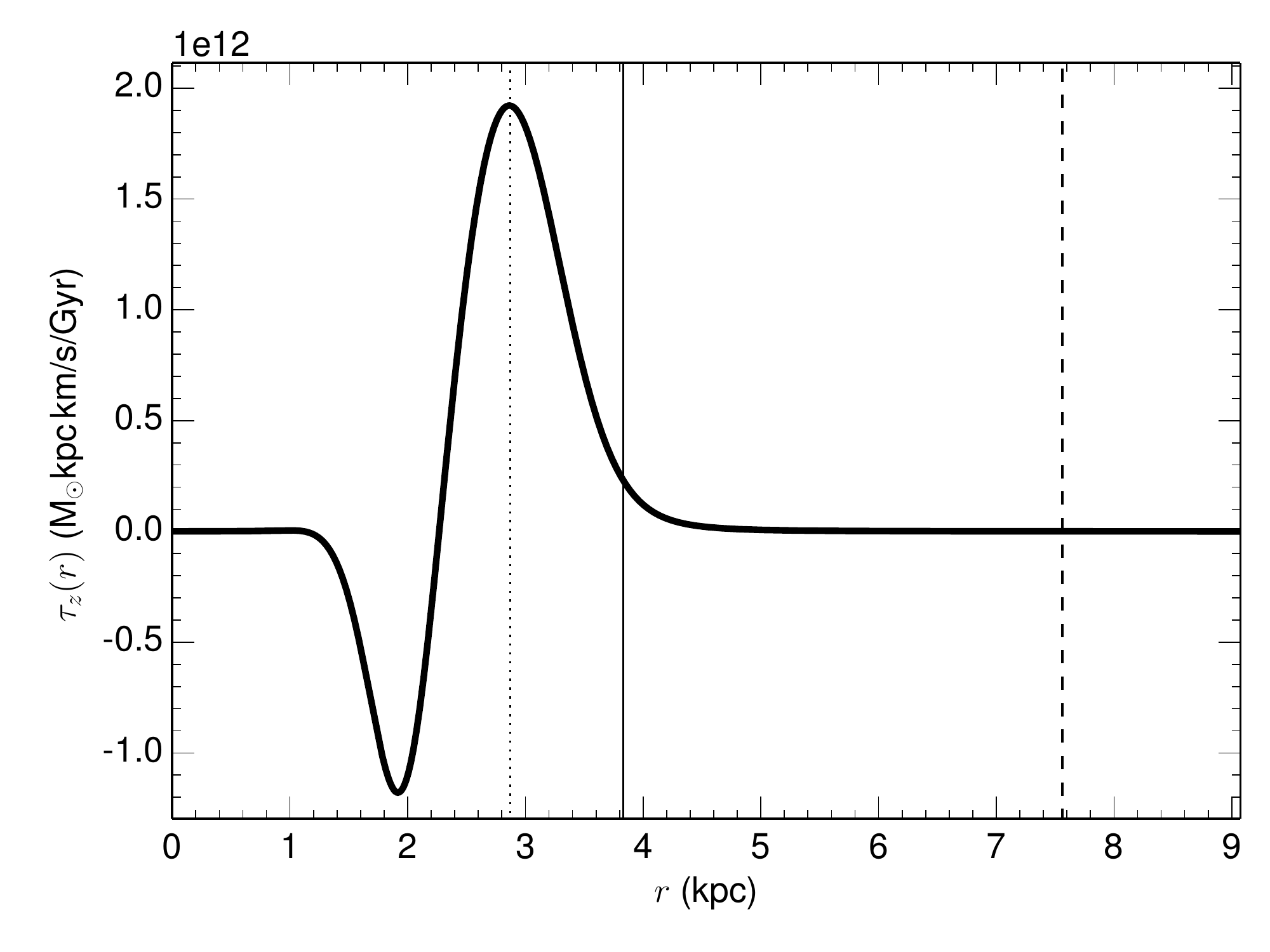}
\includegraphics[trim=0 20 0 0,clip,width=0.48\textwidth]{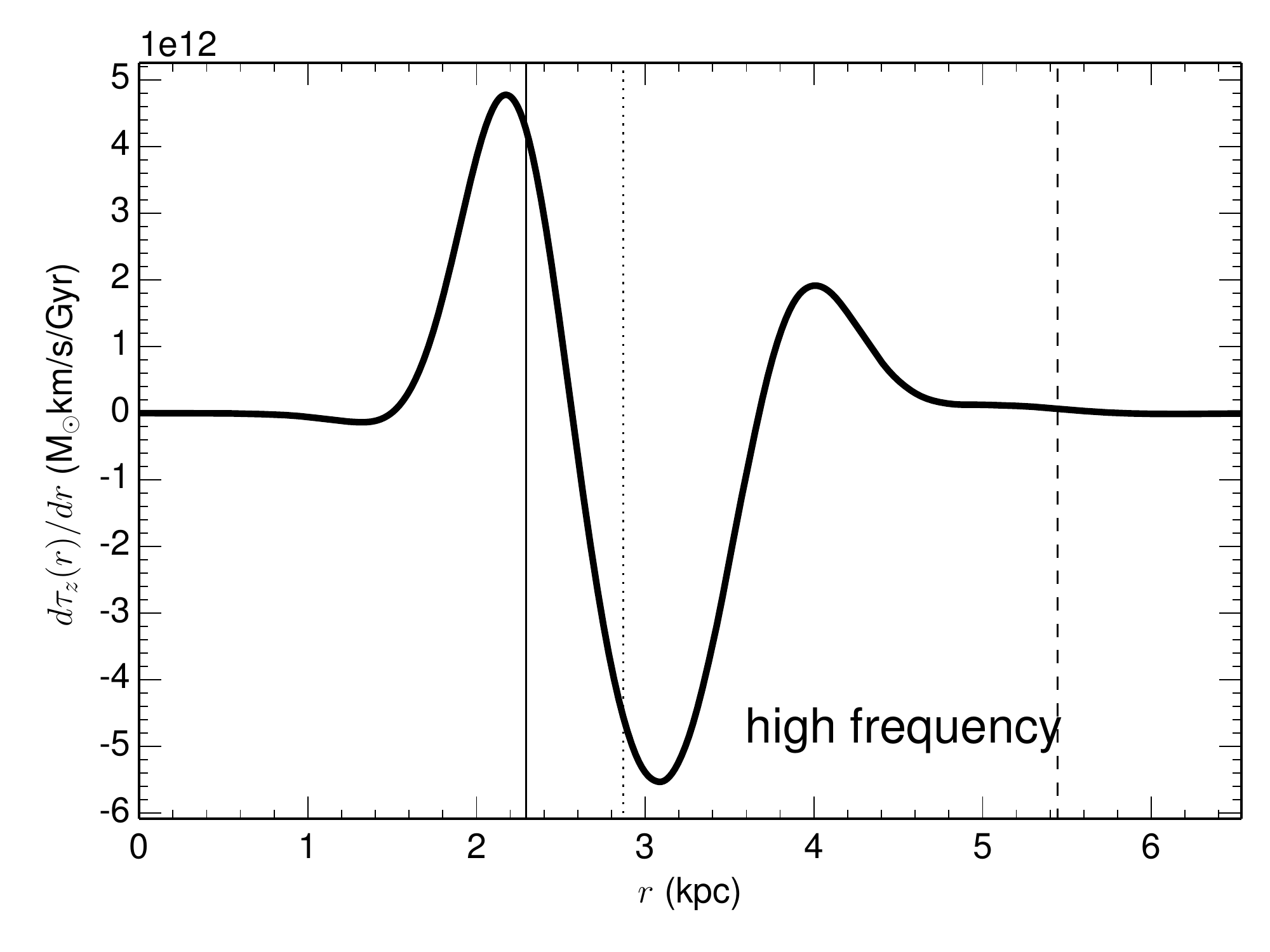}
\includegraphics[trim=0 20 0 0,clip,width=0.48\textwidth]{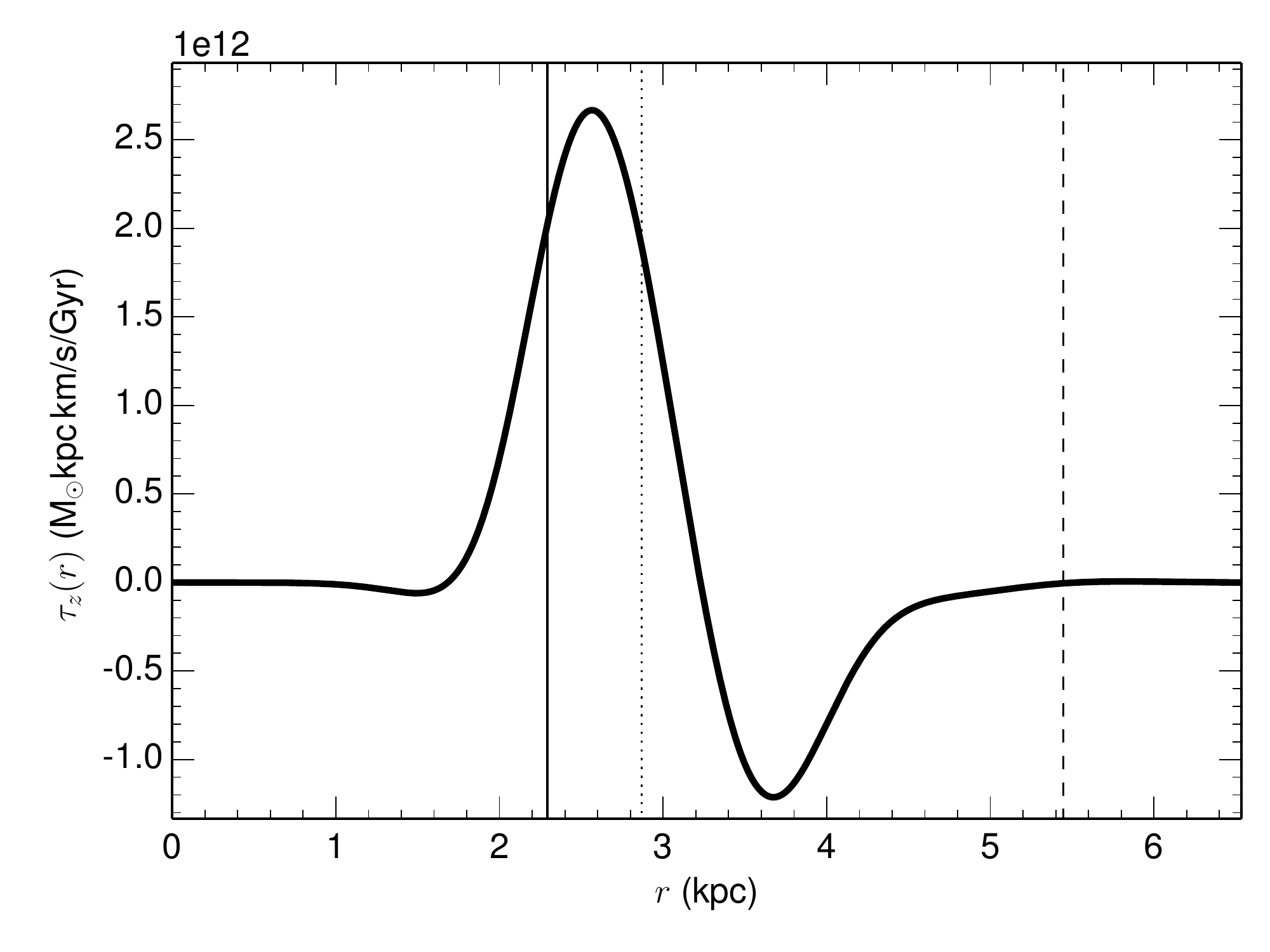}
\caption{Left column:~the torque $d\tau_z(r)$ exerted by the spiral
  pattern on an annular ring of stars with radius $r$ and width
  $dr$. Right column:~the torque $\tau_z(r)$ integrated out to radius
  $r$. Top to bottom:~the low-frequency, medium-frequency, and
  high-frequency eigenmodes of the cored exponential disc model with a
  groove at $J_{\rm groove}=433$~kpc~km~s$^{-1}$. Vertical full
  line:~the CR radius; vertical dashed line:~the OLR radius.
 \label{fig:tau2.pdf}}
\end{figure*}
As can be seen in Fig. \ref{fig:middle.pdf}, the density distribution
of a ``medium-frequency'' mode never strays significantly beyond the
groove's inner edge. This suggests that the ``medium-frequency'' mode
is a standing wave pattern formed by waves traveling between the
galaxy center and the groove's inner edge. The density distributions
presented in Fig. \ref{fig:middle.pdf} can be compared with Fig. 7 in
\citet{sellwood91}. In the latter, the density distribution of a $m=2$
mode in an $N$-body simulation of a grooved Mestel disc model is
shown. Its CR radius lies just outside of the groove and its density
is only significantly non-zero inside the groove or, equivalently, its
CR radius. In other words, it looks exactly as one would expect of a
``medium-frequency'' mode \ldots

In contrast to the modes in the ungrooved disc, the torque
  $d\tau_z(r)$ exerted by the spiral pattern on the stellar disc can
  show a complex radial dependence, with several sign changes. These
  are caused by the sudden changes in the pattern's density and
  potential connected with the groove. Still, the total torque $\tau_z
  = \int_0^\infty d\tau_z(r)$ is zero to a very good precision.

\subsection{Groove mode}

\begin{figure*}
\includegraphics[trim=55 35 95 10,clip,width=0.325\textwidth]{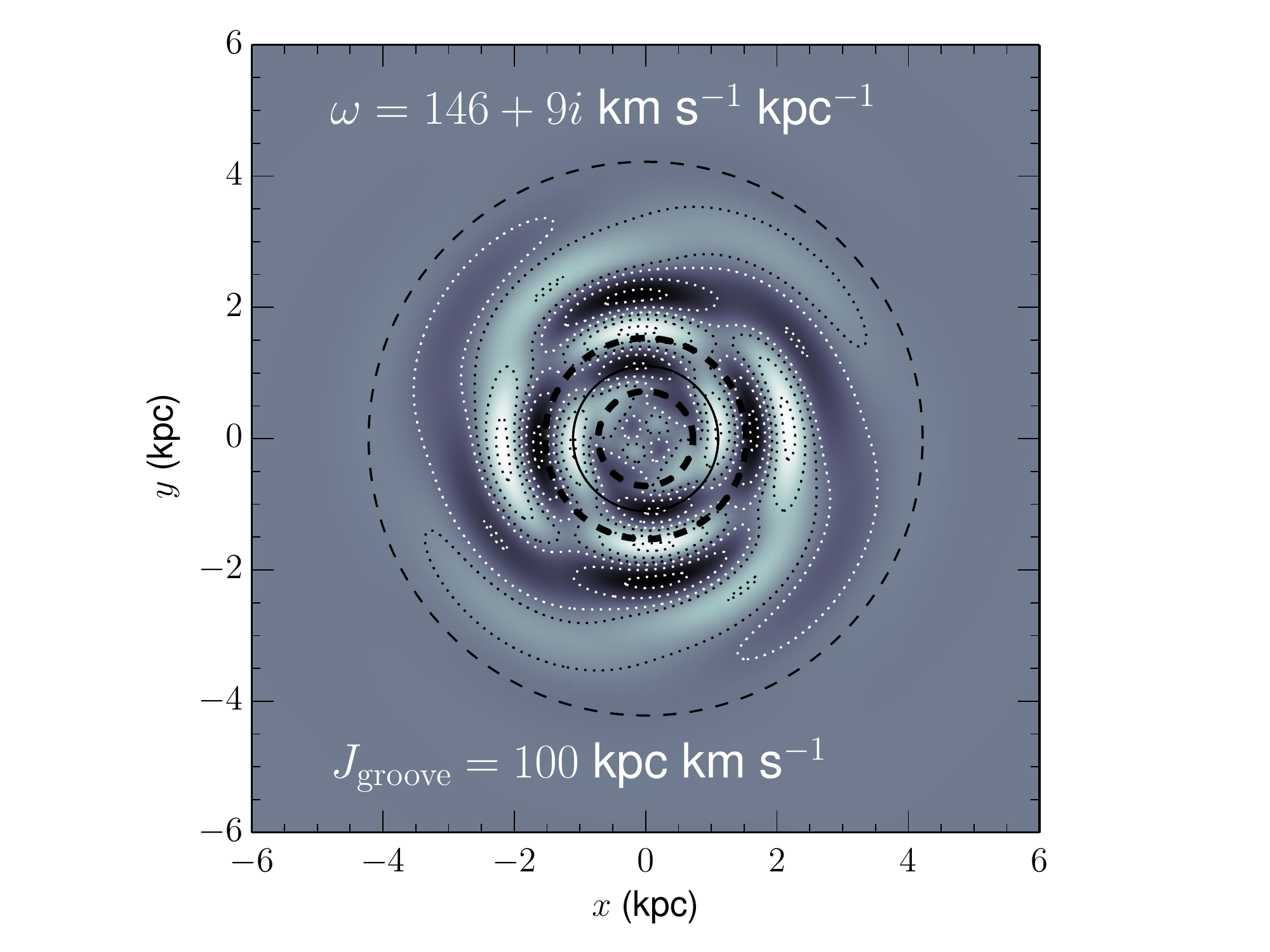}
\includegraphics[trim=55 35 95 10,clip,width=0.325\textwidth]{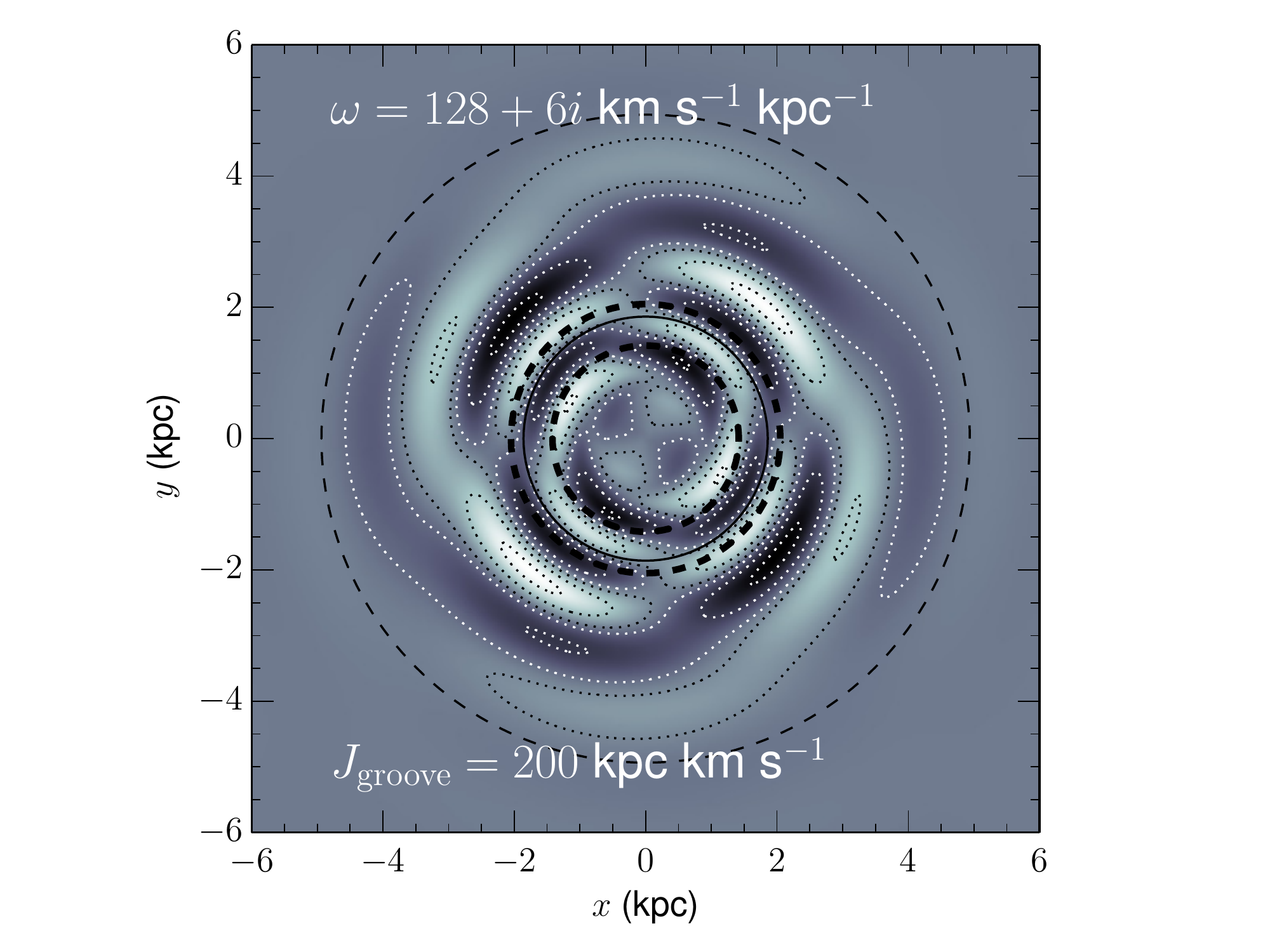}
\includegraphics[trim=55 35 95 10,clip,width=0.325\textwidth]{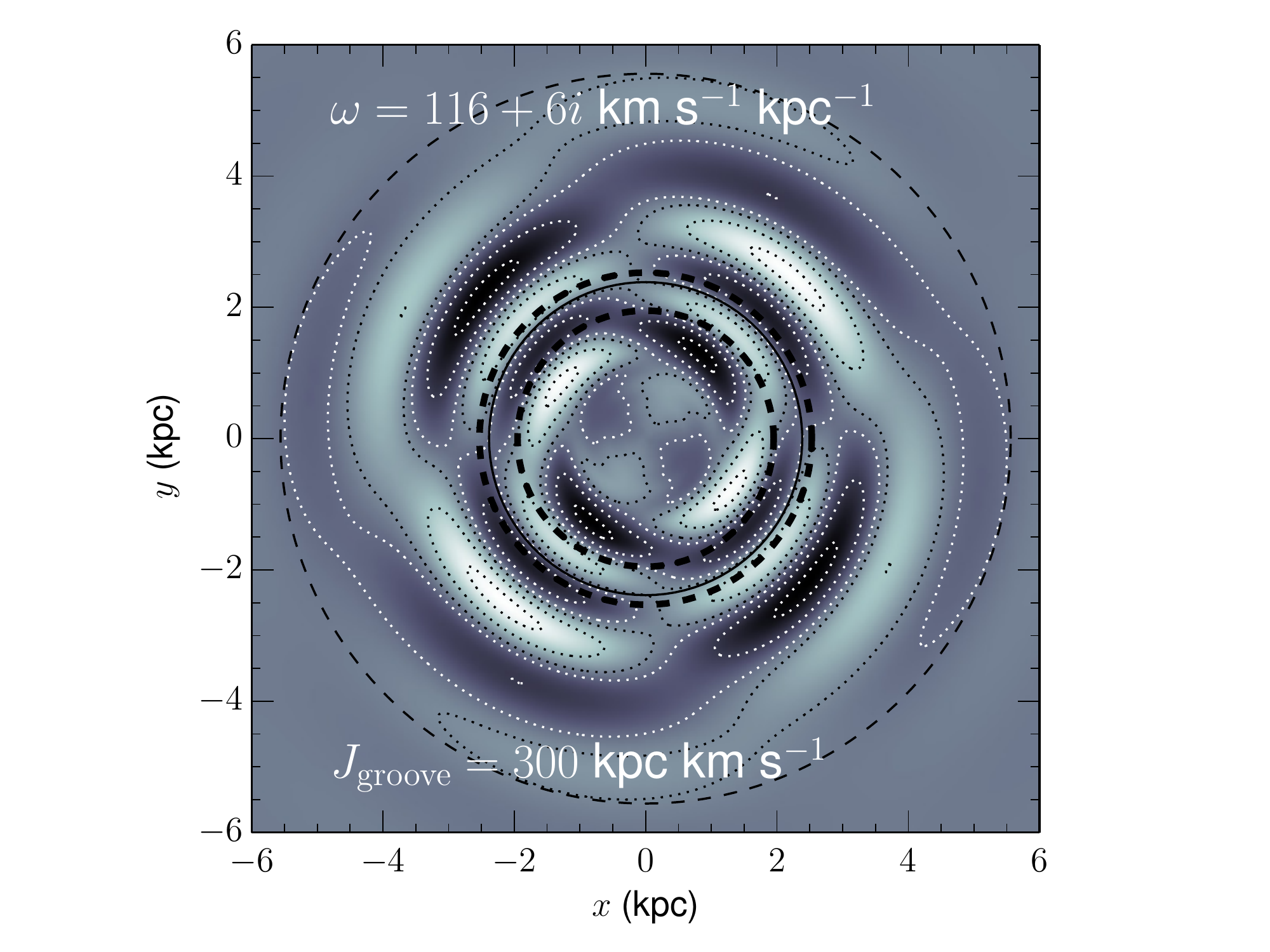}
\includegraphics[trim=55 10 95 10,clip,width=0.325\textwidth]{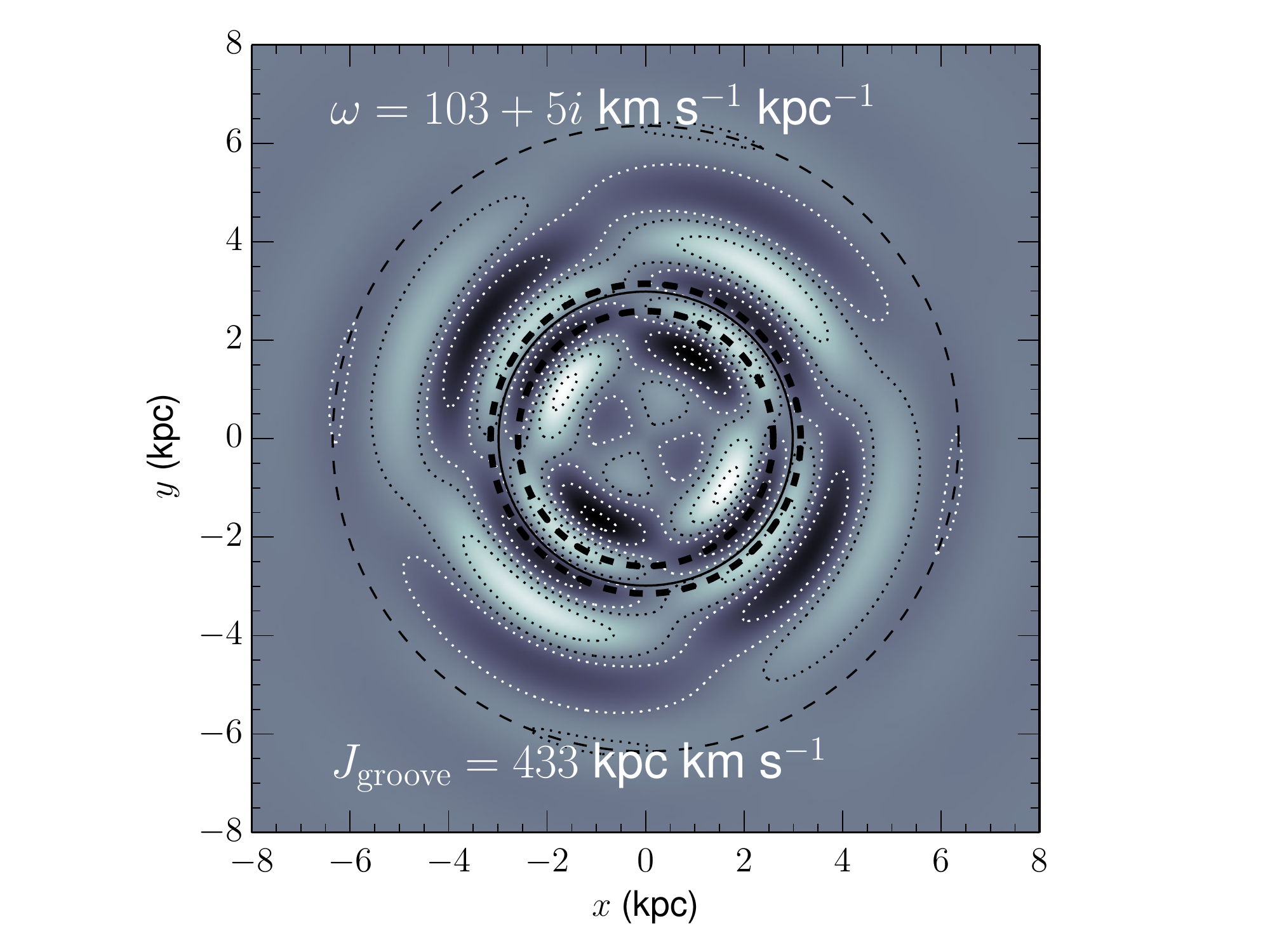}
\includegraphics[trim=55 10 95 10,clip,width=0.325\textwidth]{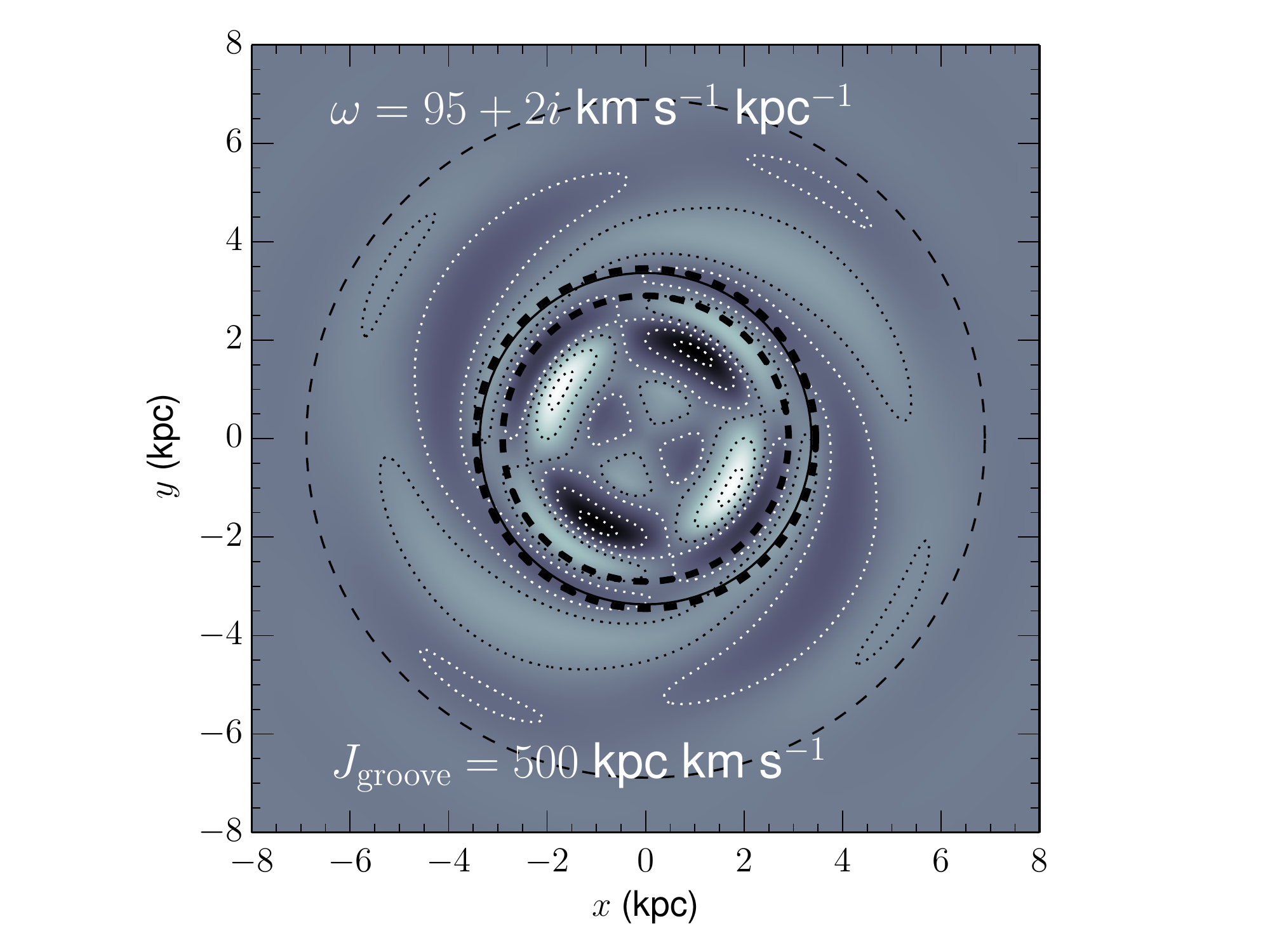}
\includegraphics[trim=55 10 95 10,clip,width=0.325\textwidth]{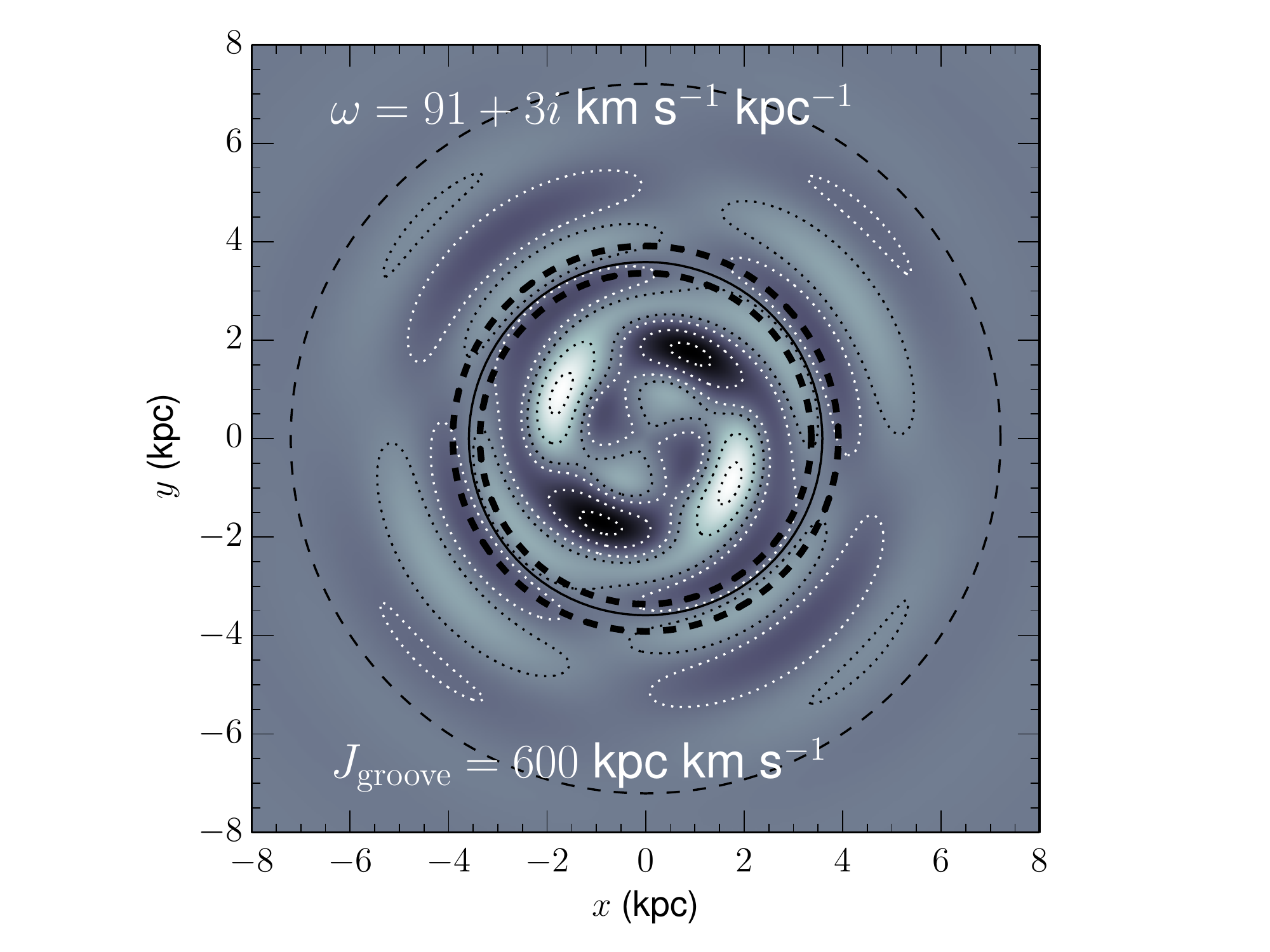}
\caption{Surface density of the $m=2$ ``groove'' modes in the cored
  exponential disc model with a groove at $J_{\rm groove}$, as
  indicated in each panel. Each panel is labeled with the complex
  frequency $\omega$ of the mode in question. The groove edges are
  indicated in thick dashed lines, the corotation radius in a thin
  full line, and the outer Lindblad resonance in a thin dashed line.
  \label{fig:groove.pdf}}
\end{figure*}

The grooves with $J_{\rm groove} \lesssim 600$~kpc~km~s$^{-1}$
destabilize a mode with its CR radius squarely within the groove. This
so-called ``groove''-mode grows much more slowly than the low-,
medium-, and high-frequency modes and is therefore dynamically less
important. As is obvious from Fig. \ref{fig:groove.pdf}, this mode
shows a tightly-wound {\em leading} spiral pattern inside the groove
edge and a {\em trailing} spiral pattern beyond the
groove. Apparently, traveling tightly-wound leading waves dominate
trailing waves in setting up these very slowly growing spiral
patterns.

As the strength of the part of the mode outside the groove's outer
edge diminishes with increasing $J_{\rm groove}$, the mode's central
part eventually changes from a leading into a trailing spiral.

\begin{figure*}
\includegraphics[trim=1 35 0 13,clip,width=0.448153\textwidth]{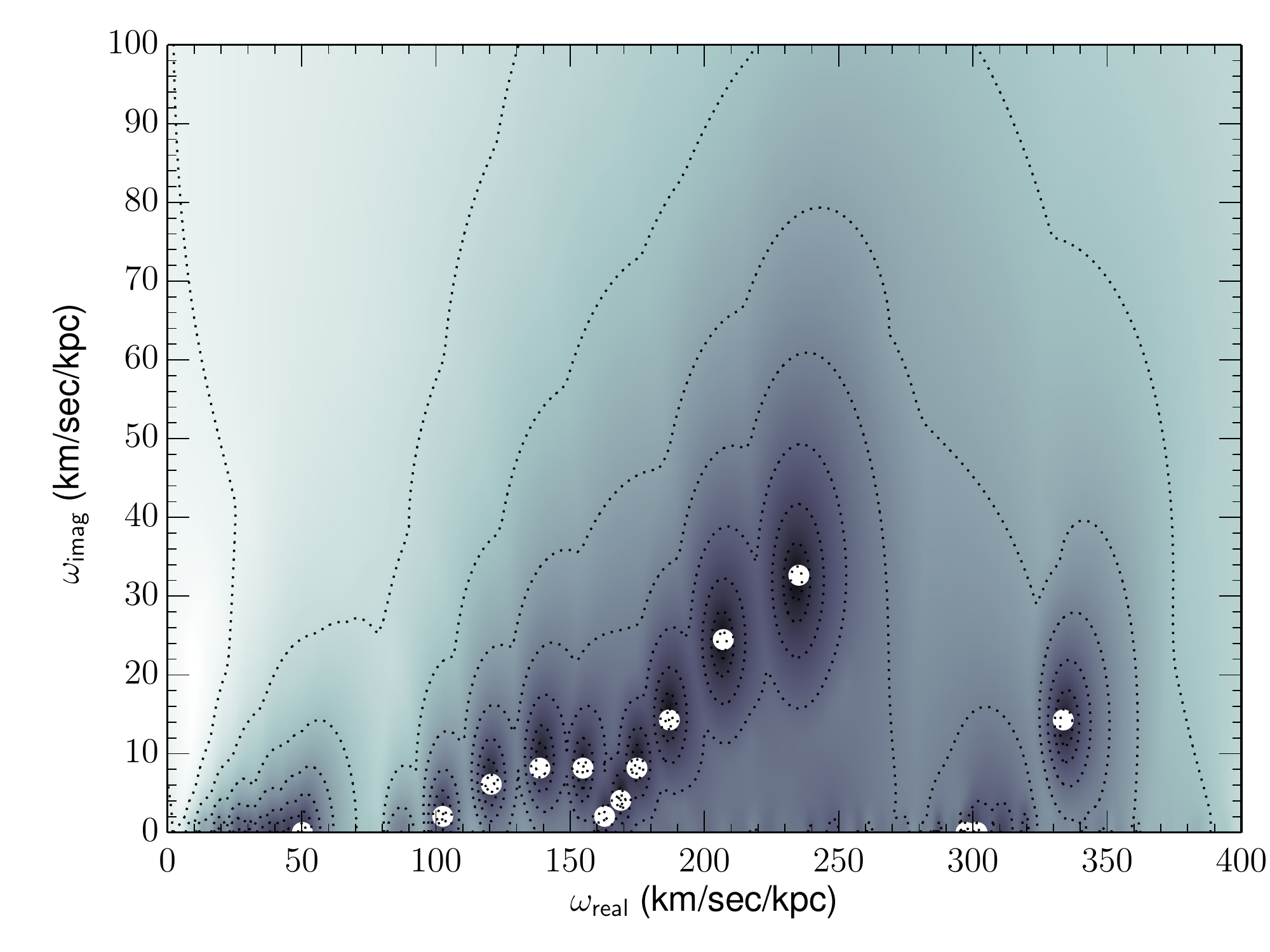}
\includegraphics[trim=45 35 0 13,clip,width=0.420755\textwidth]{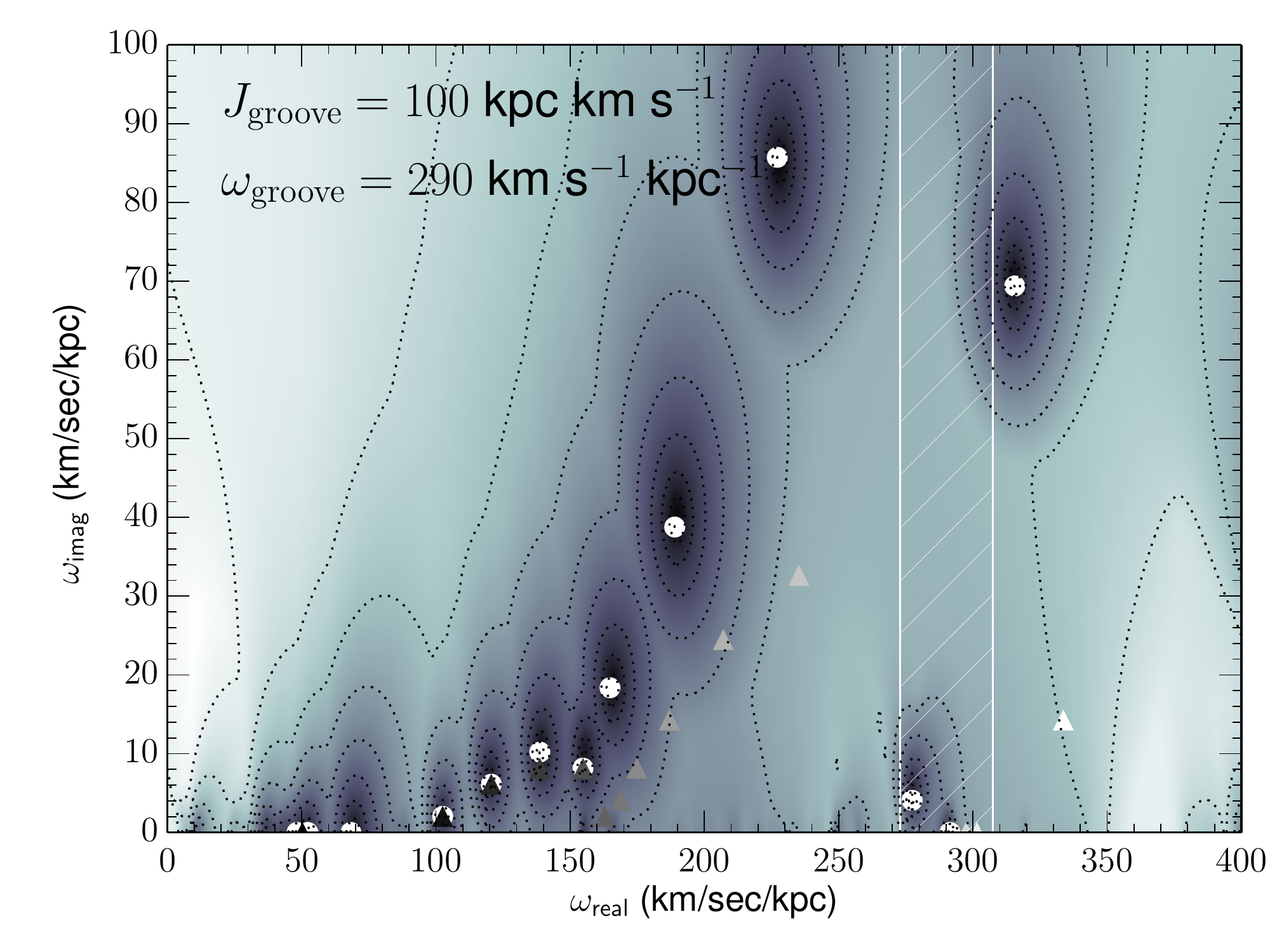}
\includegraphics[trim=1 35 0 13,clip,width=0.448153\textwidth]{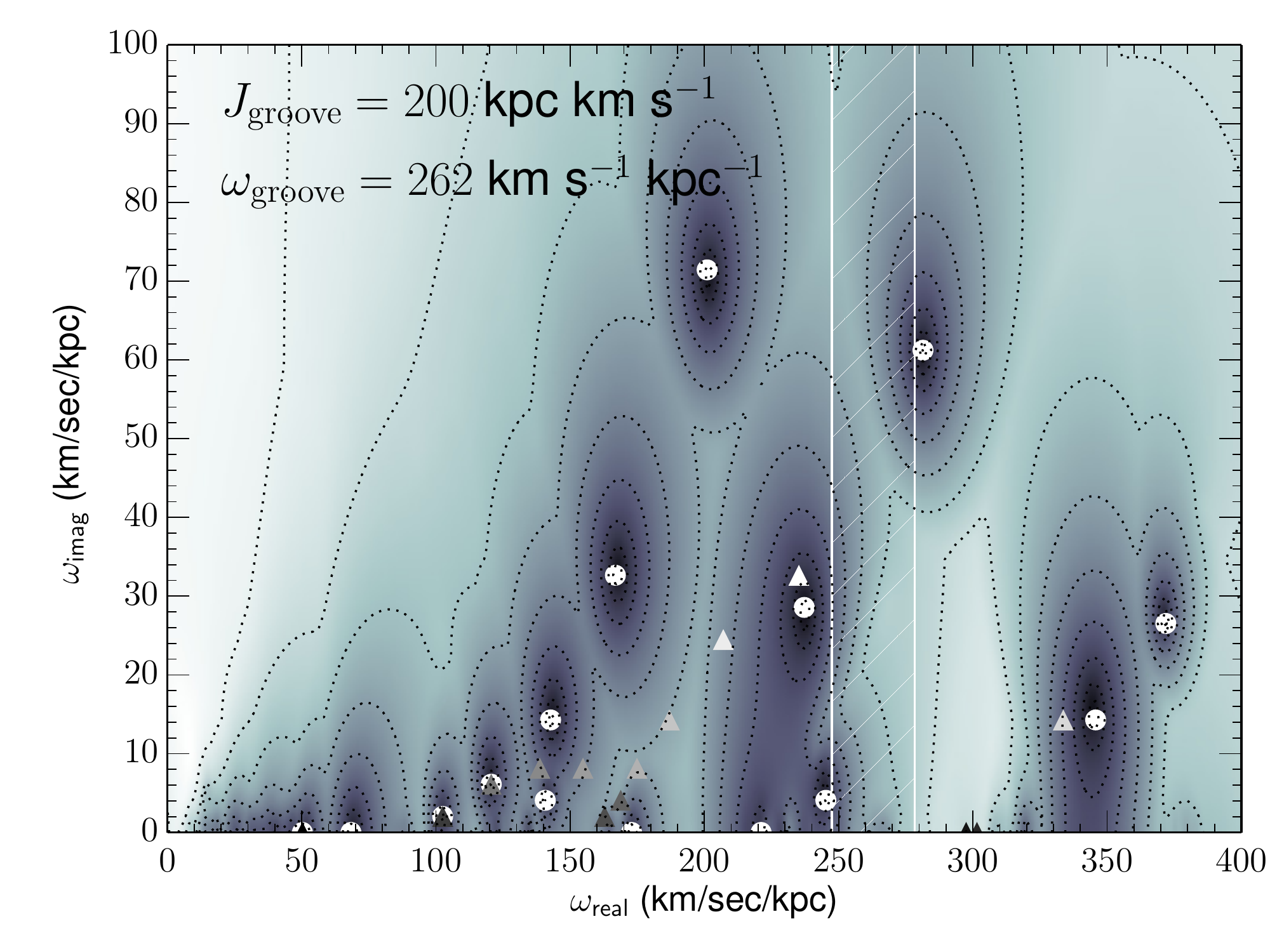}
\includegraphics[trim=45 35 0 13,clip,width=0.420755\textwidth]{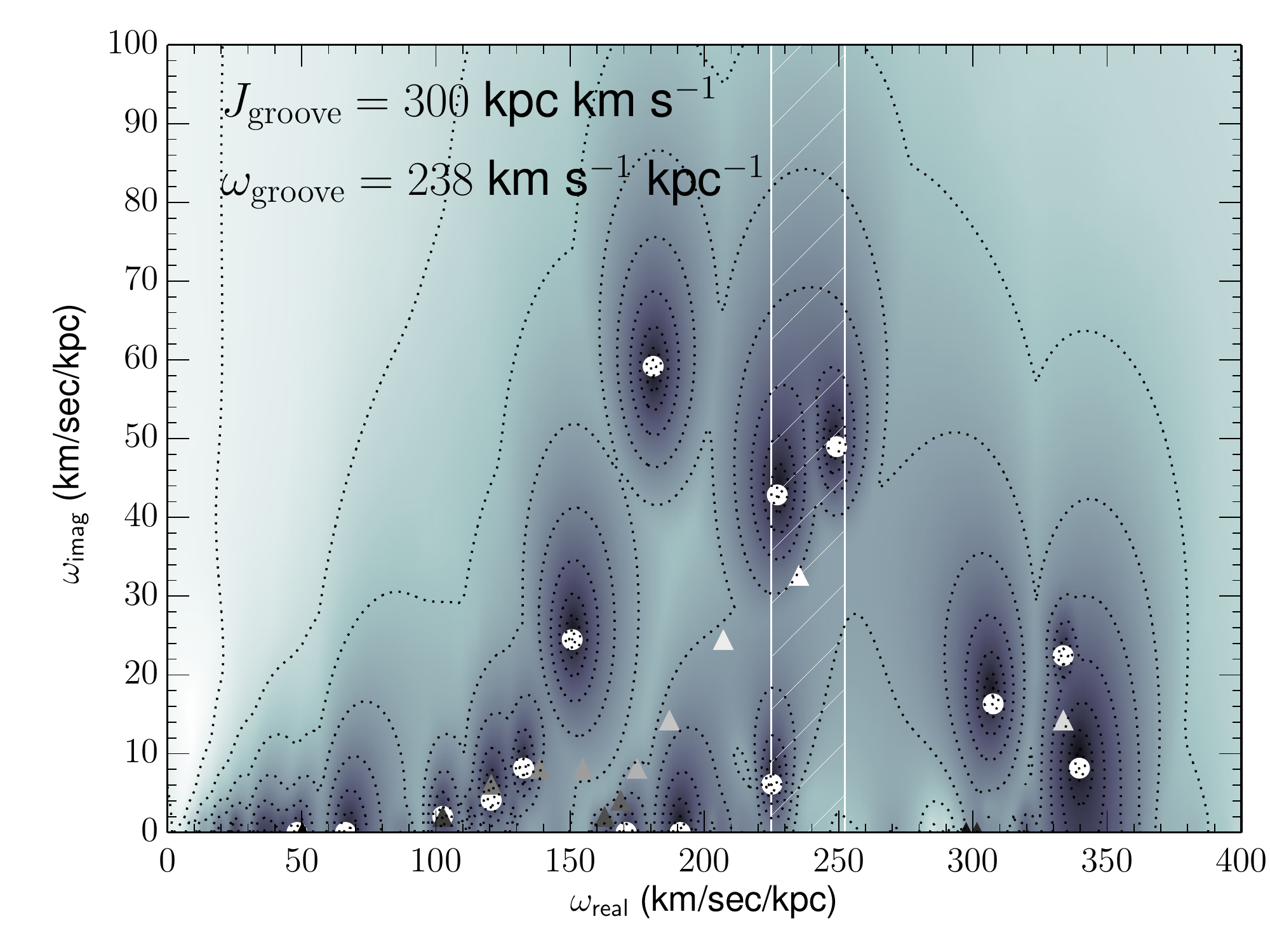}
\includegraphics[trim=1 35 0 13,clip,width=0.448153\textwidth]{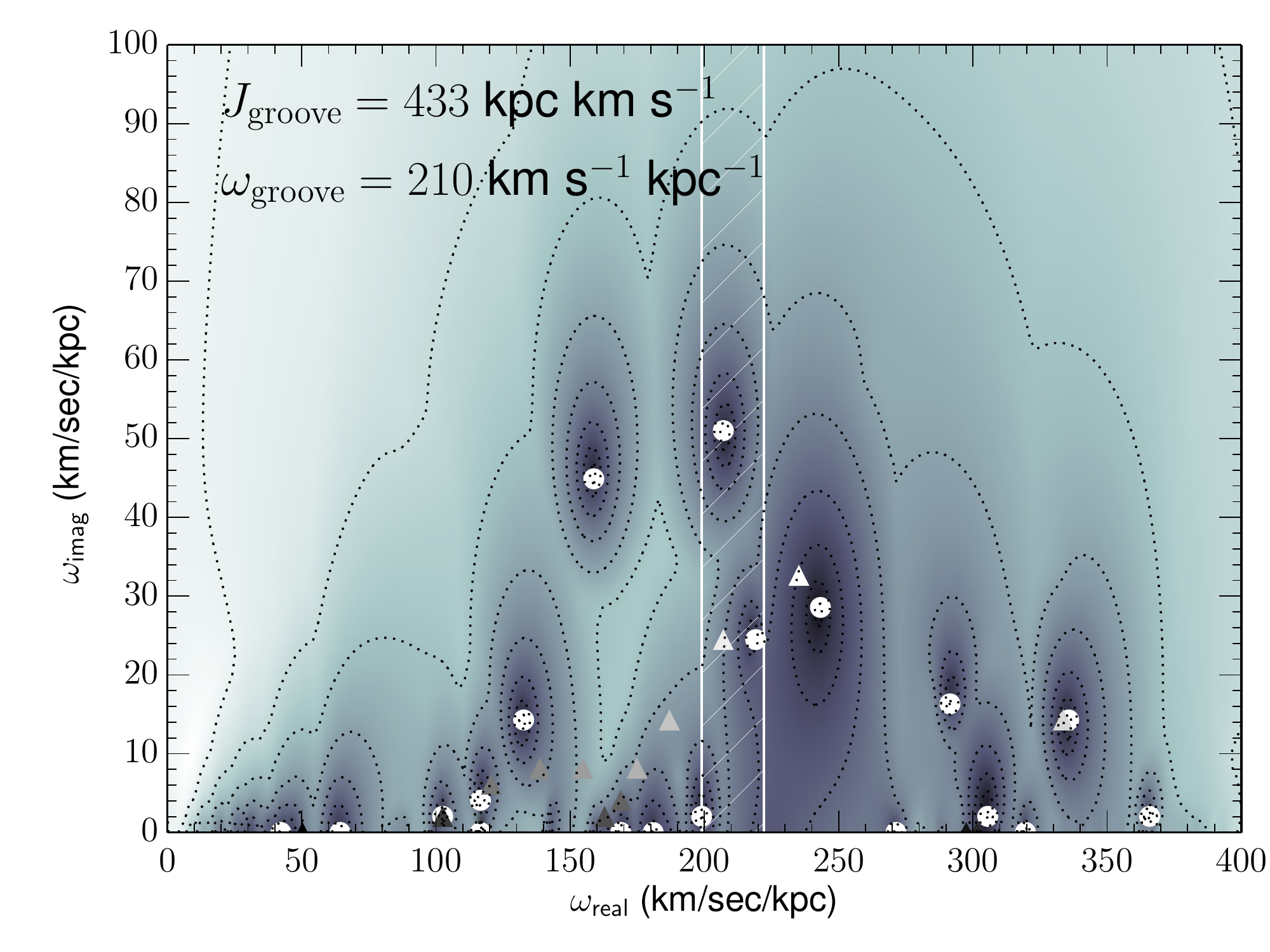}
\includegraphics[trim=45 35 0 13,clip,width=0.420755\textwidth]{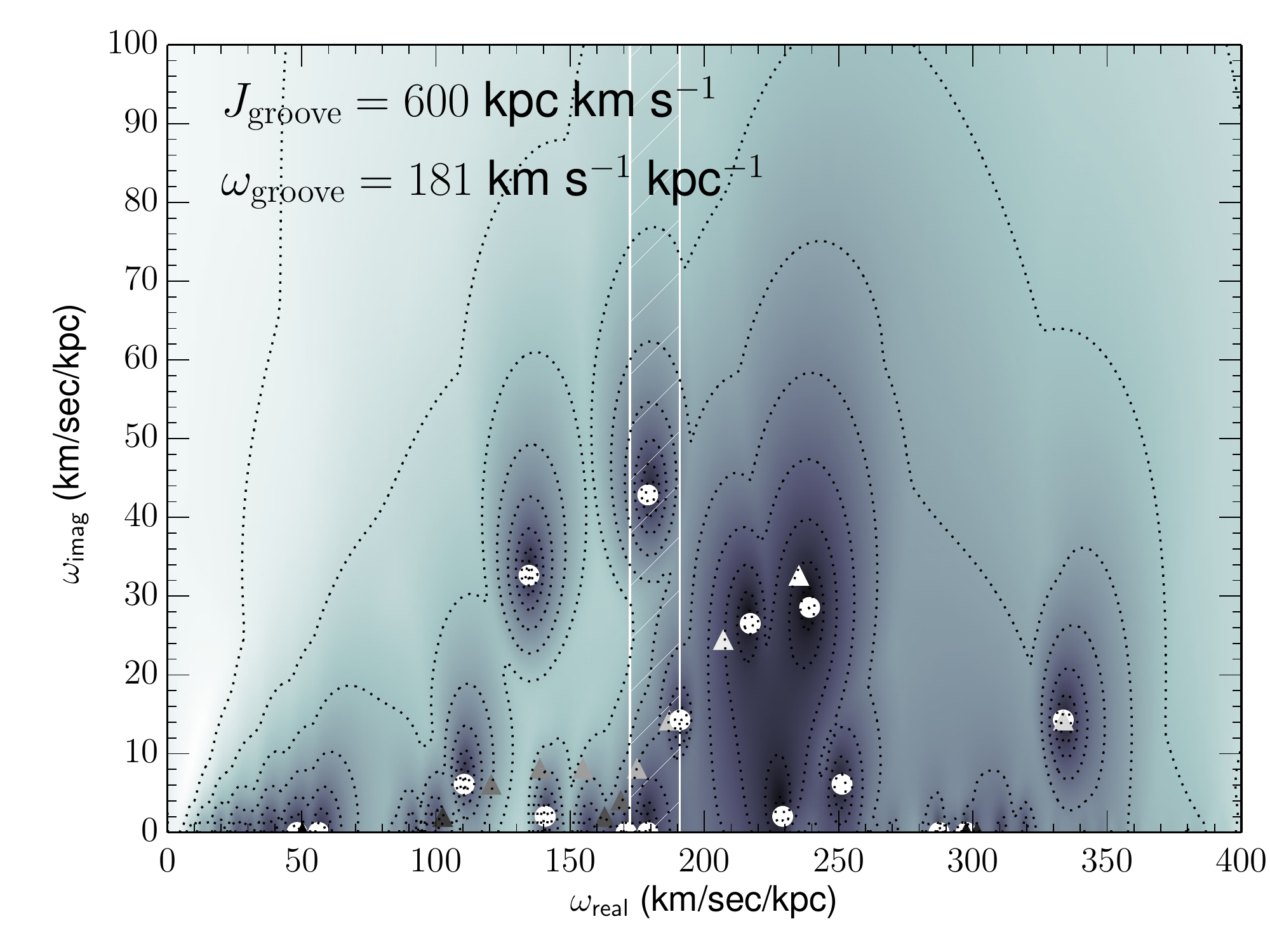}
\includegraphics[trim=1 15 0 13,clip,width=0.448153\textwidth]{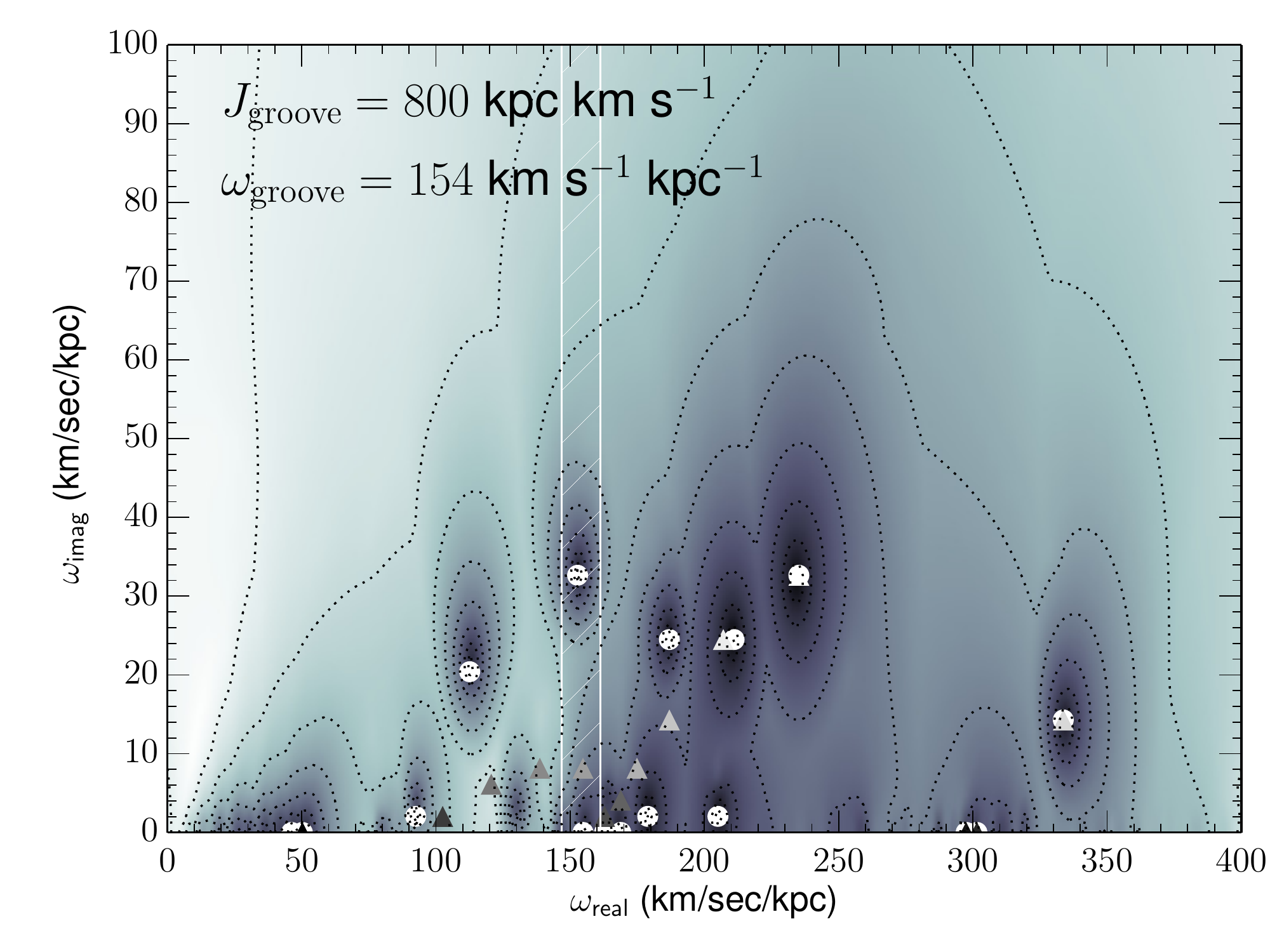}
\includegraphics[trim=45 15 0 13,clip,width=0.420755\textwidth]{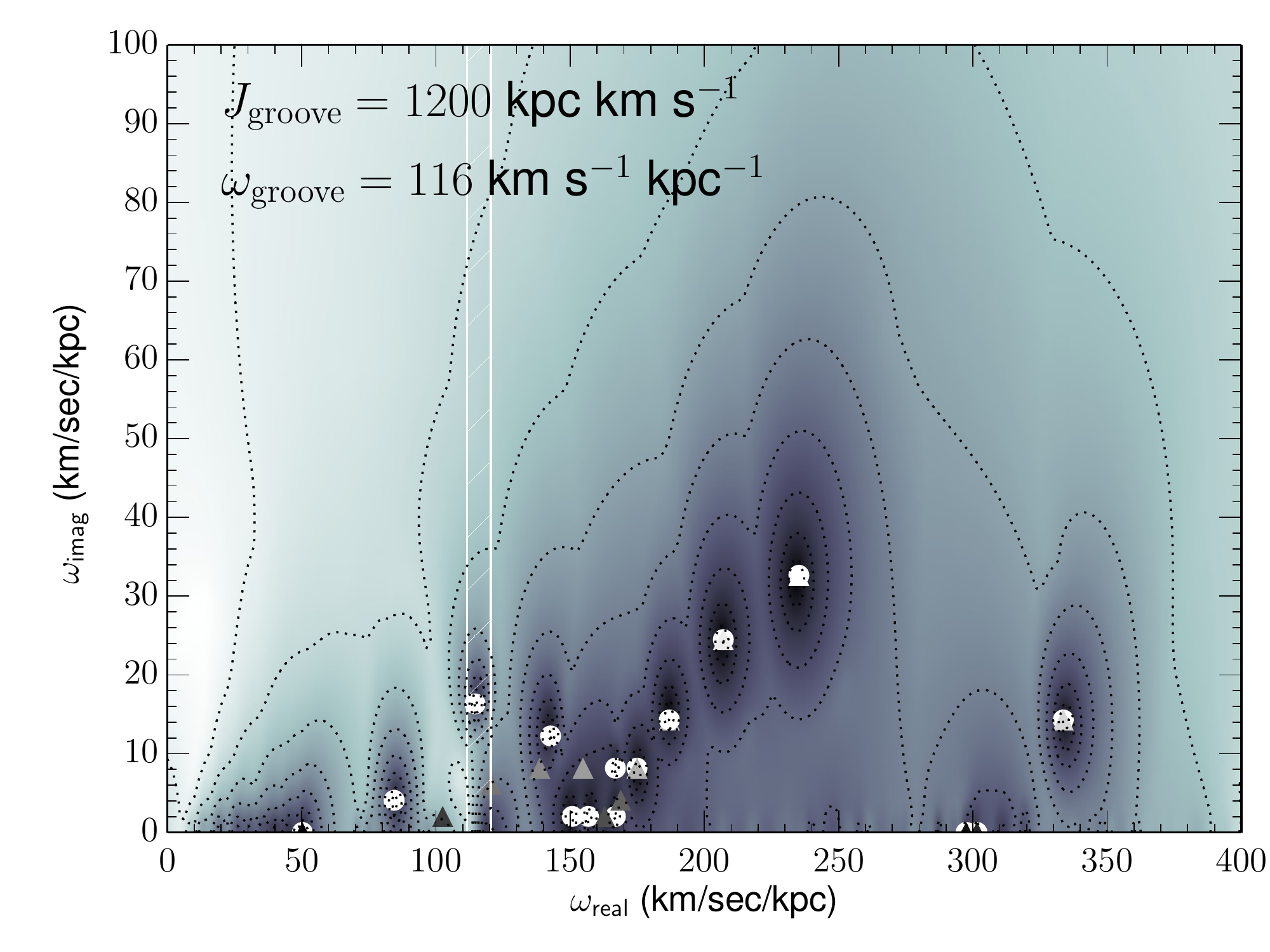}
\caption{The $m=4$ mode spectrum in the complex frequency plane of
  cored exponential disc models without a phase-space groove (top left
  panel) and with different narrow phase space grooves, centered on
  the angular momentum $J_{\rm groove}$ indicated in each panel. The
  white hatched region in each panel, centered on the frequency
  $\omega_{\rm groove}$, indicates the locus of the modes that
  corotate with stars on circular orbits inside the groove. The
  colored triangles indicate the position of the $m=4$ eigenmodes of
  the original cored exponential disc model.
 \label{fig:freqs4.pdf}}
\end{figure*}

\section{Grooves and m=4 modes} \label{m4modes}

\begin{figure*}
\includegraphics[trim=55 12 95 10,clip,width=0.323\textwidth]{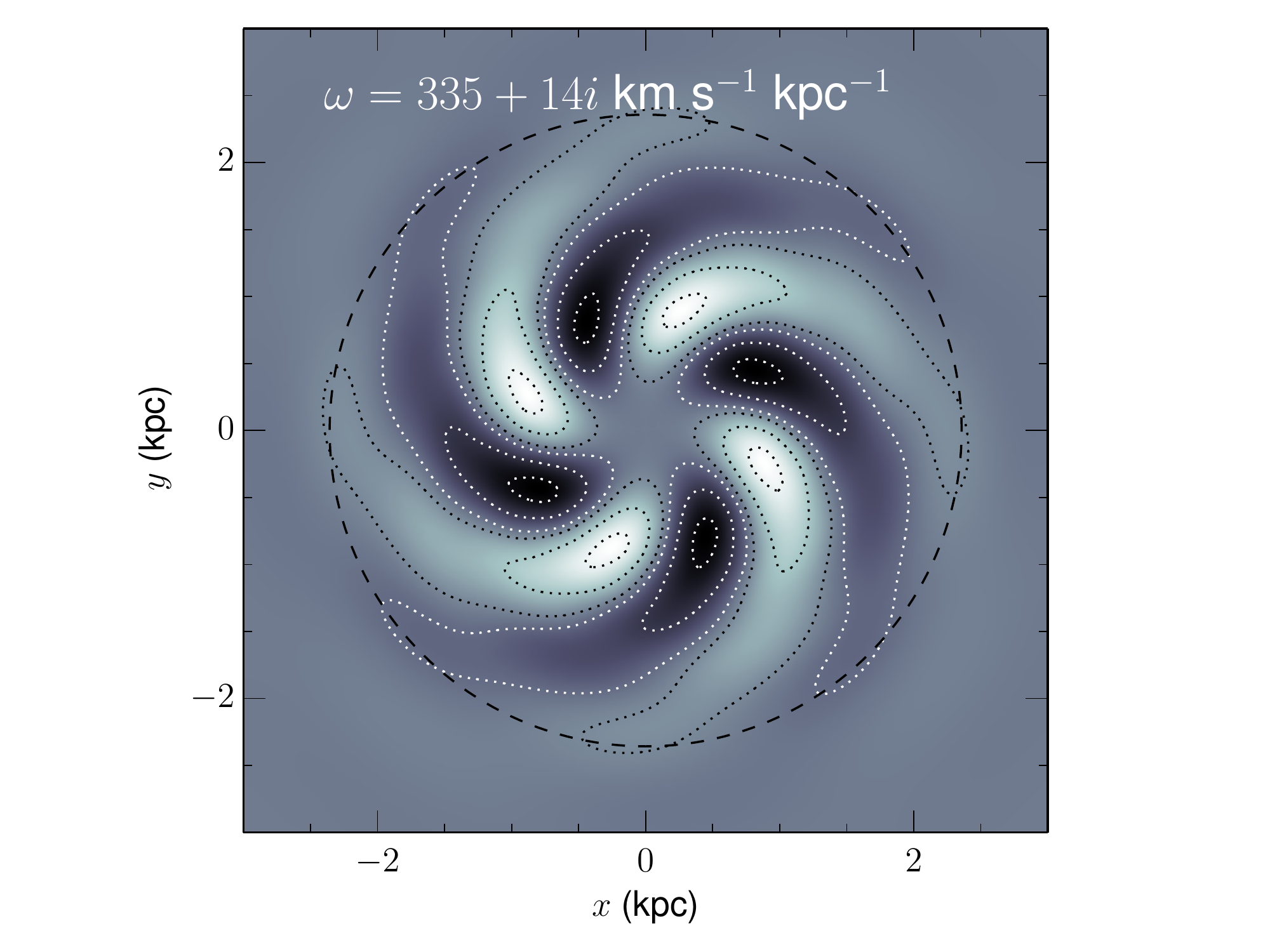}
\includegraphics[trim=55 12 95 10,clip,width=0.323\textwidth]{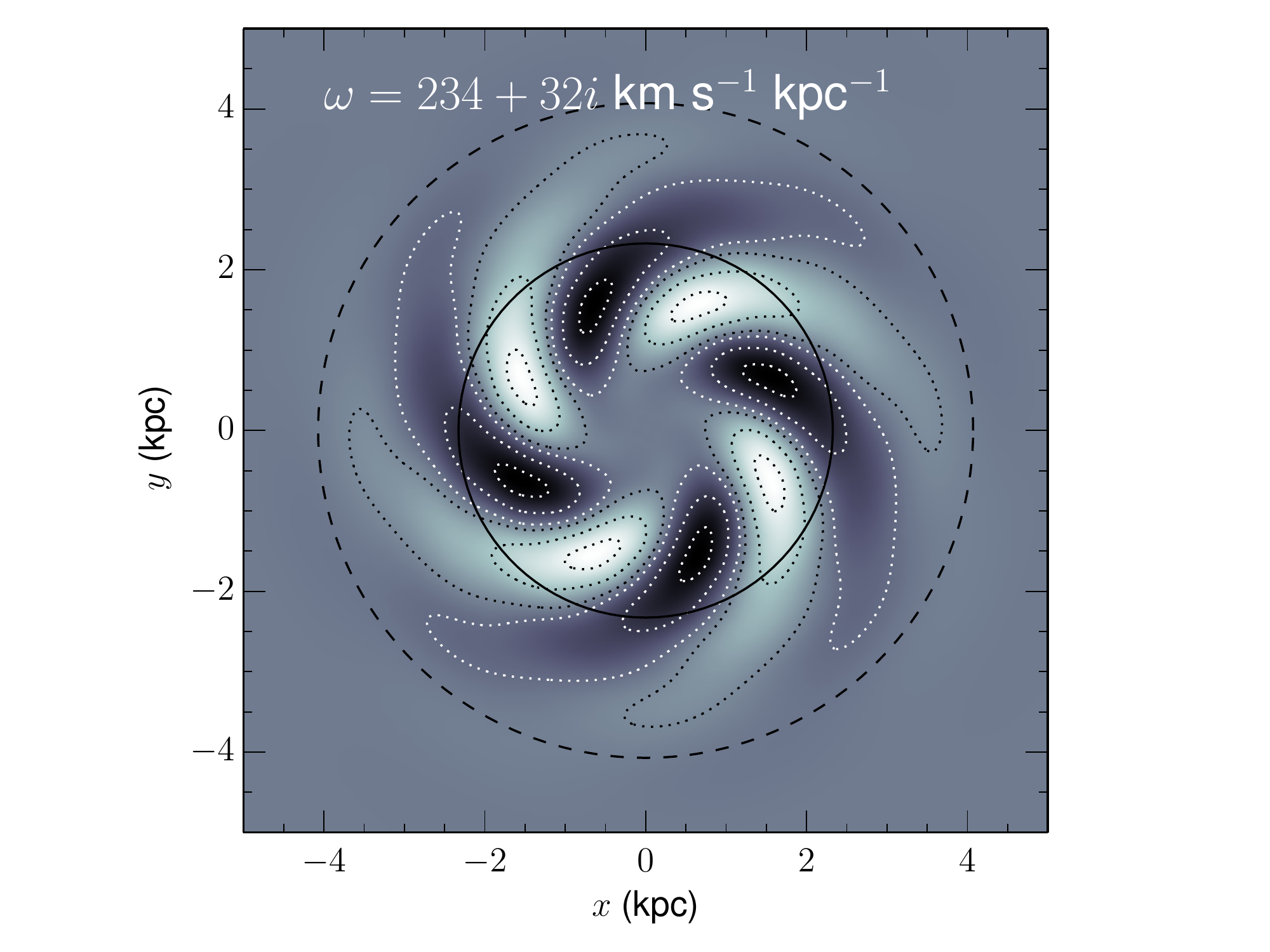}
\includegraphics[trim=55 12 95 10,clip,width=0.323\textwidth]{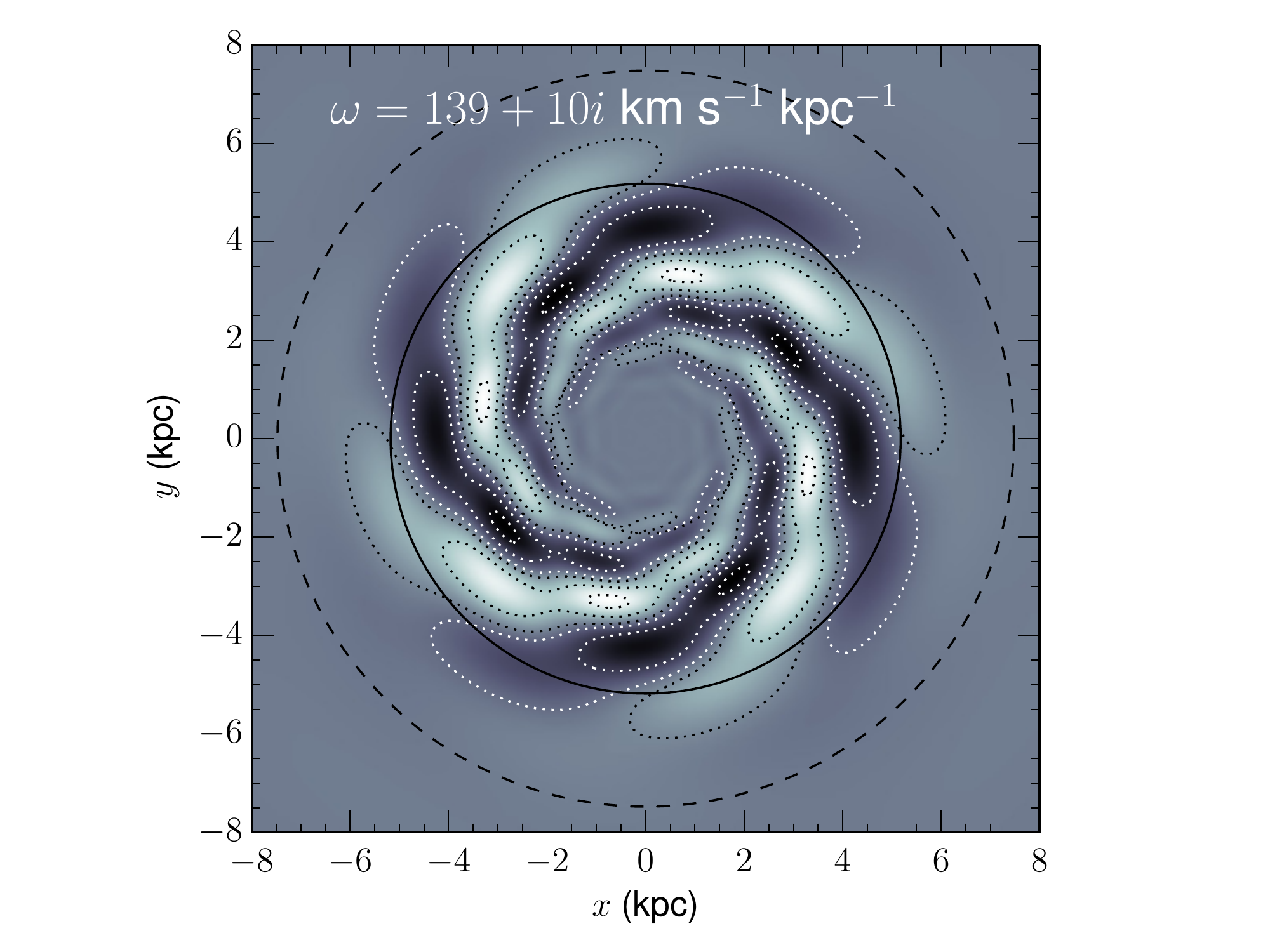}
\caption{Surface density of the most prominent $m=4$ eigenmodes of the
  ungrooved cored exponential disc model, labeled by their complex
  frequency $\omega$.  The corotation radius is indicated in a thin
  full line and the outer Lindblad resonance in a thin dashed line.
  \label{fig:mode_4.pdf}}
\end{figure*}

In the top left panel of Fig. \ref{fig:freqs4.pdf}, we show the $m=4$
mode spectrum of the ungrooved exponential disc model in the complex
frequency plane. It closely matches that presented in \citet{jalali07}
but since the latter was computed for an exponential disc without a
central hole we will not attempt a quantitative comparison. The
density distributions of three eigenmodes are presented in
Fig. \ref{fig:mode_4.pdf}. The fastest rotating modes, with
$\omega_{\rm real} \gtrsim 300$~km~s$^{-1}$~kpc$^{-1}$, lack a CR and
are confined within their OLR radius. The leftmost mode in
Fig. \ref{fig:mode_4.pdf} serves as an example. The second series of
modes, between $\omega_{\rm real} \sim 160$~km~s$^{-1}$~kpc$^{-1}$ and
$\omega_{\rm real} \sim 240$~km~s$^{-1}$~kpc$^{-1}$, comprises modes
that rotate sufficiently slowly to have a CR but fast enough to avoid
having an ILR. Their growth rate drops to zero for modes close to
having an ILR. The mode shown in the middle panel of
Fig. \ref{fig:mode_4.pdf} belongs to this series. The modes with
frequencies between $\omega_{\rm real} \sim
100$~km~s$^{-1}$~kpc$^{-1}$ and $\omega_{\rm real} \sim
150$~km~s$^{-1}$~kpc$^{-1}$ all have an ILR and have a density
distribution that exists mainly between the ILR and the CR radius,
cf. the right panel of Fig. \ref{fig:mode_4.pdf}.

The fastest rotating $m=4$ mode has an OLR radius of 2.36~kpc and,
while it is far from being a strong mode, it could be capable of
carving a groove in a very reactive region of the phase space of this
disc galaxy model (see section \ref{m2modes}), making it potentially
crucial for the existence of the second-generation of $m=2$
eigenmodes in this disc galaxy model.

As was the case for the two-armed modes, a groove has a very
destabilizing effect on the four-armed patterns as evidenced by the
eigenmode spectra shown in Fig. \ref{fig:freqs4.pdf}. For $J_{\rm
  groove} \lesssim 300$~kpc~km~s$^{-1}$, the $m=4$ eigenmode spectrum
is dominated by two modes which are the analogs of the
``high-frequency'' and ``low-frequency'' modes we encountered in the
$m=2$ spectra. As $J_{\rm groove}$ is increased, a
``medium-frequency'' mode quickly gains strength and overtakes the
``high-frequency'' and ``low-frequency'' modes. Its frequency shifts
upwards with increasing $J_{\rm groove}$ until, for $J_{\rm groove}
\gtrsim 300$~kpc~km~s$^{-1}$, it sits firmly within the groove. For
still higher, $J_{\rm groove}$ values, the $m=4$ eigenmode spectrum is
dominated by a ``low-frequency'' and a ``medium-frequency/groove''
mode.

\begin{figure*}
\includegraphics[trim=55 12 95 10,clip,width=0.323\textwidth]{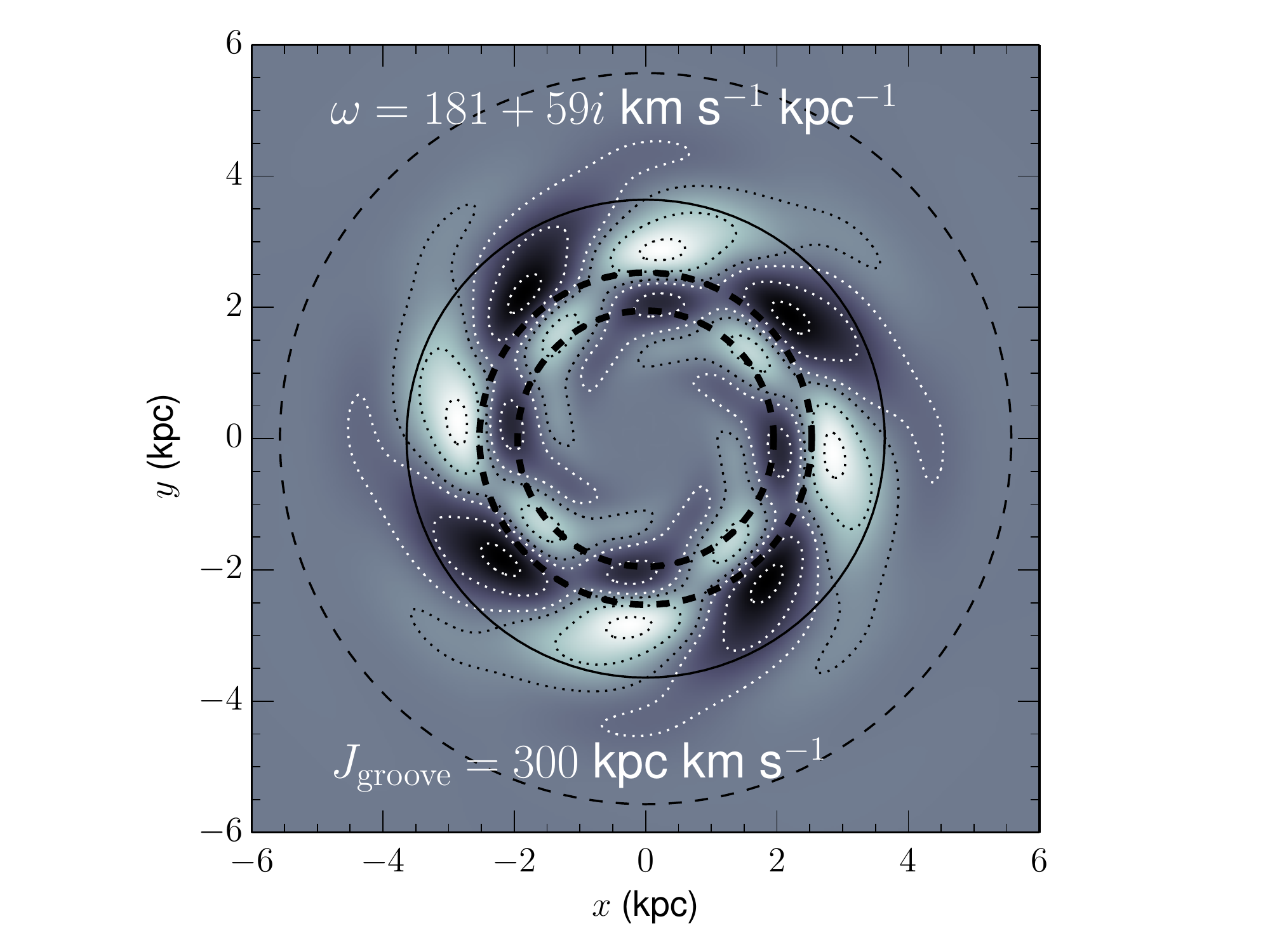}
\includegraphics[trim=55 12 95 10,clip,width=0.323\textwidth]{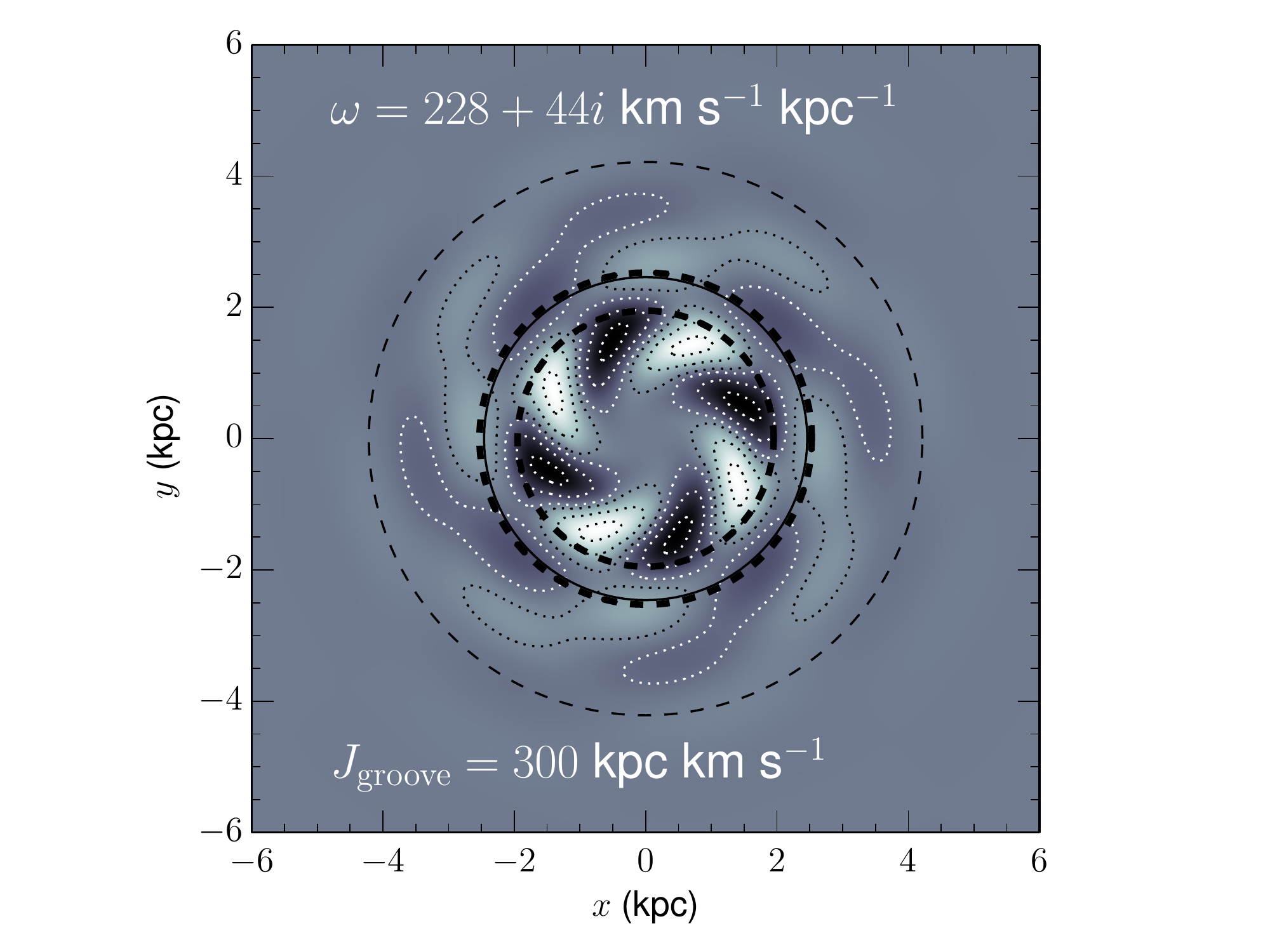}
\includegraphics[trim=55 12 95 10,clip,width=0.323\textwidth]{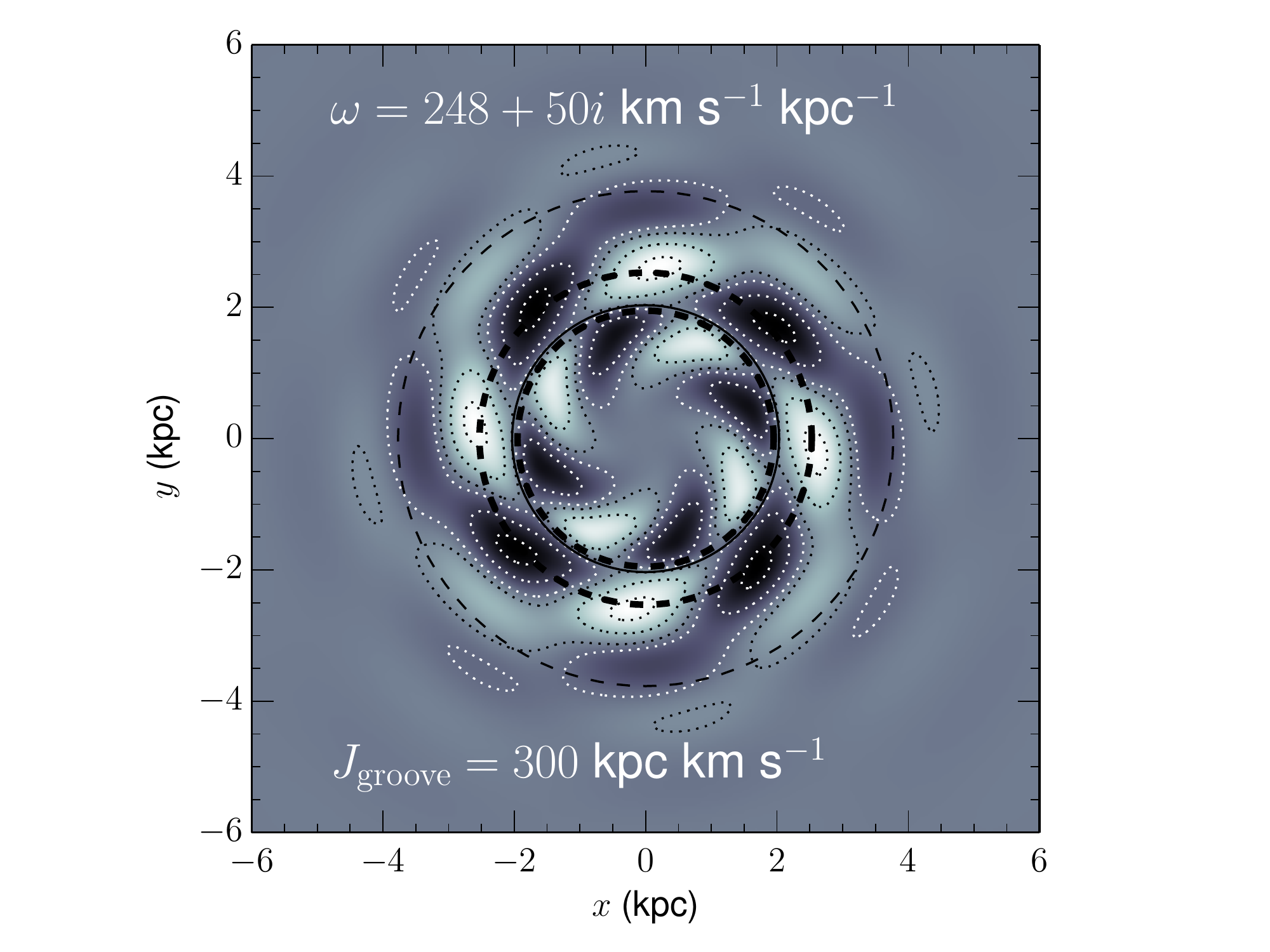}
\caption{Surface density of the most prominent $m=4$ eigenmodes of the
  cored exponential disc model with a groove at $J_{\rm groove} =
  300$~kpc~km~s$^{-1}$, labeled by their complex frequency $\omega$.
  The groove edges are indicated in thick dashed lines, the corotation
  radius in a thin full line, and the outer Lindblad resonance in a
  thin dashed line.
  \label{fig:mode_4_300.pdf}}
\end{figure*}

In Fig. \ref{fig:mode_4_300.pdf}, we show the density distributions of
the $m=4$ low-frequency mode (left panel), the medium-frequency mode
(middle panel), and the high-frequency mode (right panel) for the
cored exponential disc model with a groove at $J_{\rm groove} =
300$~kpc~km~s$^{-1}$. These show the same radial behaviour as their
two-armed analogs. The medium-frequency modes are limited to the disc
region inside the groove's inner edge while the low-frequency and
high-frequency modes have radial nodes at the groove's outer and inner
edge, respectively. Given this morphological similarity between the
two- and four-armed modes, their origin, in terms of wave dynamics, is
most likely also similar.

Using $N$-body simulations, \citet{sellwood89} found the dominant
four-armed mode to have a CR radius coincident with the groove and to
be accompanied by a small number of other modes. These observations
are in qualitative agreement with ours for a groove in the
intermediate $J_{\rm groove}$-range, where the medium-frequency mode
dominates.

\section{Groove shape} \label{grooveshape}
\begin{figure*}
\includegraphics[trim=20 15 2 13,clip,width=0.32\textwidth]{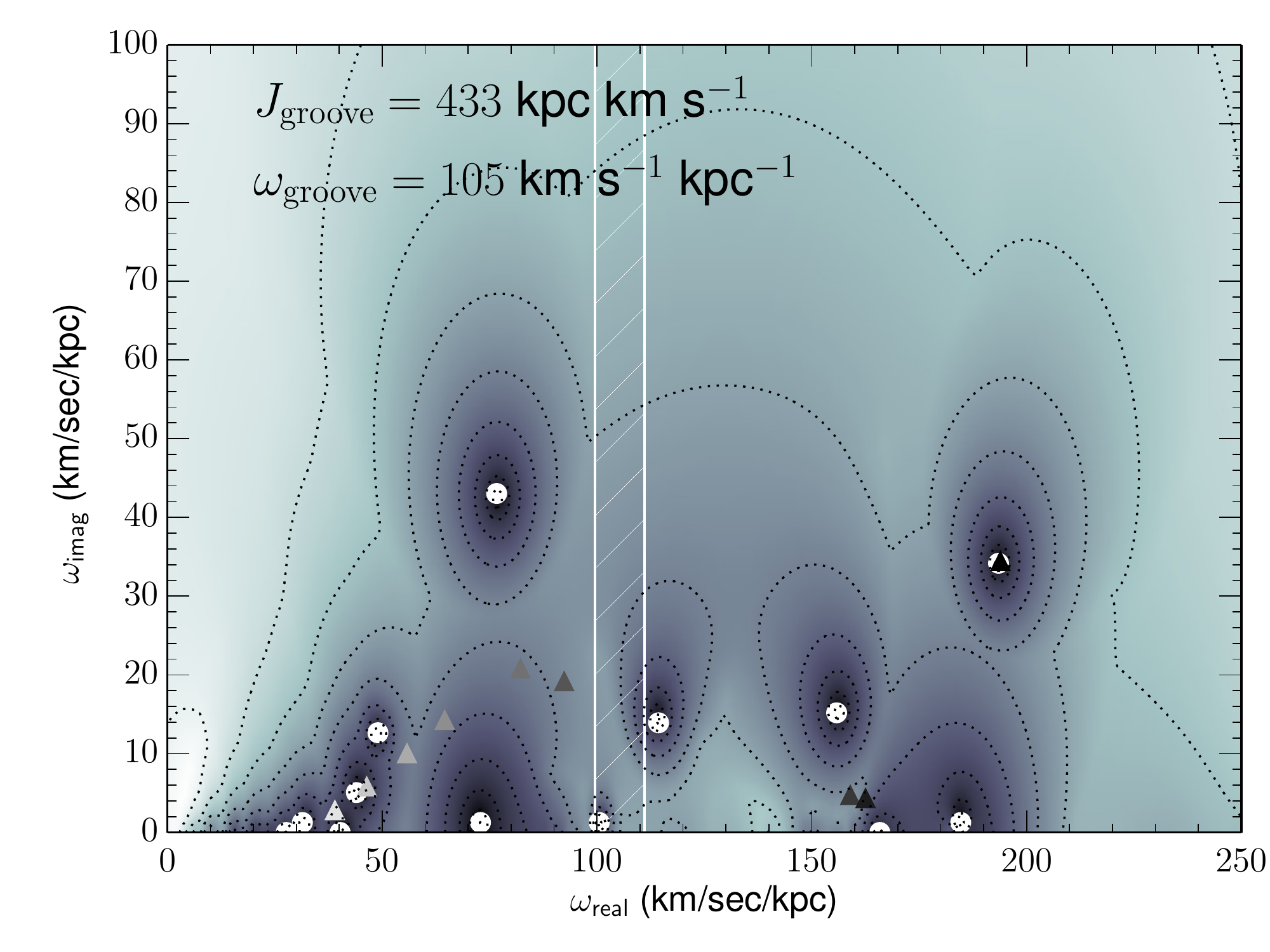}
\includegraphics[trim=20 15 2 13,clip,width=0.32\textwidth]{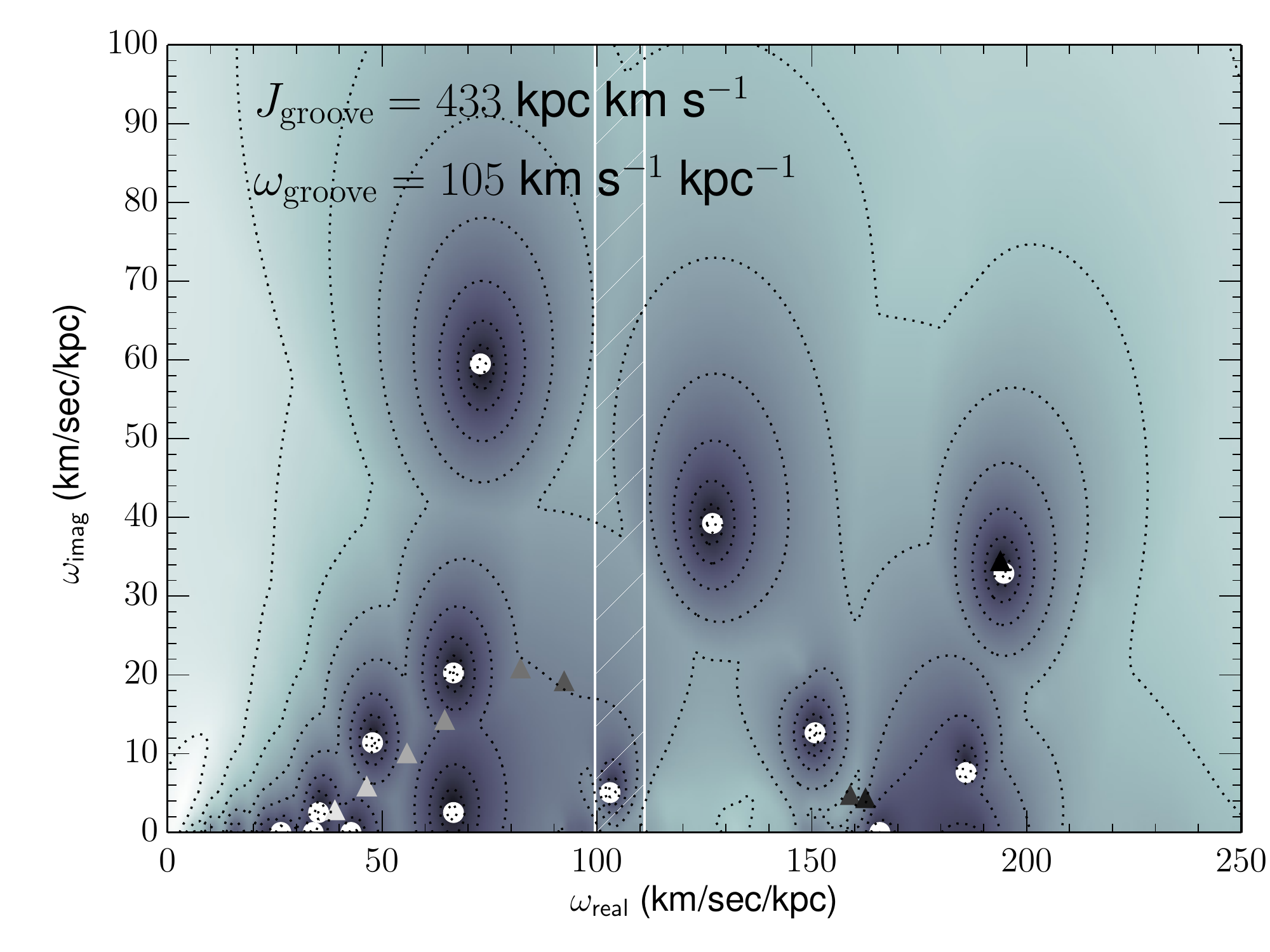}
\includegraphics[trim=20 15 2 13,clip,width=0.32\textwidth]{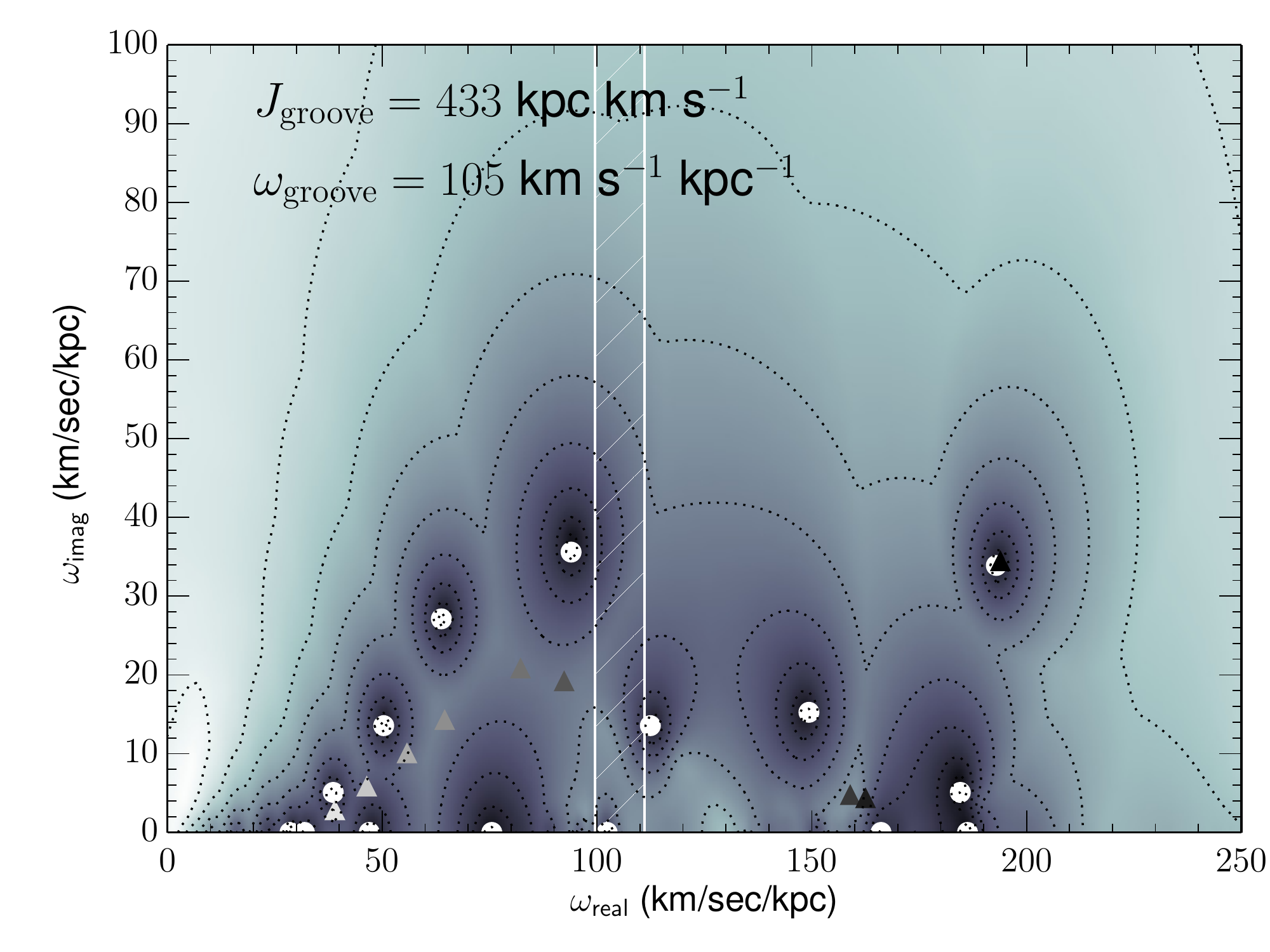}
\caption{The $m=2$ mode spectrum in the complex frequency plane of
  cored exponential disc models with a phase-space groove at $J_{\rm
    groove} = 433$~kpc~km~s$^{-1}$. Left panel:~a groove without
  compensating ridges. Middle panel:~a groove with two compensating
  ridges. Right panel:~a groove with one compensating ridge but half
  the width ($w_J=30$~kpc~km~s$^{-1}$). The white hatched region in
  each panel, centered on the frequency $\omega_{\rm groove}$,
  indicates the locus of the modes that corotate with stars on
  circular orbits inside the groove. The colored triangles indicate
  the position of the $m=2$ eigenmodes of the original cored
  exponential disc model.
 \label{fig:freqs4ridge.pdf}}
\end{figure*}

We investigated how the shape of the groove affects the mode spectrum
by calculating the latter, on the one hand, for a $J_{\rm groove} =
433$~kpc~km~s$^{-1}$ groove with no high-$J$ ridge and, on the other
hand, for the same groove but with both a compensating low-$J$ and a
high-$J$ ridge. Evidently, a groove without ridges does not conserve
the total stellar mass and therefore has a visible impact on the
model's velocity moments. To be specific:~it significantly increases
Toomre's $Q$ around the groove radius. The groove with two ridges does
conserve stellar mass and, consequently, has a much weaker effect on
the kinematics.

As can be seen in Fig. \ref{fig:freqs4ridge.pdf}, the presence or
absence of ridges has a strong effect on the mode spectrum. A groove
without ridges produces a strong medium-frequency mode but there is no
low-frequency mode and the high-frequency mode is very weak. A groove
with two ridges produces both the medium-frequency and a strong
high-frequency mode but not the low-frequency mode. From this
experiment, one can conclude that the existence of the
medium-frequency mode only depends on the presence of the groove and
{\em not} on the presence of the ridges. The ridges do affect the
interference patterns producing the low- and high-frequency modes.

We also experimented with a groove at $J_{\rm groove} =
433$~kpc~km~s$^{-1}$ with one compensating ridge but with half the
width ($w_J=30$~kpc~km~s$^{-1}$) as was used thus far. The eigenmode
spectrum of this model is shown in the right panel of
Fig. \ref{fig:freqs4ridge.pdf} and can be directly compared with the
corresponding panel of Fig. \ref{fig:freqs.pdf}. The spectra look very
much alike except for the fact that the modes associated with the
narrower groove have slightly lower growth rates. The growth rate of
the high-frequency mode seems particularly diminished. Therefore,
within the range of groove widths investigated here, the width of the
groove doesn't seem to significantly affect the position of the modes
although it does influence their exponentiation timescales.

\section{Conclusions} \label{conc}

Using {\sc pyStab}, a linear mode analysis computer code, we have
computed the $m=2$ and $m=4$ eigenmode spectra of a cored razor-thin
exponential disc galaxy model embedded in a logarithmic potential, as
introduced by \citet{jalali05}, and found excellent agreement with
previous authors \citep{jalali05,jalali07,omurkanov14}. 

We investigated how a phase-space groove at a fixed angular momentum
affects these eigenmode spectra. Our main conclusion is that a groove
has an impressive impact on the linear stability properties of a disc
galaxy. Depending on where in phase space the groove is carved,
completely new eigenmodes come into existence. We have shown that the
fastest rotating $m=2$ and $m=4$ modes of the ungrooved exponential
disc model have OLR radii in a very responsive part of the disc's
phase space and, if grooves are carved there, they would generate
vigorously growing new eigenmodes.

Using the spatial density distribution of these modes, we have
attempted to shed some light on their origin in terms of interference
patterns of traveling waves. The intermediate-frequency modes are
interference patterns that arise between the galaxy center and the
inner edge of the groove. The low- and high-frequency modes appear to
have a radial node anchored to the groove's outer and inner edge,
respectively. The very slowly growing modes with a CR radius
coincident with the groove have complex density distributions,
sometimes even with a partially leading spiral shape.  These modes
  may be related to the spiral instabilities discussed in
  \citet{poly04}. There, the author shows that if phase-space regions
  exist where the so-called Lynden-Bell derivative of the DF (a
  constrained angular-momentum derivative) is negative, unstable
  spiral modes are triggered. A narrow phase-space groove may be just
  such a region.  

Thus, we confirm the hypothesis made by previous authors
\citep{sellwood89,sellwood91} that spiral patterns beget new spiral
patterns by carving grooves in phase space at successively larger
radii. Because we are using first-order perturbation theory to
construct and compare the eigenmode spectra of grooved and ungrooved
models, we can rule out two of the possible origins of the wave
patterns observed in simulated grooved disc galaxies as suggested by
\citet{sellwood89}:~these patterns are not intrinsic modes of the
original, ungrooved disc and they are not due to non-linear mode
coupling. We do confirm the third possibility raised by these
authors:~they are true eigenmodes, particular to the grooved disc.

\section{Acknowledgements}

The authors wish to thank V. P. Debattista (UCLAN, UK) for his helpful
comments on an earlier draft of this manuscript.  We also thank
  our anonymous referee for his/her helpful remarks.

\bibliographystyle{mn2e} 
\bibliography{manuscript}

\label{lastpage}

\end{document}